\documentclass{aa}
\bibpunct{(}{)}{;}{a}{}{,} 
\usepackage[varg]{txfonts}
\usepackage{hyperref}
\usepackage[dvipsnames]{xcolor}
\hypersetup{
  colorlinks,
  linkcolor={blue!90!black},
  citecolor={blue!90!black},
  urlcolor={blue}
}
\usepackage{euclid}

\usepackage[bottom]{footmisc}
\newcommand{\LC}[1]{\textcolor{black}{#1}}
\newcommand{\RR}[1]{\textcolor{black}{#1}}

\newcommand{\orcid}[1]{\protect\href{https://orcid.org/#1}{\protect\includegraphics[width=8pt]{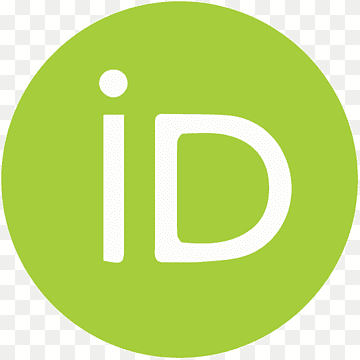}}}

\begin{document}

\title{The PAU Survey and \Euclid: Improving broadband photometric redshifts with multi-task learning\thanks{This paper is published on behalf of the Euclid Consortium}}

\author{L.~Cabayol\orcid{0000-0002-9498-2572}$^{1,2}$\thanks{\email{lcabayol@ifae.es}}, M.~Eriksen\orcid{0000-0003-0601-0990}$^{1,2}$, J.~Carretero\orcid{0000-0002-3130-0204}$^{1,2}$, R.~Casas\orcid{0000-0002-8165-5601}$^{3,4}$, F.J.~Castander\orcid{0000-0001-7316-4573}$^{4,3}$, E.~Fern\'andez$^{1}$, J.~Garcia-Bellido$^{5}$, E.~Gaztanaga$^{3,4}$, H.~Hildebrandt\orcid{0000-0002-9814-3338}$^{6}$, H.~Hoekstra\orcid{0000-0002-0641-3231}$^{7}$
% ...
}

\institute{Affiliations are listed at the end of the paper.}
   \date{Received \today; accepted XX.XX.XXXX}

 \abstract{Current and future imaging surveys require  photometric redshifts (photo-$z$s) to be estimated for millions of galaxies. Improving the photo-$z$ quality is a major challenge but is needed to advance our understanding of cosmology. In this paper we explore how the synergies between narrow-band photometric data and large imaging surveys can be exploited to improve broadband photometric redshifts. We used a multi-task learning (MTL) network to improve broadband photo-$z$ estimates by simultaneously predicting the broadband photo-$z$ and the narrow-band photometry from the broadband photometry. The narrow-band photometry is only required in the training field, which also enables better photo-$z$ predictions for the galaxies without narrow-band photometry in the wide field. This technique was tested with data from the Physics of the Accelerating Universe Survey (PAUS) in the COSMOS field. We find that the method predicts photo-$z$s that are 13\% more precise down to magnitude $i_{\rm AB} < 23$; the outlier rate is also 40\% lower when compared to the baseline network.
 Furthermore, MTL reduces the photo-$z$ bias for high-redshift galaxies, improving the redshift distributions for tomographic bins with $z>1$. Applying this technique to deeper samples is crucial for future surveys such as \Euclid or LSST. For simulated data, training on a sample with $i_{\rm AB} <23$, the method reduces the photo-$z$ scatter by 16\% for all galaxies with $i_{\rm AB}<25$. We also studied the effects of extending the training sample with photometric galaxies using PAUS high-precision photo-$z$s, which reduces the photo-$z$ scatter by 20\% in the COSMOS field.}

\keywords{Galaxies: photometry, Methods: data analysis, Surveys, Cosmology: observations}
             
\titlerunning{Improving broadband photometric redshifts with multi-task learning}
\authorrunning{Cabayol et al.}
\maketitle
   
\section{Introduction}
Over the last few decades, multi-band wide imaging surveys have been driving discoveries, demonstrating the power of large datasets to enable precision cosmology. Obtaining precise photometric redshifts is crucial for exploiting large galaxy imaging surveys \citep{Salvato}, and they are a limiting factor in the accuracy of cosmology measurements that use galaxies \citep{cosmo_photoz}. Current and upcoming imaging surveys such as the Dark Energy Survey \citep[DES;][]{DES}, the Kilo-Degree Survey \citep[KiDS;][]{kids}, \Euclid \citep[][]{Euclid}, and the Rubin Observatory Legacy Survey of Space and Time \citep[LSST;][]{LSST} critically depend on robust redshift estimates to obtain reliable science results \citep[][]{photoz_req}. 

With larger imaging surveys (as the quality and number of photometric observations increase), the photo-$z$ performance requirements, both in terms of bias and precision, have become increasingly stringent in response to a need to reduce the uncertainties in the science measurements.  As an example, the analysis of the first year of DES data (DES Y1) had a photo-$z$ precision requirement $\sigma_{z_\mathrm{p}-z_\mathrm{s}}<0.12$ \citep[][]{DES_sanchez}, with $\sigma_{z_\mathrm{p}-z_\mathrm{s}}$ being the standard deviation of the residuals between the photometric redshift, $ z_{\rm p}$, and the spectroscopic redshift, $z_{\rm s}$ (as a proxy of the true redshift). In order to exploit the constraining power of LSST, it is required that the mean fractional photo-$z$ bias $|\langle \Delta z\rangle| < 0.003$, with $\Delta z := (z_{\rm p}- z_{\rm s}) / (1+z_{\rm s})$,  and  the scaled photo-$z$ scatter $\sigma_{\Delta z}$ < 0.02 \citep[][]{Rubin_requirements}, which corresponds to photo-$z$s that are around three times more precise  than in DES Y1. Similarly, for \Euclid, the scaled photo-$z$ bias is required to be below 0.002 and  $\sigma_{\Delta z} < 0.05$ \citep[][]{Euclid}.

The increasingly stringent requirements on the photo-$z$ measurements have triggered extensive investigation efforts dedicated to improving photo-$z$ estimation methodologies. Therefore, there are many different photo-$z$ codes, which can be classified into two main approaches: the so-called template-fitting methods (e.g. \texttt{LePhare}: \citealt{Lephare}, \texttt{BPZ}: \citealt{Bpz}, and \texttt{ZEBRA}:  \citealt{Zebra}) and data-driven (machine-learning) methods  (e.g. \texttt{ANNz}: \citealt{ANNz}, \texttt{ANNz2}: \citealt{ANNz2}, \texttt{tpz}: \citealt{tpz}, \texttt{Skynet}: \citealt{skynet}, and  \texttt{spiderZ}: \citealt{spiderz}). These methods commonly only use the measured photometry to produce photo-$z$ estimates. Furthermore, there is a wealth of techniques for improving the photo-$z$ performance, such as including galaxy morphology \citep[][]{morphoz}, using Gaussian processes \citep[][]{GPz,Delight}, implementing `pseudo-labelling' semi-supervised approaches to determine the underlying structure of the data (Humphrey et al. in prep.), and directly predicting the photo-$z$ from astronomical images \citep[][]{Pasquet2,Pasquet,Chong_img}. 

The broadband photo-$z$ performance is limited by the resolution and the wavelength coverage provided by the photometric filters. Narrow-band photometric surveys are in between spectroscopy and broadband photometry \citep{Jpas,photoz-Marti,photoz-Martin}. They are imaging surveys with a higher wavelength resolution than broadband surveys, but they typically cover smaller sky areas due to the increased telescope time needed to cover the same wavelength range. In this paper we use multi-task learning \citep[MTL;][]{MTL} and narrow-band data to improve broadband photo-$z$ estimates. Multi-task learning is a machine-learning methodology in which the model benefits from predicting multiple related tasks together, for example a network that predicts the animal type (e.g. elephant, dog, dolphin, or unicorn) and its weight. In this example, the network learns the correlations between each animal class and  how heavy they are (e.g. an elephant is heavier than a dog), and such correlations are used to improve the final predictions in both tasks. 

In astronomy, data that could be helpful for improving the photo-$z$ performance often exist, for example photometry in several bands. However, such data are not always available for the complete wide field, preventing us from using it. With MTL, we can utilise these data to improve the photo-$z$ predictions without explicitly providing them as input. Particularly, we implemented an MTL neural network that predicts the photo-$z$ and the narrow-band photometry of a galaxy from its broadband photometry. The narrow-band data are used to provide ground-truth labels to train the auxiliary task of reconstructing the narrow-band photometry \citep{mtl_auxiliarytask}. Therefore, we only need it to train the network, and we can evaluate the photo-$z$ of any galaxy with only its broadband photometry. In this way, the data available in certain fields can be exploited to improve the photo-$z$ estimations in other fields. 

We tested the method with data from the Physics of the Accelerating Universe Survey (PAUS).\ It is a narrow-band imaging survey carried out with the PAUCam instrument \citep{PAUCAM_Francisco, PAUCam-Padilla,PAUcam_Padilla_2019},  a camera with 40 narrow bands that cover the optical spectrum \citep{Casas2016}. The method could also be applied to other narrow-band surveys such as the Javalambre Physics of the Accelerating Universe Survey \citep[J-PAS;][]{Jpas}.

The paper is structured as follows. In Sect.\,\ref{sec:data} we present the data used throughout the paper. Section \ref{sec:method} introduces MTL and the method developed and tested in this work. In Sect.\,\ref{sec:cosmosres} we show the performance of the photo-$z$ method in the COSMOS field, including bias, scatter, outliers, and the photo-$z$ distributions. The performance on a deeper galaxy sample is tested in Sect.\,\ref{sec:depth_tests} using simulated galaxies. Finally, we use self-organising maps (SOMs) to explore the photo-$z$ distribution of COSMOS galaxies in colour space (Sect.\,\ref{sec:colourspace}) and to gain a better understanding of the underlying mechanism of our method (Sect.\,\ref{sec:MTLmech}). 


\section{Data}
\label{sec:data}
In this section we  present the PAUS data (Sect.\,\ref{sec:data:paus_data}) and the photometric redshift galaxy sample (Sect.\,\ref{sec:data:samppz}). The broadband data and the spectroscopic sample are introduced in Sect.\,\ref{sec:data:bb_data} and Sect.\,\ref{sec:data:sampzs}, respectively, while Sect.\,\ref{sec:data:mocks} shows the galaxy simulations used in the paper.

\subsection{PAUS data}
\label{sec:data:paus_data}
PAUS data are taken at the \textit{William Herschel} Telescope (WHT), at the Observatorio del Roque de los Muchachos in La Palma (Canary Islands). Images are taken  with the PAUCam instrument \citep{PAUCAM_Francisco,PAUcam_Padilla_2019}, an optical camera equipped with 40 narrow bands covering a wavelength range from 4500\,\AA\  to 8500\,\AA\ \citep{Casas2016}. The narrow-band filters have a 130\,\AA\ full width at half maximum and a separation between consecutive bands of 100\,\AA. They are mounted in five trays with eight filters per tray that can be exchanged and placed in front of the CCDs. The narrow-band filter set effectively provides a high-resolution photometric spectrum ($R \sim 50$).  This allows PAUS to measure high-precision photo-$z$s to faint magnitudes ($i_{\rm AB}<23$) while covering a large sky area \citep{photoz-Marti}. In this work we use the full pass-band filter information\footnote{Similar filter functions to the ones used in the paper are available at the PAUS website \url{www.pausurvey.org}}.

With a template-fitting algorithm, PAUS reaches a photo\nobreakdash-$z$ precision $\sigma_z / (1+z) = 0.0035$ for the best 50\% of the sample \citep[][]{photoz-Martin}.  Similar precision is obtained with \texttt{Delight} \citep[][]{Delight}, a hybrid template-machine-learning photometric redshift algorithm that uses Gaussian processes. The PAUS photo-$z$ precision was improved further with a deep-learning algorithm that reduces the scatter by 50\% compared to the template-fitting method in \citet{deepz}. Furthermore, with a combination of PAUS narrow bands and 26 broad and intermediate bands covering the UV, visible, and near infrared spectral range, \citet{PAUSCOSMOS_alex} presented an unprecedented precise photo-$z$ catalogue for COSMOS \citep[][]{cosmos} with $\sigma_z / (1+z) = 0.0049$ for galaxies with $i_{\rm AB}<23$.  The excellent PAUS photo-$z$ precision enables studies of intrinsic galaxy alignments and three-dimensional galaxy clustering \citep[][]{IA_john}, as well as determining galaxy properties \citep[][]{Luca_ABC} and measuring the D4000\,\AA\ spectral break (Renard et al. in prep.). 

PAUS has been observing since the 2015B semester, and as of 2021B, PAUS has taken data during 160 nights. It partially covers the Canada-France-Hawaii Telescope Legacy Survey (CFHTLS) fields\footnote{\url{http://www.cfht.hawaii.edu/Science/CFHTLS\_Y\_WIRCam\\/cfhtlsdeepwidefields.html}} W1, W2, and W3, as well as the full COSMOS field\footnote{\url{http://cosmos.astro.caltech.edu/}}. In the W2 field, so far PAUS has observed in the overlapping region with the GAMA 9-hour field\footnote{\url{https://www.astro.ljmu.ac.uk/~ikb/research/gama\_fields/}} (G09). Currently, PAUS data have a 40 narrow-band coverage of 10\,deg$^2$ in each of W1 and G09,  20\,deg$^2$ in W3, and 2\,deg$^2$ in COSMOS. The PAUS data are stored at the Port d'Informaci\'o Cient\'ifica (PIC), where the data are processed and distributed \citep{Tonello}. This paper uses data from the COSMOS field \citep[][]{cosmos}, which were specifically taken in the semesters 2015B, 2016A, 2016B, and 2017B. The complete PAUS photometric catalogue in COSMOS comprises 64\,476 galaxies to $i_{\rm AB} < 23$ in 40 narrow-band filters. This corresponds to approximately 12.5 million galaxy observations (5 observations per galaxy and narrow-band filter). 

Two methods for extracting the galaxy photometry have been developed for PAUS: a forced aperture algorithm (\texttt{MEMBA}) and a deep-learning-based pipeline \citep[\texttt{Lumos};][]{bkgnet,lumos}. In this study we have found that the resulting photo-$z$ performance with both photometric approaches is very similar. In the COSMOS field, the parent detection  catalogue is provided by \citet{Laigle} and the photometry calibration is relative to the Sloan Digital Sky Survey (SDSS) stars (Castander et al. in prep.). A brief description of the photometric calibration can be found in \citet{photoz-Martin}. 

\subsection{Photometric redshift sample}
\label{sec:data:samppz}
Throughout the paper we also use the high-precision photometric redshifts from \citet[][PAUS+COSMOS hereafter]{PAUSCOSMOS_alex}. \LC{They were estimated with a template-fitting method modelling the spectral energy distributions (SEDs) as a linear combination of emission line and continuum templates to then compute the Bayes evidence by integrating over the linear combinations. In addition to the PAUS narrow bands, the PAUS+COSMOS catalogue uses 26  broad and intermediate bands covering the UV, visible, and near-infrared spectrum \citep[see Sect.\,2 in][for more details]{PAUSCOSMOS_alex}.} The PAUS+COSMOS photo-$z$s reach a precision of $\sigma_z / (1+z)  = 0.0036$ and $\sigma_z / (1+z) = 0.0049$ for galaxies at $i_{\rm AB} < 21$ and $i_{\rm AB} < 23$, respectively. These photo-$z$s are more precise and less biased than those from \citet{Laigle}, which use a combination of 30 broad-, intermediate-, and narrow-band filters.

\subsection{Broadband data}
\label{sec:data:bb_data}
The broadband data used in this paper are from \citet[][COSMOS2015 hereafter]{Laigle}, which includes the $u$-band from the Canada-France-Hawaii Telescope (CFHT)/MegaCam and the Subaru $BVriz$ filters. We carry out a spatial matching of COSMOS2015 and PAUS galaxies within $1^{\rm \prime\prime}$. Then, we apply a cut on magnitude $i_{\rm AB}<23$ and on redshift $z < 1.5$, which results in a catalogue with around 33\,000 galaxies of which approximately 9000 have spectroscopic redshifts. The redshift cut is prompted by the photo-$z$ distribution in the PAUS+COSMOS catalogue, with very few galaxies with  $z > 1.5$ (Fig. \ref{fig:cosmos_zbias}).

\subsection{Spectroscopic galaxy sample}
\label{sec:data:sampzs}

\begin{figure}
\includegraphics[width= 0.48\textwidth]{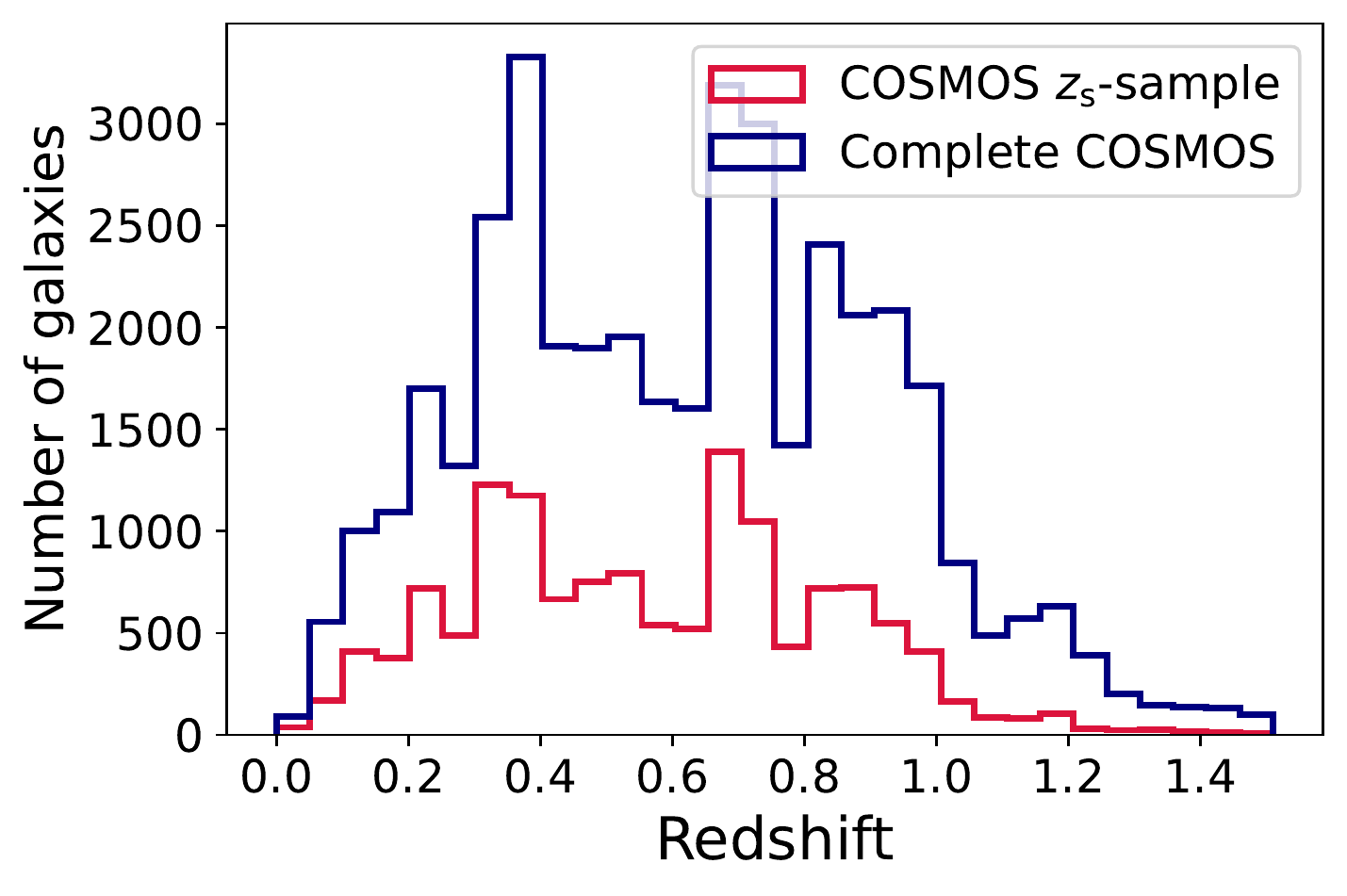}
\centering
\caption{Redshift distributions for the COSMOS spectroscopic sample (red line) and the full (spectroscopic and photo-$z$) COSMOS sample.}
\label{fig:cosmos_zdist}
\end{figure}

To train the neural network, one needs a galaxy catalogue with known redshifts. We used the zCOSMOS Data Release (DR) 3 bright spectroscopic data \citep{zcosmos}, which cover 1.7\,deg$^2$ of the COSMOS field. The catalogue covers a magnitude range of $15<i_{\rm AB}<23$ and a redshift range of $0.1<z<1.2$. We only keep redshifts with a confidence class ($conf$) of $3<conf<5$, which leads to a catalogue with  $\sim9400$ galaxies. We extended the spectroscopic sample with a compilation of 2693 redshifts from \citet{PAUSCOSMOS_alex}. This compilation includes redshifts from C3R2 DR1 and DR2 \citep{Masters1,Masters2}, 2dF \citep{Colless}, DEIMOS \citep{Deimos}, FMOS \citep{FMOS}, LRIS \citep{LRIS}, MOSFIRE \citep{MOSFIRE}, MUSE \citep{MUSE}, Magellan \citep{Magellan}, and VIS3COS \citep{VIS3COS}, with a quality cut to keep only those objects with a reliable measurement.

\RR{Figure \ref{fig:cosmos_zdist} shows the redshift distribution of the COSMOS spectroscopic sample (red) and the full PAUS sample in the COSMOS field (blue), where the redshift is defined as the spectroscopic redshift if this is available and as the PAUS+COSMOS photo-$z$ otherwise. Including the PAUS+COSMOS photo-$z$ is particularly relevant for galaxies with $z>1$, where there are very few spectroscopic measurements.}

\subsection{Galaxy mocks}
\label{sec:data:mocks}
\LC{In Sect. 5 we also use the Flagship galaxy simulations described
in Castander et al. (in prep.). The Flagship galaxy catalogue has
been developed to study the performance of the \Euclid mission.
The mock catalogue populates the halos detected in the \Euclid 
Flagship N-body simulation \citep{Potter2016}, which is a large 
 two trillion particles simulation on a box of 3780 $h^{-1}$Mpc, 
and a mass resolution of $m_p = 2.4 \times 10^9$ $h^{-1}$M$_{\odot}$.
The N-body simulation uses a cosmological model with parameters similar 
to the Planck 2015 cosmology \citep{Planck2015}. 
Halos are identified with the ROCKSTAR halo finder \citep{rockstar_Behroozi}. 
Galaxies are assigned to the halos using a hybrid halo occupation distribution 
and abundance matching technique similar to the one used for the MICE 
catalogues described in \citet{MICE_Carretero}.}
\LC{Galaxies are divided into central and satellites. 
Each halo contains a central and a number of satellites given by their halo occupation. 
Galaxies are also tagged in three colour types: blue, green, and red. 
The relative abundance of central and satellites as a function of colour type and 
absolute magnitude is constrained by the observed colour-magnitude distribution 
and the clustering as a function of colour at low redshift. 
At higher redshift only observed colour distributions are used. 
Each galaxy is assigned a SED, including its extinction, 
from the COSMOS SED library (e.g. \citealt{Ilbert2013}), which includes SED templates 
from \citet{templates_polleta} and additional blue templates 
from \citet{templates_bruzual}. In order to have a more continuous distribution 
of galaxy magnitudes and colours, the SED assigned to each galaxy is a linear combination 
of two consecutive templates in the COSMOS template library.}
\LC{Emission lines are then
added to the SED of each galaxy. The H$\alpha$ flux is computed from the rest-frame ultra-violet flux of each galaxy template following \citet{Halpha}. The H$\alpha$ fluxes are then re-adjusted to make them follow the \citet{pozzetti2016} model1 and model3 distributions. The H$\beta$ flux is computed from the H$\alpha$ flux assuming case B recombination \citep{Osterbrock}. }
\LC{The other emission line fluxes ([OII], [OIII], [NII], and [SII])  are computed 
following relations obtained from observed distributions. The emission line fluxes are added to the continuum assuming a Gaussian distribution of width given by the galaxy magnitude and the Faber-Jackson or Tully-Fisher relation.  Finally, the SEDs containing the emission lines are convolved with the filter transmission curves to produce the expected observed fluxes. This prescription is followed to generate both broad- and narrow-band photometry.}
\LC{The Flagship catalogue is a property of the
Euclid Consortium and is available at\ CosmoHub\footnote{https://cosmohub.pic.es} \citep{2017ehep.confE.488C,TALLADA2020100391}, a web application based on Hadoop to interactively distribute and explore massive cosmological datasets.}


\section{Multi-task neural network to improve broadband photo-{\it z}s}
\label{sec:method}
In this section we describe MTL (Sect.\,\ref{sec:method:MTL}) and present the networks and training procedures used throughout the paper (Sect.\,\ref{sec:method:arch}).

\subsection{Multi-task learning}
\label{sec:method:MTL}
Deep-learning algorithms consist of training a single or an ensemble of models to accurately perform a single task, for example predicting the redshift. Multi-task learning is a training methodology that aims to improve the performance on a single task by training the model on multiple related tasks simultaneously \citep{MTL}. One can think of MTL as a form of inductive transfer, where the knowledge that the network acquires from one task introduces an inductive bias to the model, making it prefer certain hypotheses over others. A simple pedagogical example is a network to classify cats and dogs. If we include a secondary task to classify the shape of the ears in, for example spiky or rounded, the network will make correlations between the ear shapes and the animal class, in such a way that the predicted ears shape will also affect the cat-dog classification. This kind of network has already been successfully applied in other fields, such as video processing \citep[][]{MTL-imageproc} and medical imaging \citep[][]{MTL-medical}, where in the latter case a single network is trained to segment six tissues in brain images, the pectoral muscle in breast images, and the coronary arteries. There are also successful implementations in astrophysics. Examples include, for example, \citet[][]{MTL_Lyalpha}, which characterises the strong HI Ly$\alpha$ absorption in quasar spectra simultaneously predicting the presence of strong HI absorption and the corresponding redshift $z_{\rm abs}$ and the HI column density. Also,  \citet[][]{MTL_photoz} describe \texttt{SHEEP}, a machine-learning pipeline for the
classification of galaxies, quasi-stellar objects, and stars from photometric data.
Broadly speaking, there are two types of MTL-network architectures, called soft- and hard-parameter sharing \citep[][]{MTL_survey}. In the former, each task has its parameters, which are regularised to be similar amongst tasks. For the latter, the hidden layers of the network are shared between tasks, while keeping task-specific layers separate. Hard-parameter sharing is the most common MTL architecture and it is the one used in this paper.

\subsection{Model architecture and training procedures}
\label{sec:method:arch}

\begin{figure}
\includegraphics[width= 0.47\textwidth]{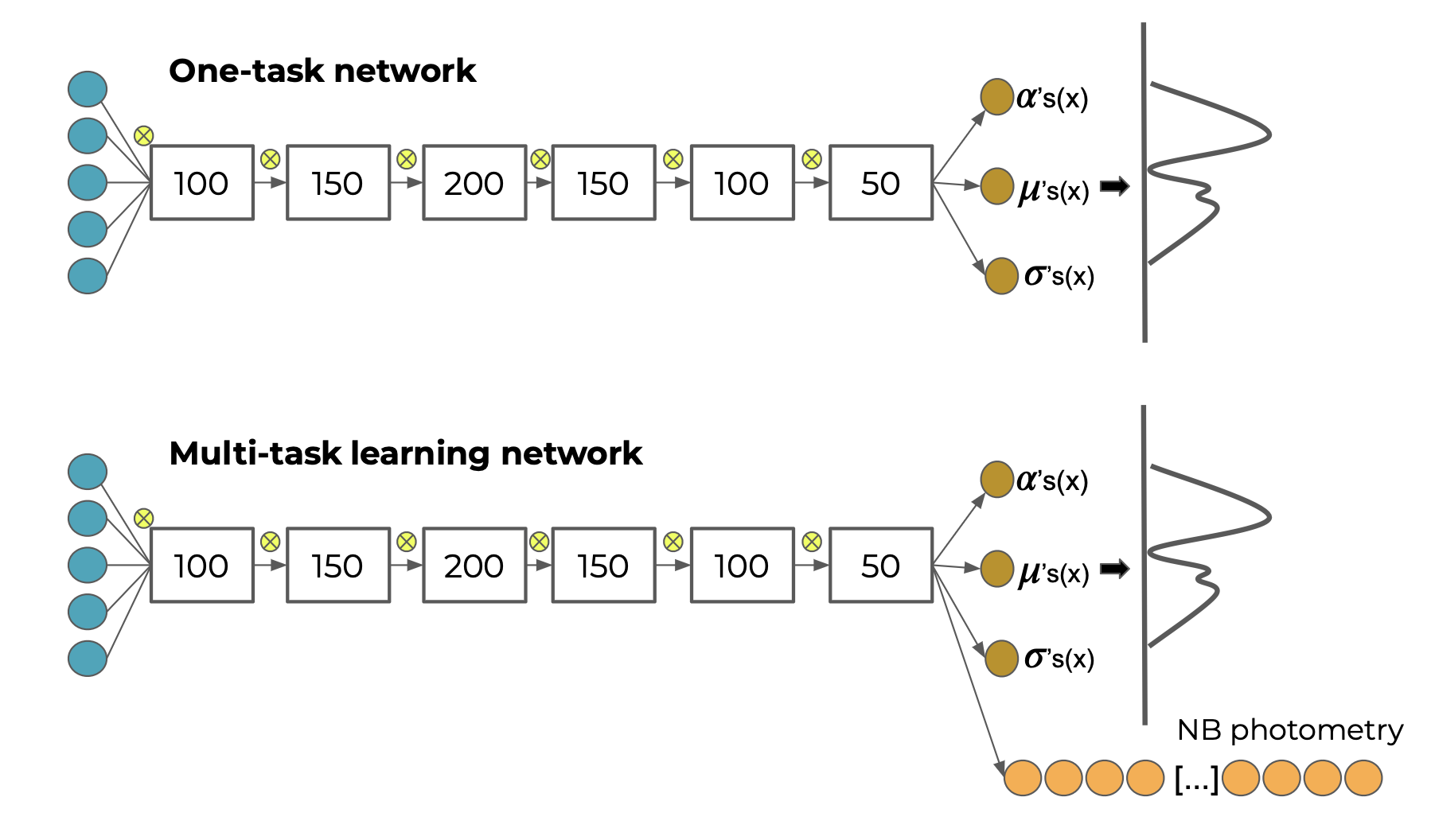}
\centering
\caption{\emph{Top:} Baseline network architecture. The input contains five colours that propagate through six fully connected layers. Each layer is followed by a dropout layer, which is represented by a yellow-crossed circle. \emph{Bottom:} MTL network. This builds on the baseline network and adds an extra output layer for the additional task of predicting the narrow-band photometry.}
\label{fig:architecture}
\end{figure}
\LC{In our analysis we used mixture density networks (MDNs) to predict the photo-$z$ probability distribution as a linear combination of $N$ independent Gaussians  \citep[][]{photoz_isanto,deepz}. The network predicts the mean and the standard deviation of $N$ distributions, together with $N$ additional mixing coefficients ($\alpha$) weighting the relative importance of each Gaussian component to the combined probability distribution, so that $\sum_{i=0}^{i=N}\alpha_{\rm i} = 1$.} 
 
 \LC{Figure \ref{fig:architecture} shows the two MDNs used in this paper, both of them predicting the photo-$z$ probability distribution $p(z)$ as the combination of three independent Gaussian distributions. The top panel presents the baseline network, a single-task network mapping the broadband photometry to the photometric redshifts. It concatenates six fully connected layers with parameters 5:300:500:1000:500:300:9, where the numbers correspond to the number of nodes in the layers. Therefore, the first contains five nodes, corresponding to the $uBVriz$ broadband colours. The last layer consists of nine output parameters  corresponding to the mean ($z$), the standard deviation ($\sigma_{\rm i}$), and the mixing coefficients, $\alpha$, of the three Gaussians building the $p(z)$.}  Each layer is followed by a 2\% dropout layer \citep[][]{dropout}, a regularisation method in which several nodes are randomly ignored during the training phase. 

The bottom panel in Fig.\,\ref{fig:architecture} represents the MTL network introduced in this paper, which includes the additional task of predicting the PAUS narrow-band photometry using a hard parameter-sharing architecture (Fig. \ref{fig:architecture}). The core architecture is the same as that of the baseline network (upper panel) but this network contains an extra output layer for the additional task of predicting the narrow-band photometry. 

\LC{The photo-$z$ loss function of both networks is the negative log-likelihood:
    \begin{equation}
        \mathcal{L}_{\rm z} \coloneqq   \sum_{\rm i=1}^{N} \left [  \log(\alpha_{\rm i}) - \frac{(z_{\rm i} - z_{\rm s})^2}{\sigma_{\rm i}^2} - 2\log{(\sigma_{\rm i})}\right]  \,.
        \label{eq:lossz}
    \end{equation} The ground-truth redshift labels are the spectroscopic redshifts ($z_{\rm s}$) as defined in Sect.\,\ref{sec:data:sampzs} and the summation is over the Gaussian components.} \RR{For some training configurations, we also used high-precision photo-$z$s (Sect.\,\ref{sec:data:samppz}) as ground-truth labels to extend the photo-$z$ training sample beyond the spectroscopic sample.}
    
    The MTL network enables including information from the galaxy SED, while extending the training sample to galaxies without spectroscopic redshift but with narrow-band photometry. The two tasks share internal representations when predicting the photo-$z$ and the narrow-band photometry simultaneously; thus, the non-spectroscopic galaxies indirectly affect the training of the photo-$z$ prediction.     
    
    The training of the narrow-band is addressed with a least absolute deviation loss function,  
    \begin{equation}
        \mathcal{L}_{\rm NB} \coloneqq  \frac{\sum_{i}\,\left|{\rm NB}^{\rm pred}_i - {\rm NB}^{\rm obs}_i\right|}{N-1}\,,
        \label{eq:lossNB}
    \end{equation} where ${\rm NB}^{\rm pred}_i$ and ${\rm NB}^{\rm obs}_i$ are the predicted and observed narrow-band colours in the $i$-th filter, respectively, and $N$ is the number of narrow bands. We also tested other alternatives, for example the mean-squared error, but this was hindering the network's convergence and we decided on the absolute-mean error. \LC{Another alternative was to predict the probability distribution of the narrow-band fluxes using a MDN as well, but this did not resulted in better photo-$z$ estimations}.
    
    Consequently, there are the following two training methodologies. The first is $z_{\rm s}$: This is the usual training that maps the broadband photometry to photo-$z$ using spectroscopic redshifts as ground-truth redshifts and a negative log-likelihood loss function (Eq.\,\ref{eq:lossz}). 

The second is $z_{\rm s}$+NB: This methodology includes MTL. It maps the broadband photometry to photo-$z$ and narrow-band photometry, and therefore the loss function is the mean of the combined negative log-likelihood  loss (Eq.\,\ref{eq:lossz}) and narrow-band reconstruction (Eq.\,\ref{eq:lossNB}) tasks for all galaxies ($N$) for which the loss is computed as
      \begin{equation}
       \mathcal{L}_{\rm NB+z_{\rm s}} \coloneqq \frac{1}{N}\sum_{j=1}^{N}\left[ \mathcal{L}_{z}^{j} + \mathcal{L}_{\rm NB}^{j}\right]\, .
   \end{equation}
   We only used galaxies with spectroscopic redshift to train the photo-$z$ predictions, while all galaxies with narrow-band observations trained the narrow-band reconstruction. In general, one can also weight the two terms in the loss functions. Testing different values, we found the photo-$z$ scatter to have a minimum in a wide range of values around equal weighting.

Furthermore, we considered two variants in the training procedure to explore the possibility of using high-precision photometric redshifts (Sect.\,\ref{sec:data:samppz}) to train the networks:
The first is $z_{\rm s} + z_{\rm PAUS}$: This is a variation of the $z_{\rm s}$ method. The training sample extends to galaxies having a high-precision photo-$z$ estimate in the PAUS+COSMOS catalogue. For galaxies with spectroscopy, we use the spectroscopic redshift as ground-truth while for the rest of the training sample, the PAUS+COSMOS photo-$z$ is used to train the network. 

The second is $z_{\rm s}$+NB+$z_{\rm PAUS}$:  This is a variation of the $z_{\rm s}$+NB method, and it also extends the training sample with galaxies with a high-precision photo-$z$ estimate in the PAUS+COSMOS catalogue. In contrast to the $z_{\rm s}$+NB method, here all galaxies are used to train the photo-$z$ prediction and the narrow-band photometry reconstruction. The ground-truth redshift labels are the spectroscopic redshifts if available and otherwise, the PAUS+COSMOS photo-$z$.  

The networks are implemented in \texttt{PyTorch} \citep[][]{pytorch}.\ All the training procedures use an \texttt{Adam} optimiser \citep{ADAM} for 100 epochs with an initial learning rate of $10^{-3}$ that reduces by a factor of ten every 50 epochs.


\section{Photo-{\it z} performance in the COSMOS field}
\label{sec:cosmosres}
In this section we show the photo-$z$ performance of our method on galaxies with $i_{\rm AB}<23$ and $z<1.5$ in the COSMOS field. We study the effect that MTL has on the dispersion (Sect.\,\ref{sec:cosmosres:dispersion}) and the bias (Sect.\,\ref{sec:cosmosres:bias}) of the predicted photo-$z$s.

\subsection{Photo-{\it z} performance metrics}
To evaluate the accuracy and precision of the photo-$z$ estimates, we define 
\begin{equation}
    \Delta z \coloneqq (z_{\rm p} - z_{\rm t})\ /\ ( 1 + z_{\rm t})\,,
\end{equation} where $z_{\rm p}$ and $z_{\rm t}$ are the mean predicted photo-$z$ and the ground-truth redshift, respectively. The bias and the dispersion are defined as the median and $\sigma_{\rm 68}$ of $\Delta z$, respectively, where we define $\sigma_{68}$ as 
\begin{equation} 
    \sigma_{68} \coloneqq  \frac{1}{2} \left[ Q_{84}(\Delta z) - Q_{16}(\Delta z) \right]\,,
    \label{eq:bias_disp}
\end{equation}  and $Q_{16}(\Delta z)$, $ Q_{84}(\Delta z)$ are the 16th and 84th percentiles of the $\Delta z$ distribution. We also include the metric 
\begin{equation}
    \sigma_{\rm NMAD} \coloneqq 1.4826 \times \rm{median}\left[\,|\Delta z - \rm{median}(\Delta z)|\,\right]
\end{equation} used in the \Euclid photo-$z$ challenge paper \citep[][]{photoz_challenge}.

To evaluate the performance on the full COSMOS catalogue, we define the ground-truth redshift as the spectroscopic redshift if available and otherwise, as the PAUS+COSMOS photo-$z$ (Sect.\,\ref{sec:data:samppz})\footnote{The PAUS+COSMOS photo-$z$s used to evaluate the precision of non-spectroscopic galaxies (Sect.\,\ref{sec:data:samppz}) also have an associated dispersion. This corresponds to approximately 4\% lower photo-$z$ scatter than that obtained for very bright galaxies and around 1\% lower at the faintest end.}. If it is not specified by the method, our networks are trained with spectroscopic redshifts only. For the performance evaluation, however, the PAUS+COSMOS photo-$z$s are also used, but only to evaluate the photo-$z$ of galaxies from the full COSMOS catalogue that do not have a spectroscopic redshift estimate.  The predicted photo-$z$s are defined as the mean of the redshift probability distribution provided by the network  (Sect.\,\ref{sec:method:arch}).

In order to estimate the photo-$z$s of the complete COSMOS catalogue, the networks are trained independently ten times with $\sim$11\,000 spectroscopic galaxies in each iteration, which roughly corresponds to 90\% of the sample. Each network is used to evaluate the corresponding 10\% of excluded galaxies in such a way that the ensemble of networks evaluates the full COSMOS catalogue. 

Including MTL extends the training sample to about 40\,000 galaxies, which corresponds approximately 3.5 times more galaxies than in the spectroscopic sample. In order to evaluate the full COSMOS sample, we trained the network seven independent times with 85\% of the spectroscopic galaxies and 85\% of the non-spectroscopic sample. This corresponds to around 11\,000 galaxies with spectroscopy and  25\,000 without. We ensured that the fraction of galaxies with spectroscopic redshifts in each iteration is similar by sampling without replacement the same number of spectroscopic galaxies in each iteration. 

\subsection{Photo-{\it z} dispersion}
\label{sec:cosmosres:dispersion}

\begin{figure}
\includegraphics[width= 0.47\textwidth]{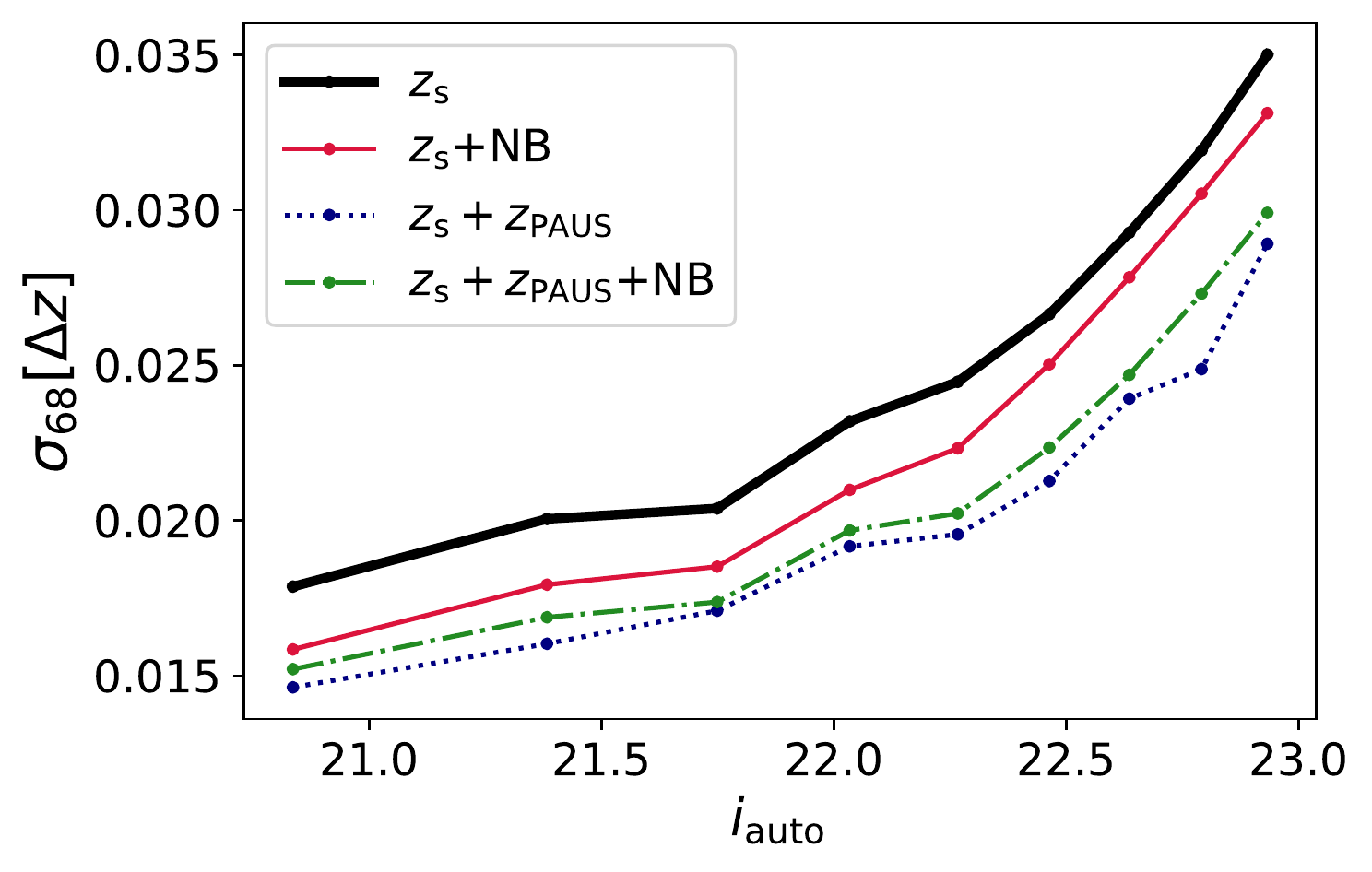}
\includegraphics[width= 0.47\textwidth]{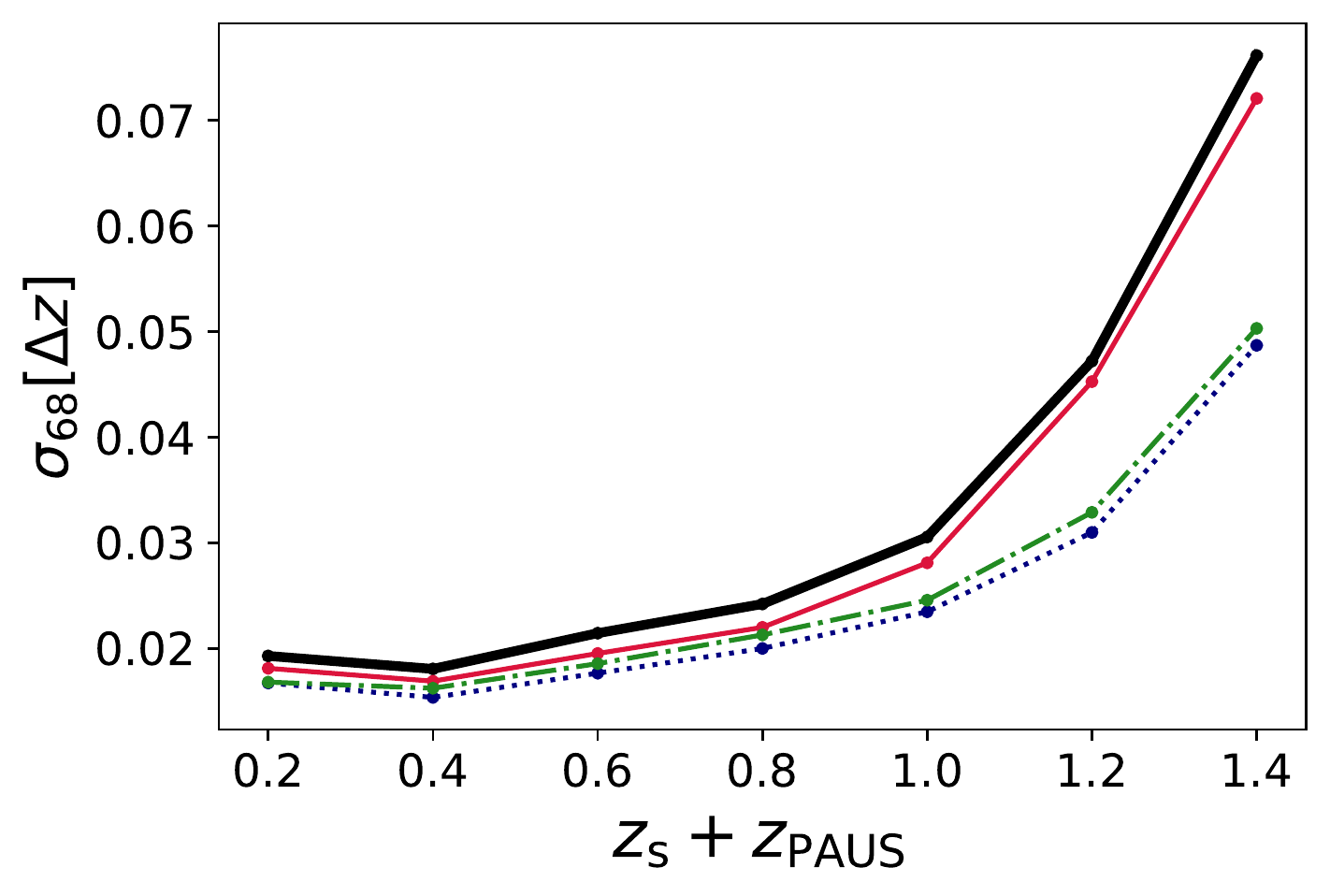}
\centering
\caption{Photo-$z$ dispersion in equally populated magnitude differential bins to $i_{\rm AB} < 23$ (top) and {{} equally spaced redshift bins to $z<1.5$ (bottom).} Each line corresponds to a different training procedure (see Sect.\,\ref{sec:method:arch}). While the black line corresponds to a baseline training, the other coloured lines include MTL (red and green lines) and data augmentation with photo-$z$s from the PAUS+COSMOS catalogue as ground-truth redshifts (blue and green lines).}
\label{fig:main_cosmos_dispersion}
\end{figure}

Table \ref{tab:photoz_results}  presents the photo-$z$ precision for the COSMOS spectroscopic sample and the complete COSMOS sample using the four different training procedures presented in Sect.\,\ref{sec:method:arch}. These results are presented in more detail in Fig.\,\ref{fig:main_cosmos_dispersion}, which shows the photo-$z$ dispersion in equally populated magnitude and redshift bins with the same four methodologies. The solid black line  corresponds to the baseline network mapping broadband photometry to photo-$z$ (method $z_{\rm s}$ in Sect.\,\ref{sec:method:arch}).  This method is trained on the spectroscopic sample and provides a $\sigma_{\rm 68} = 0.020$ for the full sample. These are quite precise and accurate redshifts compared to other broadband redshift estimates in the same field. In \citealt{Hendrik-photozs}, redshifts in the D2 CFHT deep field \citep[][]{D2}, which overlaps with COSMOS, were estimated with the template-fitting code \texttt{BPz} \citep[][]{Bpz} using the CFHT $ugriz$ filter set. Their photo-$z$ precision is $\sigma_{\rm 68} = 0.0498$, while for the same galaxy sub-sample our network provides $\sigma_{\rm 68} = 0.0187$. Here neither the methodology nor the input data are the same, but having these CFHT photo-$z$ estimates as a reference improves our photo-$z$ baseline network performance. Here neither the methodology nor the input data are the same, but the CFHT photo-$zs$ are a reference to compare the performance of our baseline network with.

\LC{In Fig.\,\ref{fig:main_cosmos_dispersion} we show the MTL training (method $z_{\rm s}$+NB in Sect.\,\ref{sec:method:arch}) that uses all galaxies with PAUS photometry to train the narrow-band reconstruction and only those with spectroscopy to train the photo-$z$ prediction. This extends the training sample of the shared layers (see the bottom panel of Fig. \ref{fig:architecture}) from around 12\,000 to 30\,000 galaxies. This method provides a precision of $\sigma_{\rm 68} = 0.0176$, corresponding to a 13\% improvement with respect to the baseline methodology (solid black line). Moreover, the additional PAUS galaxies for the narrow-band reconstruction loss  includes a more homogeneous colour-space coverage in the training sample.} In Sect.\,\ref{sec:colourspace} we discuss the underlying mechanism that causes MTL with PAUS to improve the photo-$z$s.

The blue dotted line in Fig.\,\ref{fig:main_cosmos_dispersion} also corresponds to a direct mapping of the broadband photometry to photo-$z$s. However, in contrast to the solid black line, this case is trained on an extended sample including galaxies without spectroscopic redshifts (method $z_{\rm s}+z_{\rm PAUS}$ in Sect.\,\ref{sec:method:arch}), for which the PAUS+COSMOS photo-$z$ measurement is used as a ground-truth redshift label in the training. It shows a precision of  $\sigma_{\rm 68} = 0.0168$, which corresponds to a 18\% improvement with respect to the baseline training.   

The best photo-$z$ performance is achieved combining MTL and photo-$z$ data augmentation with PAUS+COSMOS data (method $z_{\rm s}$+NB+$z_{\rm PAUS}$ in Sect.\,\ref{sec:method:arch}), which corresponds to the dotted green line in Fig.\,\ref{fig:main_cosmos_dispersion}. This method gives a 22\% improvement with respect to the baseline network, with a precision of  $\sigma_{\rm 68}= 0.0163$. 

\RR{In addition to uncertainties due to limited sample size (sample variance), our findings could also be affected by the intrinsic galaxy distribution being different at different parts of the sky (cosmic variance). To ensure that our results are not due to imprinting cosmic variance from the training to the test field, we tested our methods on two independent and spatially separated fields. These two fields are  $\sim 2\,{\rm deg}^2$ and contain galaxies from the Flagship simulations to $i<23$ (Sect.\,\ref{sec:data:mocks}). 
All the networks have been trained with 30\,000 galaxies from the train field, and later evaluated on 20\,000 galaxies different galaxies from the train and test fields (making sure that there is no overlap between the training and test galaxies in the train field).  
We estimated the sample variance of each of these fields by making 1000 bootstrap realisations and it is a sub-percent error. With the baseline $z_{\rm s}$ method, we obtain a 2\% change in the photo-$z$ precision between the train and test fields. Repeating the same test with the $z_{\rm s}$+$z_{\rm PAUS}$+NB method, we obtain a 3\% change in the photo-$z$ precision between fields, similar to the baseline case. These changes are much lower than the photo-$z$ improvement we obtain with the MTL implementations (e.g. 22\% for the $z_{\rm s}$+$z_{\rm PAUS}$+NB method), suggesting that such improvements are not caused by cosmic variance.}

\begin{table*}
\centering
\caption{Photo-$z$ dispersion $\sigma_{68}\times 100$ for the different network configurations. The second column displays results restricted to the spectroscopic sample, while the third column shows the results for the full COSMOS to $i_{\rm AB}<23$. For the full COSMOS sample results, the PAUS+COSMOS high-precision photo-$z$s are used as ground-truth redshifts when spectroscopy is not available. The numbers in parenthesis corresponds to the $\sigma_{\rm NMAD}$. The fourth and fifth columns present the percentage of photo-$z$ outliers on the COSMOS spectroscopic and complete sample, respectively.}
\begin{tabular*}{0.885\textwidth}{|c|cccc|}
\hline
\textbf{\,} & Precision ($z_{\rm s}$ sample) & Precision (COSMOS) & Outliers ($z_{\rm s}$ sample) &Outliers (COSMOS) \\ \hline
$z_{\rm s}$  & 2.00\,(1.92) & 2.43\,(2.30)&0.6& 1.1  \\ 
$z_{\rm s}$+NB  & 1.76\,(1.68) & 2.21\,(2.04)&0.5&0.8\\ 
$z_{\rm s}+z_{\rm PAUS}$  & 1.68\,(1.62) & 2.03\,(1.91) &0.5&0.6 \\
$z_{\rm s}+z_{\rm PAUS}$+NB & 1.63\,(1.58) & 1.99\,(1.86)  &0.5&0.6  \\ \hline
\end{tabular*}
\label{tab:photoz_results}
\end{table*}

\subsection{Photo-{\it z} bias and outlier rate}
\label{sec:cosmosres:bias}
\begin{figure*}
\includegraphics[width= 0.48\textwidth]{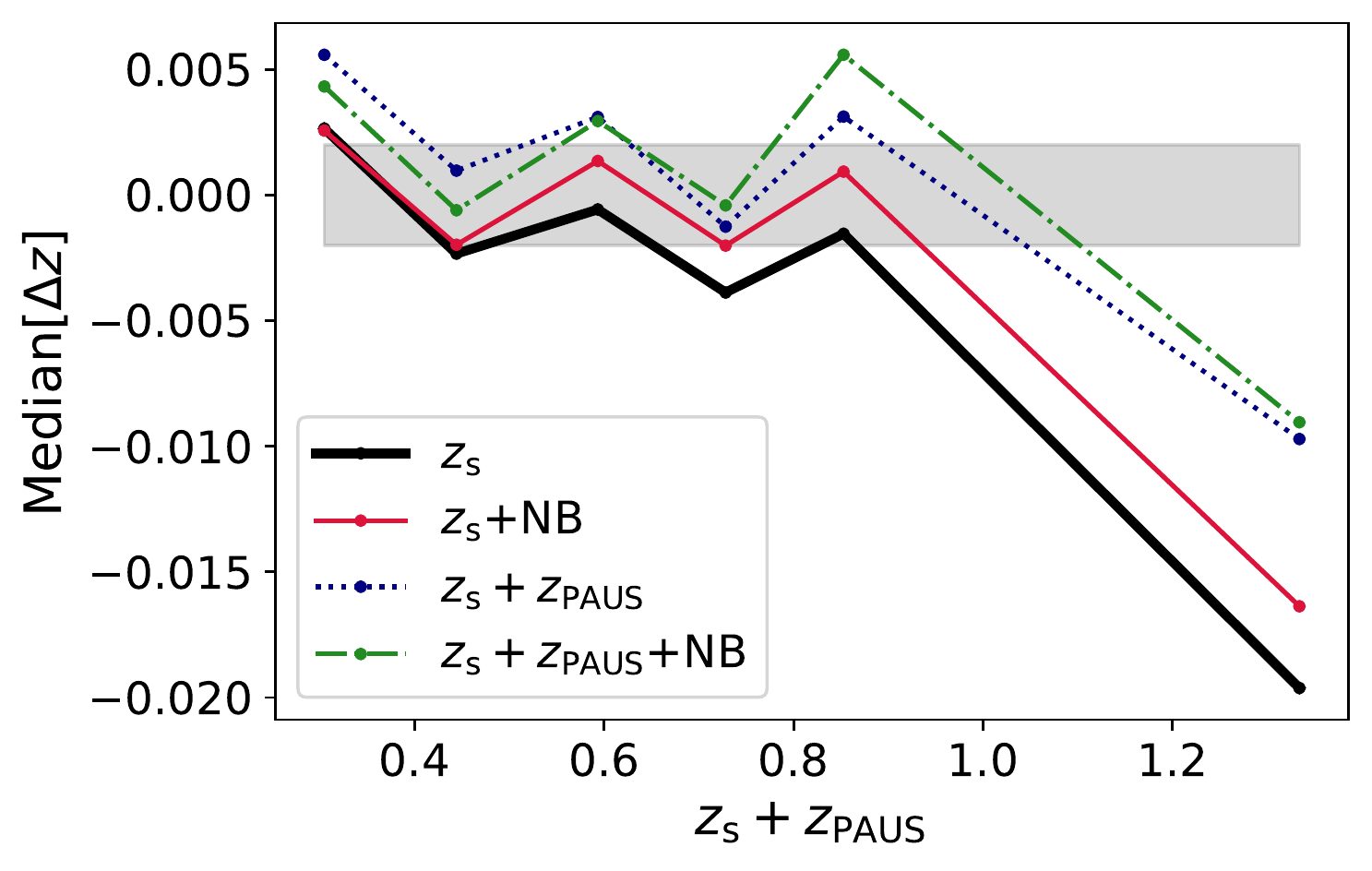}
\includegraphics[width= 0.48\textwidth]{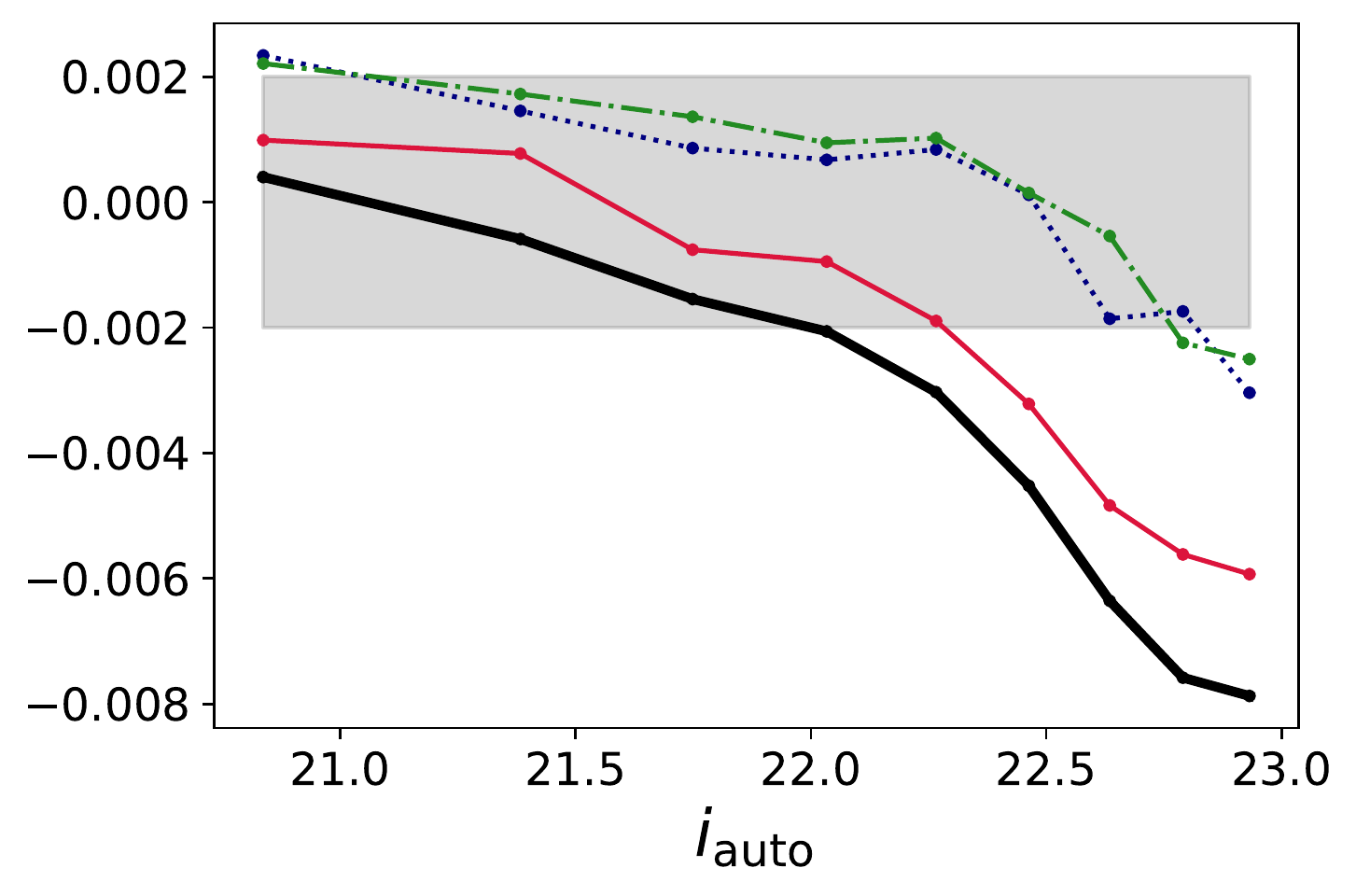}
\centering
\caption{Photo-$z$ bias in equally populated redshift bins (\textit{left}) and equally populated $i$-band magnitude bins (\textit{right}). The grey area corresponds to the \Euclid photo-$z$ bias requirement of $\Delta z$ = 0.002.}
\label{fig:cosmos_zbias}
\end{figure*}

In this subsection we show the bias and the outlier rate for the photo-$z$ predictions with the MTL networks and the baseline broadband network. The left panel in Fig.\,\ref{fig:cosmos_zbias} shows the photo-$z$ bias in equally populated redshift bins in the redshift range $0.1<z_{\rm t} <1.5$. We excluded the first redshift bin from the analysis since there are almost no galaxies with $z_{\rm t} <0.07$, which caused a bias at very low redshift\footnote{There are training mechanisms to deal with unbalanced training samples such as up-weighting the contribution of unbalanced class objects in the training or oversampling synthetic data from the unbalanced original ones \citep[][]{class_imbalance}. However, the number of objects with $z<0.07$ is too small to efficiently apply these techniques and there are very few galaxies affected.}. \RR{The shaded area corresponds to the \Euclid photo-$z$ bias requirement $<0.002$ \citep[][]{Euclid}. Overall, for $z_{\rm t}<1$ the four methods presented in Sect.\,\ref{sec:method:arch}  are unbiased at the level of $<0.002$.} \RR{However, the $z_{\rm s}$ and the $z_{\rm s}$+NB are still showing a trend within the 0.2\% bias range, where low-redshift galaxies tend to be biased positive and the high-redshift ones, biased negative. In contrast, the $z_{\rm s}+z_{\rm PAUS}$ and the $z_{\rm s}+z_{\rm PAUS}$+NB methods display a flatter bias with redshift.} 

\LC{At higher redshifts ($z_{\rm t} > 1$), the baseline network photo-$z$s show a $\sim$2\% bias. Implementing MTL without increasing the photo-$z$ training sample (solid red line) moderately improves the bias, but it is still far from the \Euclid requirement. On the other hand, increasing the training sample with PAUS+COSMOS photo-$z$s produces a strong bias reduction (blue and green lines), decreasing the bias to $\sim$1\% for the highest-redshift galaxies.  Figure \ref{fig:cosmos_zdist} suggests that this is likely to be caused by a lack of training examples with spectroscopy at $z_{\rm t} > 1$. The training sample at high-redshift is increased with the PAUS+COSMOS photo-$z$s.}

\RR{The right panel of Fig.\,\ref{fig:cosmos_zbias} shows the photo-$z$ bias in equally populated $i$-band magnitude bins. Comparing to the right panel in the same figure, the bias binning in $i$-band magnitude is lower than that binning in redshift. For instance, binning in redshift the largest bias that the $z_{\rm s}$ method obtains is a $\sim2.5\%$ for the highest-redshift galaxy bin. In contrast, binning in magnitude, galaxies in the faintest bin reach a 0.8\% bias with the same method. This is partly because binning in magnitude, positive and negative biases in redshift cancel each other out. The photo-$z$s of galaxies with $i<22$ are unbiased with the four methods. For galaxies with $i>22$, the $z_{\rm s}$ method displays the largest bias, which is already reduced with the MTL method without data augmentation ($z_{\rm s}$+NB). The methods extending the sample using the PAUS+COSMOS photo-$z$s (green and blue lines) reduce the bias of the $z_{\rm s}$ and the $z_{\rm s}$+NB methods.}

In this paper, we consider a galaxy to be an outlier if
 \begin{equation}
   |z_{\rm p}  - z_{\rm t}|\ /\ (1 + z_{\rm t}) > 0.15\,.
   \label{eq:outlier_rate}
 \end{equation}
 In the spectroscopic sample, the baseline network yields 0.6\% outliers, which reduces to 0.5\% with the MTL using PAUS photometry, the training sample extension with PAUS+COSMOS photo-$z$, and the combination of both. The fraction of outliers in the PAUS sample in COSMOS is 1.1\% for the baseline network and for the training sample extension with PAUS+COSMOS photo-$z$s ($z_{\rm s} + z_{\rm PAUS}$). The methodologies including MTL reduce the outlier fraction to 0.8\% ($z_{\rm s}$+NB) and 0.6\% ($z_{\rm s}+z_{\rm PAUS}$+NB). While in the spectroscopic sample extending the training sample and including MTL have a similar effect on the outlier fraction, in the full PAUS sample in COSMOS MTL has a stronger impact. \LC{The MTL methodologies are particularly reducing the number of high-redshift photo-$z$ outliers.}

 \LC{In order to validate the predicted photo-$z$ probability distributions $p(z)$, we use the probability integral transform \citep[PIT;][]{PIT1,PIT2,PIT3}, which is defined as 
\begin{equation}
    {\rm PIT} \equiv \int_{\rm -\infty}^{z_{\rm true}} {\rm d}z\,{p}(z)\, 
,\end{equation}
where $z_{\rm true}$ is the true redshift. When the $p(z)$ faithfully represents the true redshift, the PIT distribution is the uniform distribution U[0,1]. Contrary, PIT histograms with peaks at the edges (i.e. around zero and unity) indicate the presence of outlier measurements. Also, PIT histograms more populated at the centres than on the edges denote over-dispersed probability distributions, while valleys at the centre of the histogram correspond to under-dispersed ones.}

\begin{figure}
\includegraphics[width= 0.48\textwidth]{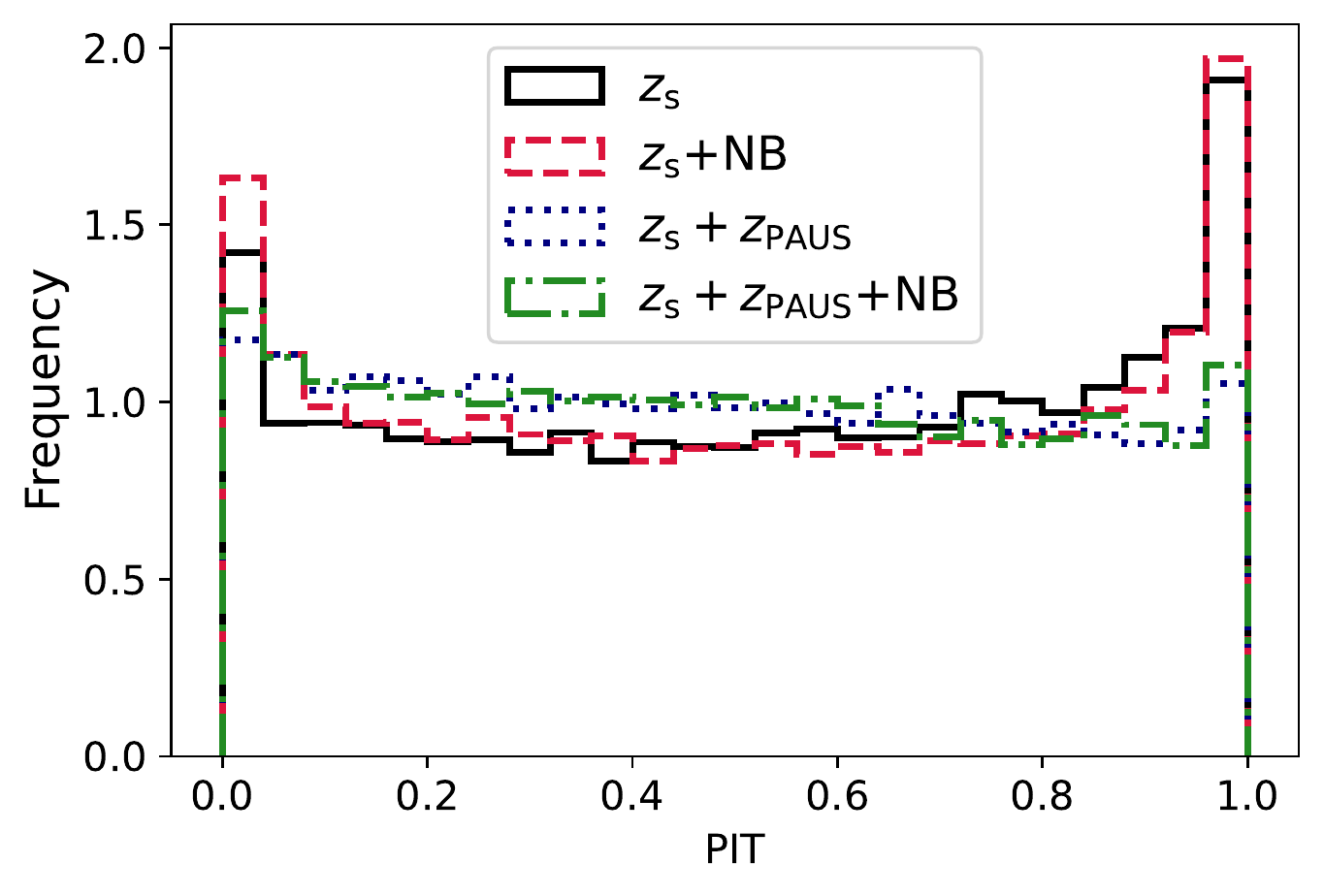}
\centering
\caption{PIT distribution for the COSMOS photo-$z$s predicted with the baseline $z_{\rm s}$ method (black), the $z_{\rm s}$+NB method (red), the $z_{\rm s}$+$z_{\rm PAUS}$ method (blue), and the $z_{\rm s}$+$z_{\rm PAUS}$+NB method (green). Including the PAUS+COSMOS photo-$z$s in the training reduces the number of outliers on the edges of the distribution.}
\label{fig:PIT}
\end{figure}

\LC{We measure the PIT distribution for the complete COSMOS sample using a combination of spectroscopic redshifts and high-precision photo-$z$ as true redshift. Figure \ref{fig:PIT} shows the PIT distributions for the $p(z)$ measured with the baseline $z_{\rm s}$ method (black line), the MTL method (dashed red line), extending the training sample with high-precision photo-$z$ (dotted blue line), and combining the training sample augmentations and MTL (green line). In all cases, the PIT distribution is approximately a U[0,1] distribution, indicating that our networks predict robust probability distributions with reliable uncertainty measurements. The baseline and the $z_{\rm s}$+NB methods display peaks on the edges of the distribution corresponding to outliers in the probability distributions. These peaks are reduced with the two methods using PAUS+COSMOS photo-$z$s in the training sample. }

\section{Photo-{\it z} performance on  deeper galaxy simulations}
 \label{sec:depth_tests}
 \begin{figure*}
 \centering
 \textbf{Photo-$z$s in Flagship}\par\medskip
\includegraphics[width= 0.48\textwidth]{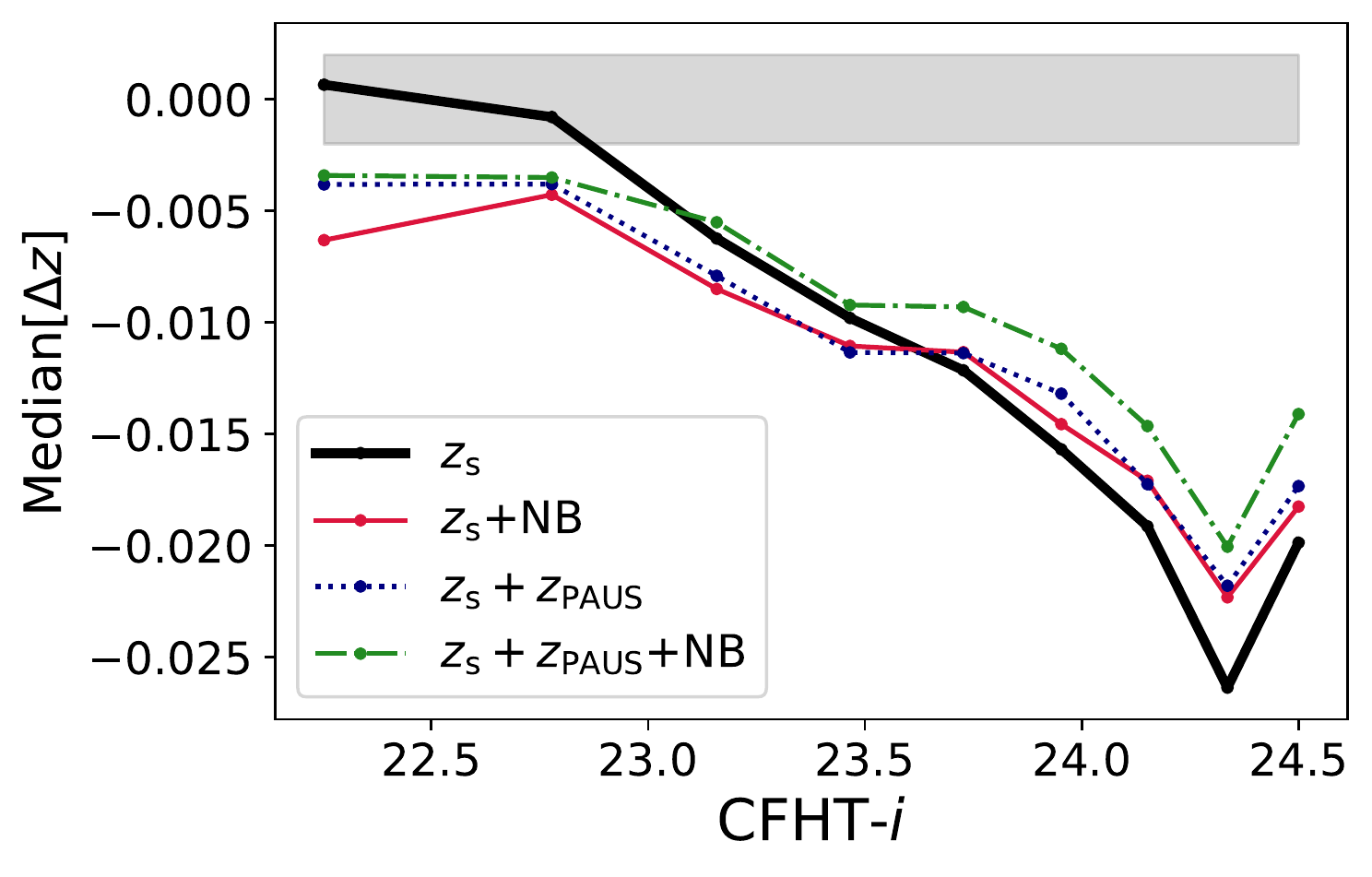}
\includegraphics[width= 0.46\textwidth]{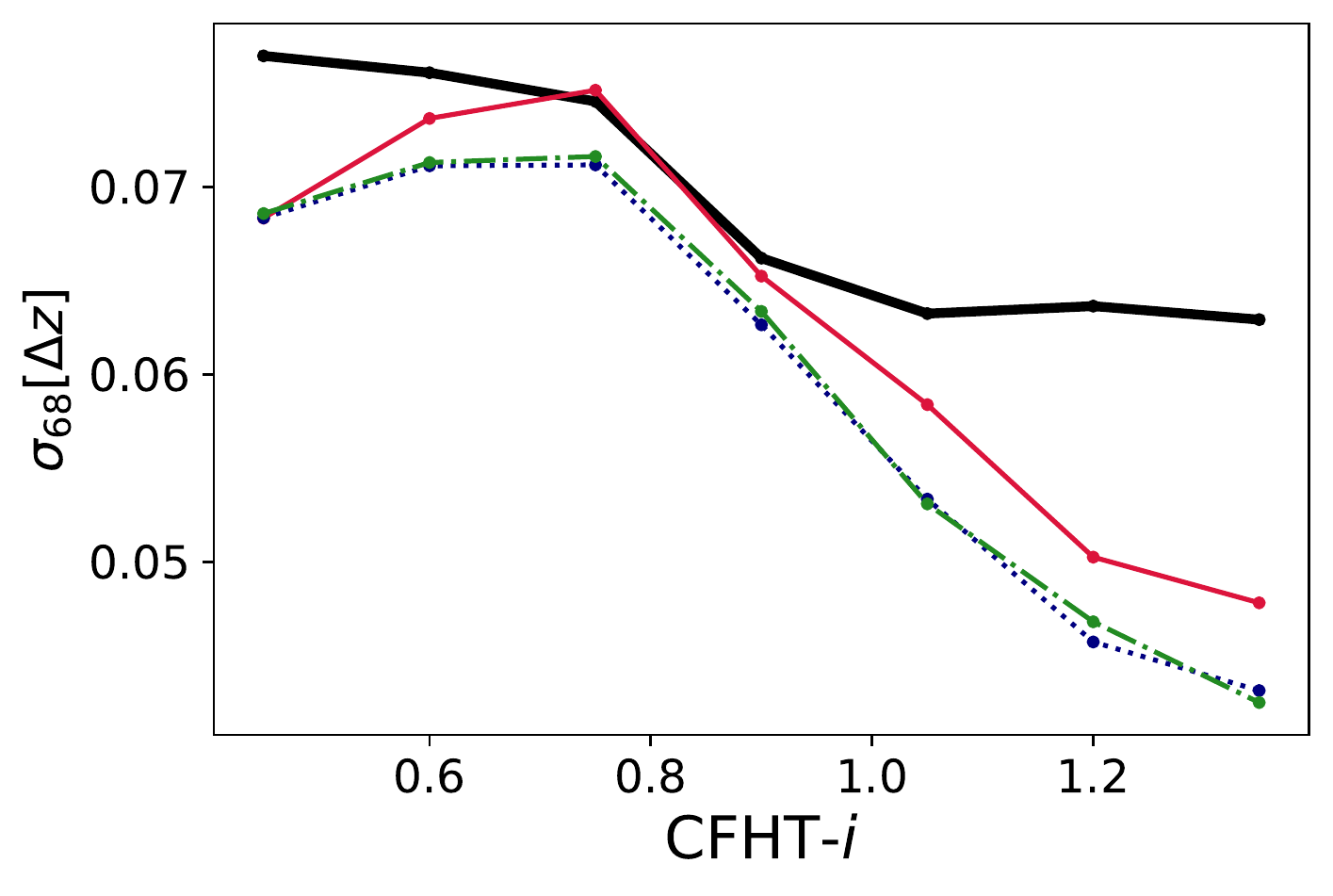}
\includegraphics[width= 0.48\textwidth]{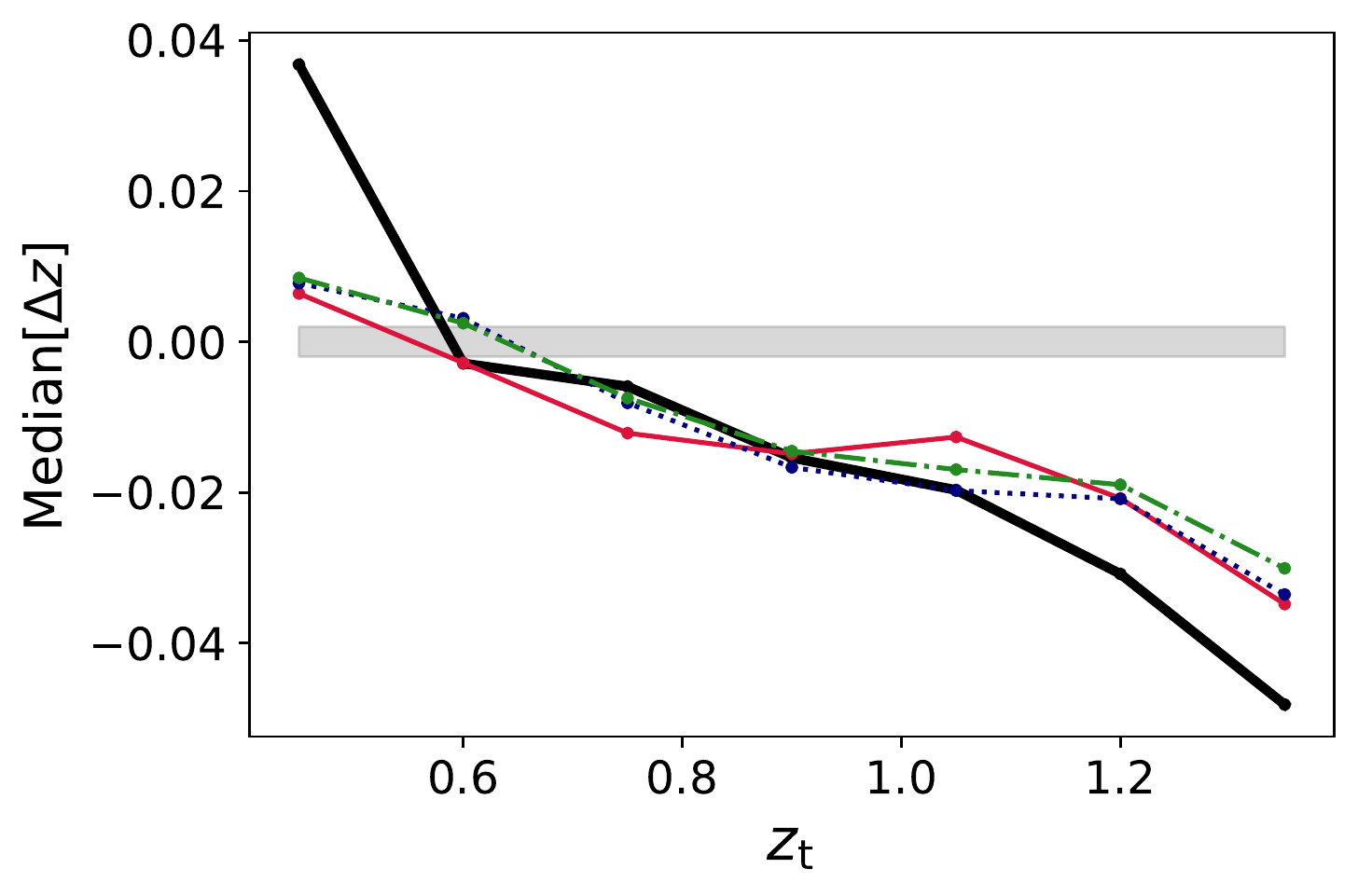}
\includegraphics[width= 0.47\textwidth]{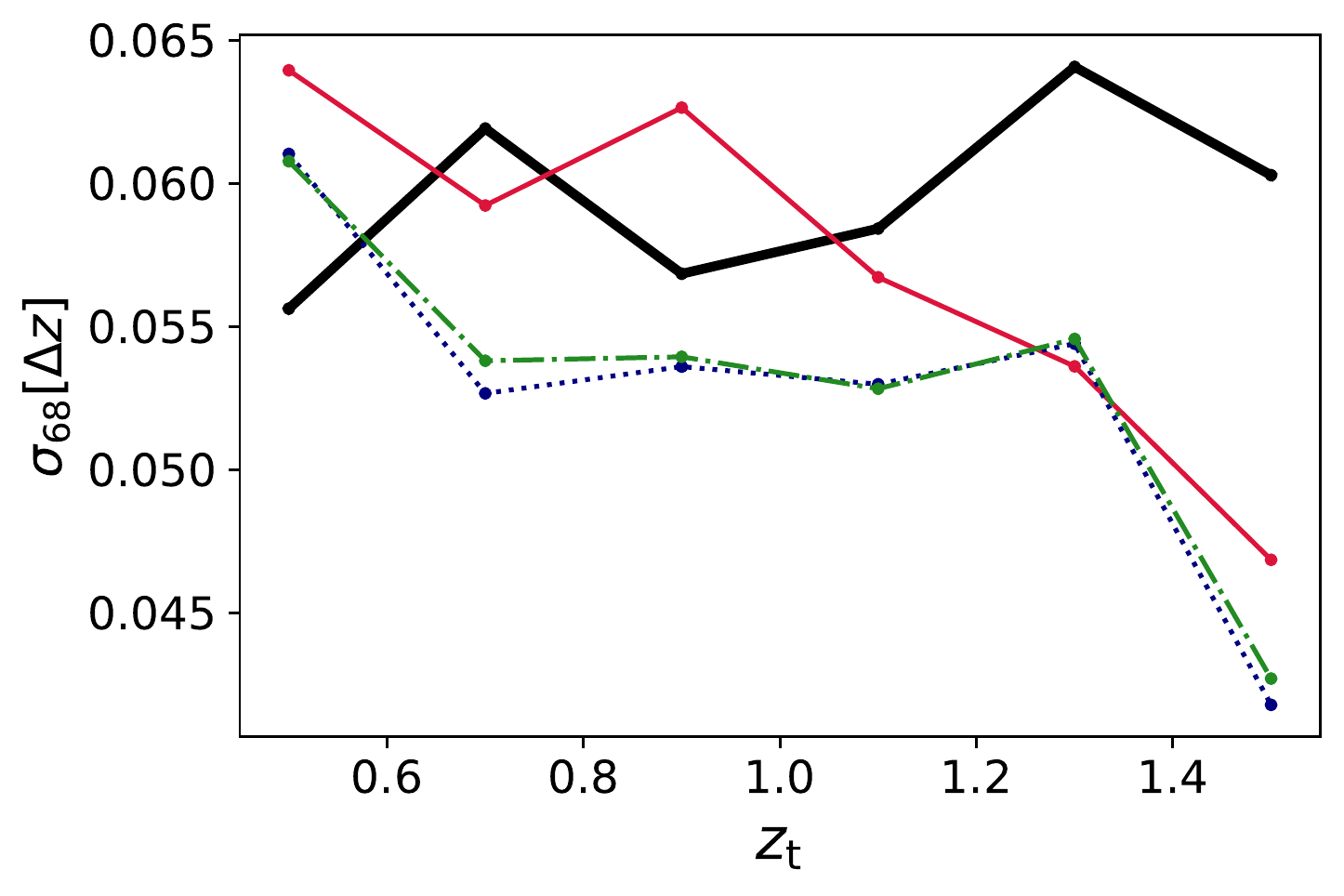}
\caption{\RR{\emph{Top}: Photo-$z$ bias (\textit{left}) and precision (\textit{right}) in equally populated magnitude bins. \emph{Bottom}: Photo-$z$ bias (\textit{left}) and dispersion (\textit{right}) in equally spaced spectroscopic redshift bins. The shaded grey areas indicate $\Delta z$ > 0.002, corresponding to the \Euclid requirement for the photo-$z$ bias.} All plots are for 30\,000 Flagship test galaxies with magnitudes $i_{\rm AB}$ < 24.5 for the methods presented in Sect.\,\ref{sec:method:arch}. The training sample contains around 15\,000 spectroscopic galaxies, extended to 30\,000 with PAUS-like galaxies without spectroscopy, all of them to $i_{\rm AB}$ < 23.}
\label{fig:depth_disp}
\end{figure*}

 So far, all the networks have been trained and evaluated on samples within the same magnitude range $i_{\rm AB} < 23$ (see Sect.\,\ref{sec:cosmosres}). However, if the MTL network developed in this paper aims to improve the photo-$z$ estimates of future deeper broadband surveys such as \Euclid or LSST, the photo-$z$ improvement it provides must hold for fainter galaxies. In the case of \Euclid, observations will reach a limiting magnitude of 24.5 for the VIS instrument \citep[][]{VIS1,VIS2} with $10\, \sigma$ depth for extended sources, which corresponds to a similar depth in the $i$-band filter. Rubin will observe to a single exposure depth of $r_{\rm AB} \sim 24.5$ and a co-added survey depth of $r_{\rm AB} \sim 27.5$ \citep[][]{LSST2}, where the depth in the $r$ band and the $i$ band are also similar.
 
 Currently, there are no PAUS measurements beyond $i_{\rm AB}=23$, thus limiting the magnitude range of the MTL training sample. Although observing deeper with PAUS is technically feasible, it would require considerably more observing time. Therefore, the MTL network must provide reliable photo-$z$ predictions for deep data samples, while it is trained on a shallower data sample. Nevertheless, we note that this problem is not  exclusive to our MTL network, but it affects all photo-$z$ machine-learning algorithms. These are usually trained on relatively shallow spectroscopic samples and used to predict the photo-$z$s for much deeper data samples \citep[][]{Masters1}. 

In this section we explore how the MTL network performs for deep samples ($i_{\rm AB }<25$), while the training is limited to galaxies with $i_{\rm AB}<23$ using Flagship \LC{simulated} galaxy mocks (see Sect.\,\ref{sec:data:mocks}). The broad bands used for this test are the CFHT $u$ band, the \textit{griz} bands from DECam \citep{Decam}, and the \Euclid Near-infrared spectrometer and photometer (NISP) near-infrared bands $H_{\rm E}$, $J_{\rm E}$ and $Y_{\rm E}$  \citep{NISP}\footnote{With the following $5\, \sigma$ limiting magnitudes: $u$: 25.25; $g$: 24.65; $r$: 24.15; $i$: 24.35; $z$: 23.95; $Y_{\rm E}$: 24.0, $J_{\rm E}$: 24, $H_{\rm E}$: 24.}. These are not the same bands that were used in the tests of the COSMOS field (see Sect.\,\ref{sec:data:bb_data} and Sect.\,\ref{sec:cosmosres}), but these bands were chosen to demonstrate the potential benefits for the \Euclid photo-$z$ estimation. 

We trained the four methods presented in Sect.\,\ref{sec:method:arch} on a sample with 10\,000 spectroscopic galaxies, which are augmented to 30\,000 with PAUS-like galaxies without spectroscopic redshifts and limited to $i_{\rm AB} < 23$. These numbers were chosen to approximately match the number of spectroscopic and PAUS-like galaxies in the COSMOS field (see Sect.\,\ref{sec:cosmosres}). To simulate the performance of the approaches that extend the training sample with high-precision photo-$z$s (methods $z_{\rm s}+z_{\rm PAUS}$ and $z_{\rm s}$+NB+$z_{\rm PAUS}$ in Sect.\,\ref{sec:method:MTL}), we added a scatter to the true redshifts of the PAUS-like simulated galaxies, so that  the precision resembles that of the PAUS+COSMOS photo-$z$s.

\LC{The left panels in Fig.\,\ref{fig:depth_disp} show the photo-$z$ bias of 30\,000 simulated test galaxies to magnitude $i_{\rm AB}< 25$ in equally populated magnitude bins (top) and in equally spaced redshift bins (bottom). The shaded areas correspond to the \Euclid photo-$z$ requirement of $\Delta z <0.002$. We obtain a larger bias than the \Euclid requirement with all methods, although those including MTL reduce the bias of fainter galaxies. Although we are not meeting the \Euclid bias requirement, our aim is to advance the usage of machine-learning photo-$z$ developing novel methodology, rather than providing the final pipeline. We hope the improvement and ideas seen in this paper can be helpful for further development of \Euclid machine-learning algorithms. }

\LC{The right panels in Fig.\,\ref{fig:depth_disp} show the photo-$z$ precision for the same 30\,000 simulated test galaxies to magnitude $i_{\rm AB}< 25$ in magnitude (top) and redshift (bottom) bins. The baseline network (black thick line) achieves an overall precision of $\sigma_{\rm 68} = 0.076$, which increases to $\sigma_{\rm 68} = 0.085$ for galaxies with $i_{\rm AB}> 23$. Training using photo-$z$s but without MTL ($z_{\rm s}+z_{\rm PAUS}$, dotted blue line) improves the precision to $\sigma_{\rm 68} = 0.0654$ and $\sigma_{\rm 68} = 0.080$ for galaxies with $i_{\rm AB} > 23$. With $z_{\rm s}$+NB, the overall precision is $\sigma_{\rm 68} = 0.067$, which degrades to $\sigma_{\rm 68} = 0.082$ for galaxies with $i_{\rm AB} > 23$. Finally, combining MTL and the photo-$z$  data augmentation ($z_{\rm s}$+NB+$z_{\rm PAUS}$, solid green line) provides the best photo-$z$ performance with $\sigma_{\rm 68} = 0.065$ for the full sample, which increases to $\sigma_{\rm 68} = 0.079$ for galaxies with $i_{\rm AB} > 23$ The best performance in terms of bias and precision is obtained with the $z_{\rm s}$+$z_{\rm PAUS}$+NB method, which provides 16\% more precise photo-$z$s than the baseline network for galaxies with $i_{\rm AB} <25$, which increases to 20\% for $i_{\rm AB}<24$.} 

 \section{Photo-{\it z} in colour space}
 \label{sec:colourspace}
 
While the effect of increasing the training sample in machine-learning algorithms has been extensively studied, we still need to understand  why MTL with narrow-band photometry improves the photo-$z$ estimates. In this section we use SOMs 
to explore the COSMOS photo-$z$ performance in colour space (Sect.\,\ref{sec:colorspace_cosmos}). Furthermore, in Sect.\,\ref{sec:colourspace:eldeg} and Sect.\,\ref{sec:colourspace:elconf} we identify colour-space regions with strong emission lines where the broadband photo-$z$s  precision is lower.

\subsection{MTL photo-{\it z} in colour space}
\label{sec:colorspace_cosmos}

\begin{figure*}
\includegraphics[width= 0.9\textwidth]{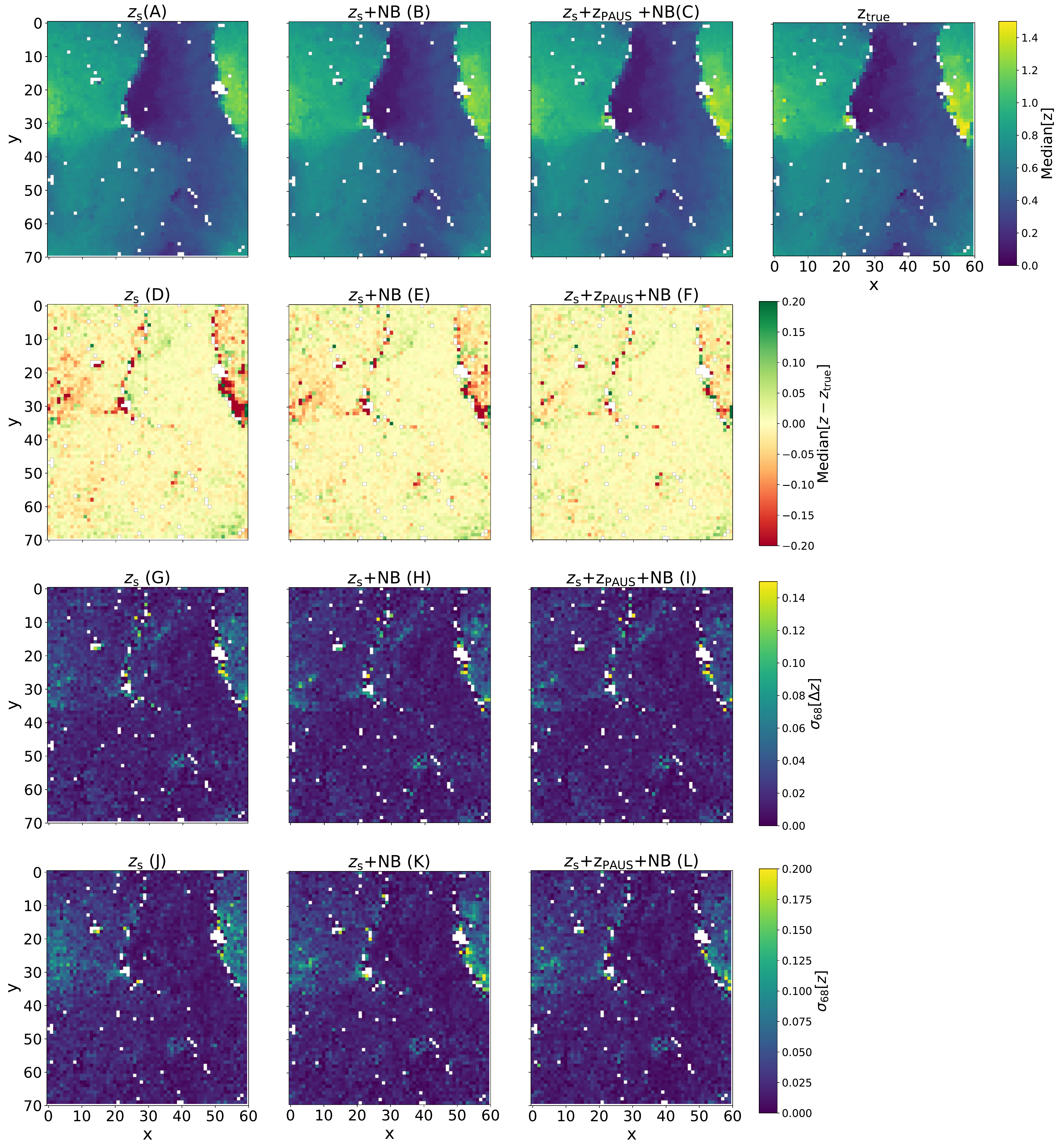}
\centering
\caption{SOMs showing the photo-$z$ performance in the COSMOS field.
The first row exhibits the median predicted photo-$z$ in colour space for the baseline network (first panel), including MTL training (second panel), with MTL and data augmentation with PAUS+COSMOS photo-$z$s (third panel), and the ground-truth redshift (fourth panel). The second row shows the bias in the photo-$z$ predictions for the three training methods of the first row (first three panels). The third row follows the same scheme as the second but displays the photo-$z$ precision. Finally, the fourth row shows the photo-$z$ cell dispersion also following the same scheme. White cells correspond to empty cells, that is, cells without any galaxy.}
\label{fig:som_map_cosmos}
\end{figure*}

\begin{figure}
\includegraphics[width= 0.47\textwidth]{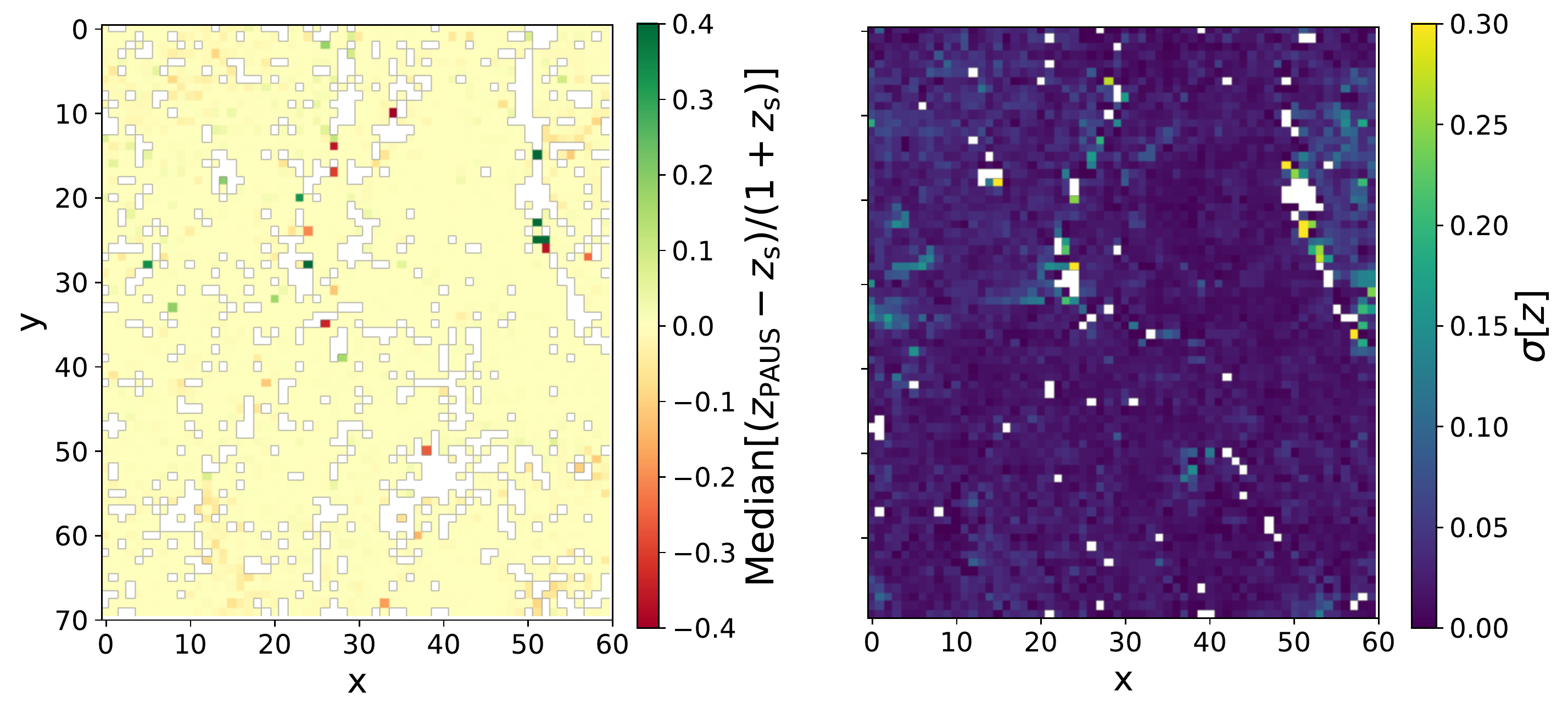}
\centering
\caption{Bias (\textit{left}) and precision (\textit{right}) of the PAUS+COSMOS photo-$z$s in the COSMOS spectroscopic sample.}
\label{fig:p+c_analysis}
\end{figure}

A SOM \citep{soms}\footnote{https://github.com/lauracabayol/SOM} is an unsupervised machine-learning algorithm trained to produce a low-dimensional (typically two-dimensional) representation of a multi-dimensional space. A two-dimensional SOM contains $(N_{x}, N_{y})$ cells, each of them with an associated vector of attributes, in our case colour vectors.
Initially, each cell is represented with random colours, which during the training phase are optimised to represent the colour space of the training sample. The SOM training also groups together cells representing similar colours, creating a colour-space map. Once trained, each galaxy is assigned to its closest cell in colour space. Moreover, since the SOM clusters galaxies with similar galaxy colours it also clusters galaxies with similar redshifts \citep{colourspacez_1,colourspacez_2}. 
The appendices contain a more detailed explanation of SOM algorithms. Self-organising maps have already been used in different astronomical applications, such
as the correction for systematic effects in angular galaxy clustering measurements \citep{SOM_clustering} and for estimation and calibration of photometric redshifts \citep[][]{SOMz,SOM_angus,wright2020a,SOM_hendrik}.

To show the MTL performance in colour space we trained a $60\times70$ SOM on the $uBVriz$ photometry from the COSMOS2015 catalogue (see Sect.\,\ref{sec:data:bb_data}), and subsequently assigned a SOM cell to each galaxy in the catalogue. The choice of SOM dimension is based on previous works, where $60\times70$ cells was found to give a good balance between resolution in colour space and the number of galaxies per cell. Figure \ref{fig:som_map_cosmos} shows the predicted photo-$z$s in colour space, with each column corresponding to a photo-$z$ estimation method described in Sect.\,\ref{sec:method:arch}. The first row shows the photo-$z$ distribution, where each cell is coloured with the median photo-$z$ of the galaxies it contains. The leftmost panel ($z_{\rm s}$, panel A) displays the photo-$z$s with the baseline network ($z_{\rm s}$ method), and the second (B) and third (C) panels include MTL in the training (i.e. $z_{\rm s}$+NB and NB+$z_{\rm s}+z_{\rm PAUS}$ methods, respectively; bottom panel of Fig.\,\ref{fig:architecture}). The rightmost panel shows the ground-truth redshift distribution.

The three methods show a photo-$z$ distribution in colour space that is similar to that of the ground-truth redshifts. However, some differences can be seen in the plots in the second row (panels D, E, and F), which show the differences between the predicted and true-redshift colour maps (e.g. panel D = panel A - $z_{\rm t}$). The network trained with only broad bands (panel D) exhibits two regions with less accurate photo-$z$s. These regions are centred around coordinates $(5,35)$ and $(55,25)$, and the redshift accuracy improves when MTL (panel E) or $z_{\rm PAUS}$+MTL (panel F) are included in the training. 

These regions are also spotted in the third row of Fig.\,\ref{fig:som_map_cosmos}, which shows the photo-$z$ precision ($\sigma_{\rm 68}$, Eq.\,\ref{eq:bias_disp}). Comparing panels G and D, we note that the photo-$z$ precision worsens in the same regions where photo-$z$s are less accurate, but this moderately improves with MTL ($z_{\rm s}$+NB, panel H) and including the PAUS+COSMOS photo-$z$s (NB+$z_{\rm s}+z_{\rm PAUS}$, panel F). Finally, the fourth row shows the dispersion of the redshift distribution (i.e. the width of the $N(z)$) within SOM cells. This quantity is also higher for the clusters pointed out in panels D and G. However, contrary to the previous panels, the $z_{\rm s}$+NB training (panel K), or the $z_{\rm s}$+NB+$z_{\rm PAUS}$ (panel L) do not narrow the redshift distributions. 

The fact that the photo-$z$ accuracy and precision improve with MTL, while the width of the redshift distribution does not, suggests that galaxies from different populations, that is, galaxies with different redshifts, are assigned to these cells. Figure \ref{fig:p+c_analysis} supports this hypothesis by showing that the PAUS+COSMOS photo-$z$s also exhibit a higher redshift dispersion (\LC{right panel}) in the SOM cells within the problematic regions, while the PAUS+COSMOS photo-$z$ accuracy is smooth across colour space (left panel). Therefore, there are galaxies with different redshifts clustered together in broadband colour space.

\subsection{Broadband degeneracies in colour space}
\label{sec:colourspace:eldeg}
\begin{figure}
\includegraphics[width= 0.47\textwidth]{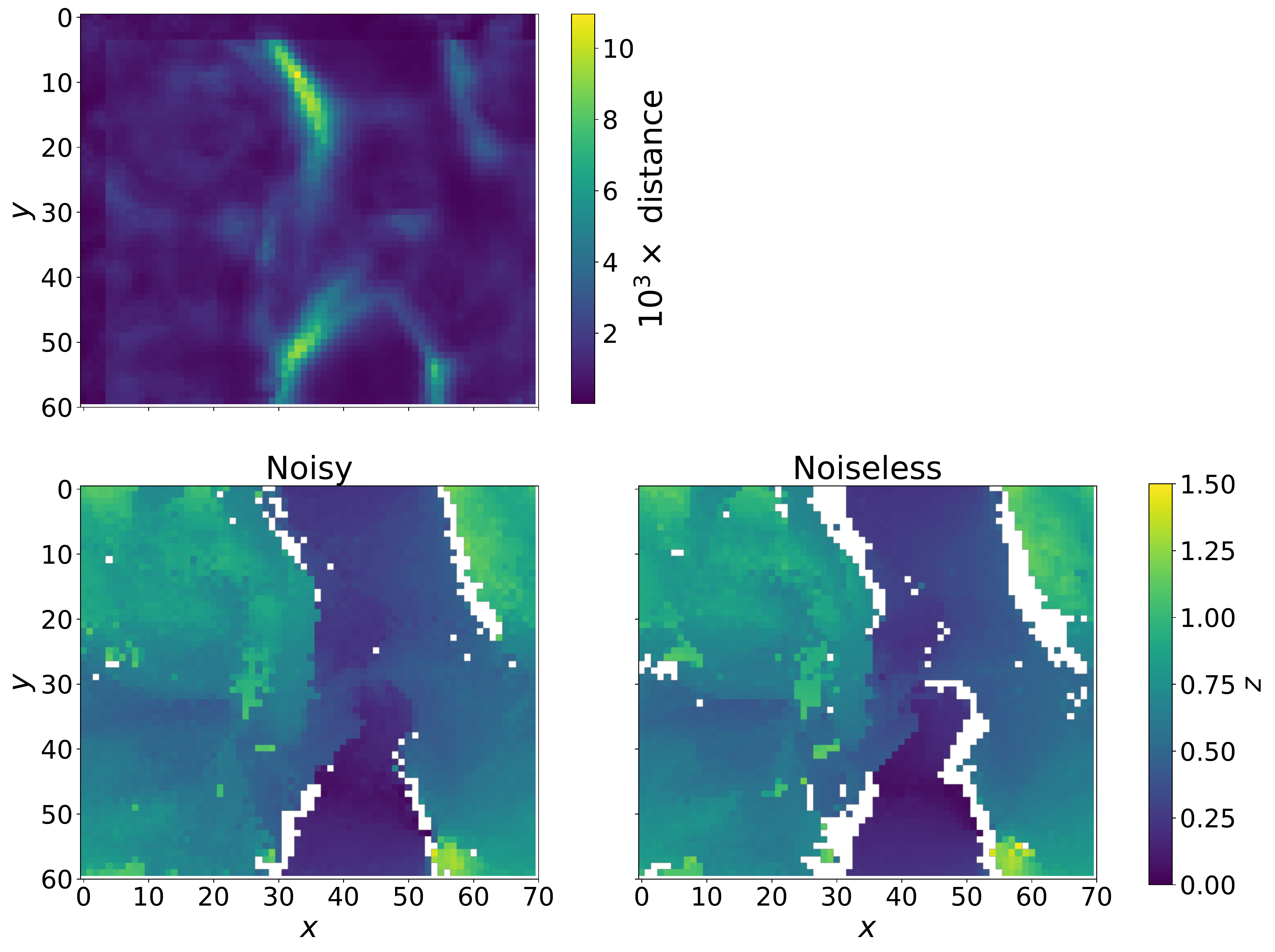}
\centering
\caption{SOM trained on a galaxy simulated mock with the $uBVriz$ broad bands. \emph{Top:} Distance between every SOM cell vector and its 3$\times$3 neighbours. \emph{Bottom left}: Median photo-$z$ in each SOM cell for noisy simulated galaxies. \emph{Bottom right}: Median photo-$z$ in each SOM cell for noiseless simulated galaxies.}
\label{fig:som_dist_empty}
\end{figure}

\begin{figure}
\includegraphics[width= 0.47\textwidth]{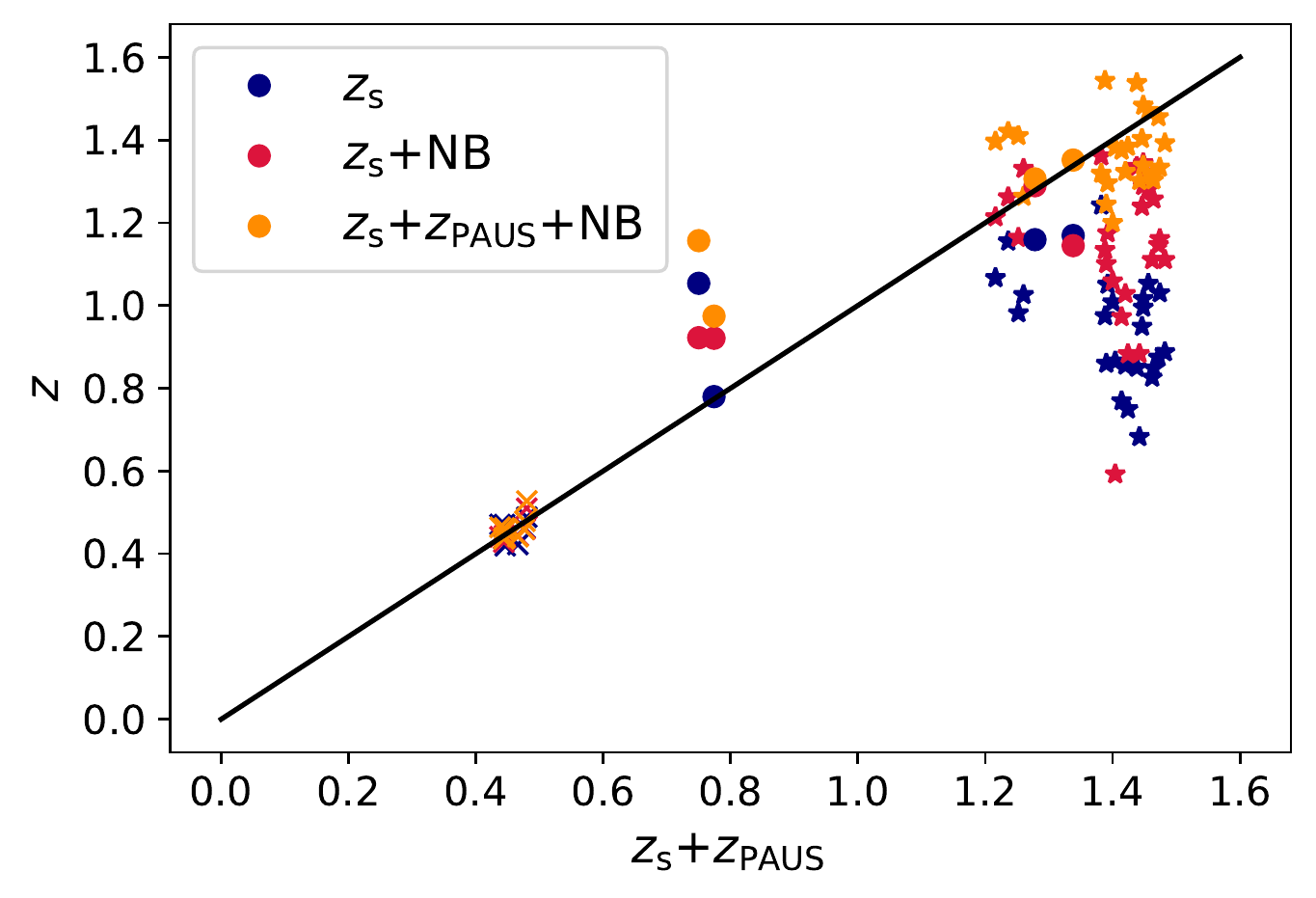}
\centering
\caption{Photo-$z$ scatter for galaxies in three independent SOM cells. The galaxies in each cell are represented with a different marker (stars, crosses, and circles).}
\label{fig:cluster_cells}
\end{figure}

\begin{figure}
\includegraphics[width= 0.47\textwidth]{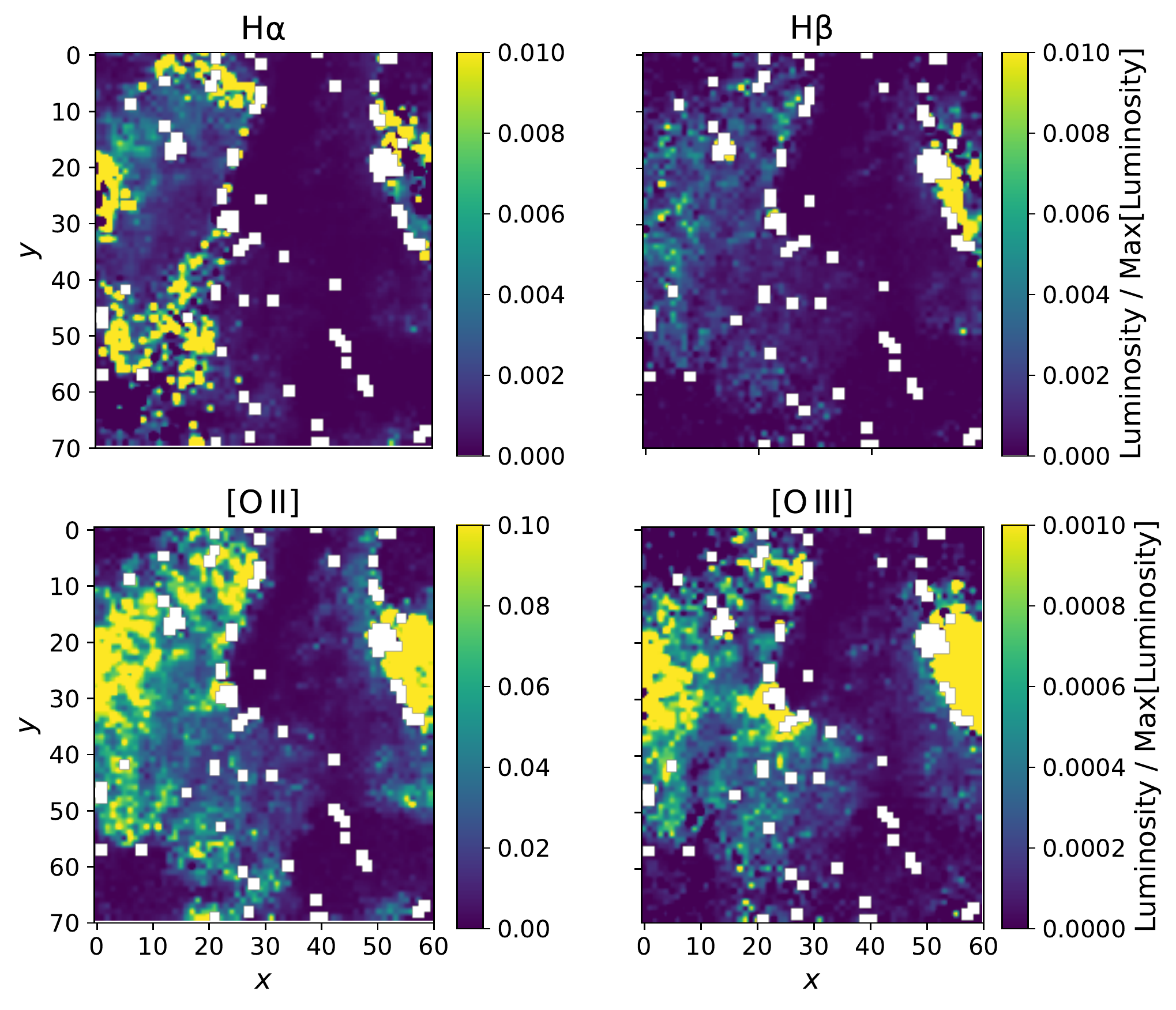}
\centering
\caption{Emission-line luminosity in colour space for H${\rm \alpha}$, H${\rm \beta}$, [\ion{O}{ii}], and [\ion{O}{iii}], as indicated in the title.}
\label{fig:EL_colourspace}
\end{figure}

Self-organising map cells that contain different galaxy populations can be the result of colour-redshift degeneracies in the broadband photometry. Such broadband degeneracies also cause the worse photo-$z$ performance of the baseline network in the problematic colour-space regions. The photo-$z$ performance improves with the MTL training (panel E in Fig.\,\ref{fig:som_map_cosmos}).

The inaccurate photo-$z$ cluster in Fig.\,\ref{fig:som_map_cosmos} is adjacent to an empty colour-space region, which shows up as a blank stripe separating two neighbouring galaxy populations. To understand which galaxies populate cells next to empty regions, we trained a SOM on a simulated galaxy sample (see Sect.\,\ref{sec:data:mocks} for details on the mock) using the $uBVriz$ broadband photometry. The top panel in Fig.\,\ref{fig:som_dist_empty} shows the median distance among the SOM vectors characterising each cell and its directly neighbouring cells (within a 3$\times$3 square). Compared with the bottom panel in the same figure (where we have assigned each galaxy in the mock to a SOM cell), one can visually see that
regions showing larger distances in the upper plot coincide with empty regions (blank stripes) in the bottom ones. Therefore, cells neighbouring empty colour-space regions represent noisier or outlier galaxies, whose colours differ from the rest of the galaxy sample.

To directly see the effect of noise in the SOM, the bottom row in Fig.\,\ref{fig:som_dist_empty} shows the colour-space redshift distribution for the noisy (left) and noiseless (right) colours of the same galaxies. Comparing the two panels demonstrates that the blank region between galaxy populations is broader in the noiseless case. When noise is included, cells on the edges of the empty regions in the right panel are populated. This, together with such cells being located further from the other cells in colour space (top panel), indicates that cells neighbouring empty spaces describe a colour-space region that is not representative of the majority of the galaxy sample (e.g. very noisy galaxies or outliers), which can potentially cause broadband colour-redshift degeneracies.

\subsection{Emission-line confusions}
\label{sec:colourspace:elconf}
The SOM in Fig.\,\ref{fig:p+c_analysis} shows a region in colour space that contains different galaxy populations, which indicates the potential presence of colour-redshift degeneracies. Figure \ref{fig:cluster_cells} shows the photo-$z$s of the galaxies assigned to three different cells within such a colour-space region. \LC{There are four different redshift populations assigned to the region: $z\sim 1.4$, $z\sim0.4$, and $z\sim 1.2$, which is many times confused with galaxies at $z\sim 0.8$.} For the three cells (each of them represented with a different style marker), we plotted the predicted photo-$z$ ($z_{\rm p}$) and the true one ($z_{\rm t}$) with the baseline network ($z_{\rm s}$, blue), the network including MTL ($z_{\rm s}$+NB, red), and that including MTL and photo-$z$ data augmentation ($z_{\rm s}$+NB+$z_{\rm PAUS}$, orange).

The first cell (marked with stars) contains galaxies with $z_{\rm t}\sim0.4$ and the three networks predict the correct redshift. The second cell (marked with crosses) contains galaxies with $z_{\rm true} \sim 0.8$ and $z_{\rm true} \sim 1.2$. In general, the MTL network improve the photo-$z$ prediction of these galaxies. Lastly, the third cell (marked with dots) contains galaxies with redshifts $z_{\rm t} \sim 1.4$. The baseline network predicts these photo-$z$s around $z_{\rm p} \sim 0.8$, and again  the $z_{\rm s}$+NB and the MTL+z$_{\rm PAUS}$ training approaches are able to improve the photo-$z$s. Photo-$z$ confusions from $z_{\rm t} \sim 0.8$ to $z_{\rm t} \sim 1.2$ and from $z_{\rm t} \sim 1.45$ to $z_{\rm t} \sim 1.25$ are recurrent, showing up at several SOM cells within the low photo-$z$ performance cluster. 

Figure \ref{fig:EL_colourspace} explores the mean H${\rm \alpha}$, H${\rm \beta}$, [\ion{O}{II}], and [\ion{O}{III}] emission-line luminosity in colour space. The emission-line luminosity is estimated as 
\begin{equation}
    L_{\rm el} \coloneqq 4\pi\ F_{\rm el}\ D_{\rm L}^2\,,
\end{equation} where $F_{\rm el}$ is the emission-line flux and $D_{\rm L}$ is the luminosity distance, which is estimated assuming Planck 2020 cosmology \citep[][]{planck}. Emission-line fluxes are taken from the photometry catalogue used for the PAUS+COSMOS photo-$z$ \citep[][]{PAUSCOSMOS_alex}, which were estimated by fitting the galaxy photometry to a template that modelled the emission-line fluxes as a 10\,\AA\ wide Gaussian distribution.

Figure \ref{fig:EL_colourspace} shows strong emission lines at the low photo-$z$ performance colour-space regions, for example the regions centred at $(5,30)$ and $(55,25)$. These results, together with the redshift confusions seen in Fig.\,\ref{fig:cluster_cells}, suggest that emission lines are likely to cause degeneracies in broadband data.

Since a high ratio of [\ion{O}{III}] to H${\rm \beta}$ lines may indicate the presence of active galactic nuclei, we first verified that our galaxies do not host a Seyfert nucleus. The distribution of our sample on the `blue' emission-line diagnostic diagram \citep[][]{BPT} classify our sources as star-forming galaxies.  Looking at the correlation of star-formation rates and stellar masses, often called the main sequence \citep[][]{mainsequence}, galaxies showing a photo-$z$ mismatch from $z_{\rm t} \sim 0.8$ to $z_{\rm p} \sim 1.2$ occupy the starburst region \citep[i.e. galaxies with enhanced star formation,][]{gosiacite}. Furthermore, these two emission lines overlap at wavelengths between the $i$- and $z$-broadband filters, which makes the emission line harder to detect. 

Our findings suggest that some photometric features cause the photo-$z$ mismatches. Emission lines have proven helpful to break colour-colour degeneracies and to improve the photo-$z$ estimation \citep{EL_degeneracies}. Despite this, in some regions of colour parameter space emission-line confusion is a potential cause for colour-redshift degeneracies.

\section{Understanding the MTL underlying mechanism}
\label{sec:MTLmech}

In this section we aim to understand the underlying mechanism of MTL that improves the photo-$z$ estimation. In Sect.\,\ref{sec:MTLmech:encoder}, we use a variation of our fiducial network to encode the galaxy photometry in a 2-dimensional space similar to a SOM, while in Sect.\,\ref{sec:MTLmech:extra} we study the impact of using other auxiliary tasks (other than predicting the narrow-band photometry) in the MTL network.

\subsection{Underlying data representation in colour space with MTL}
\label{sec:MTLmech:encoder}

\begin{figure*}
\includegraphics[width= 0.45\textwidth]{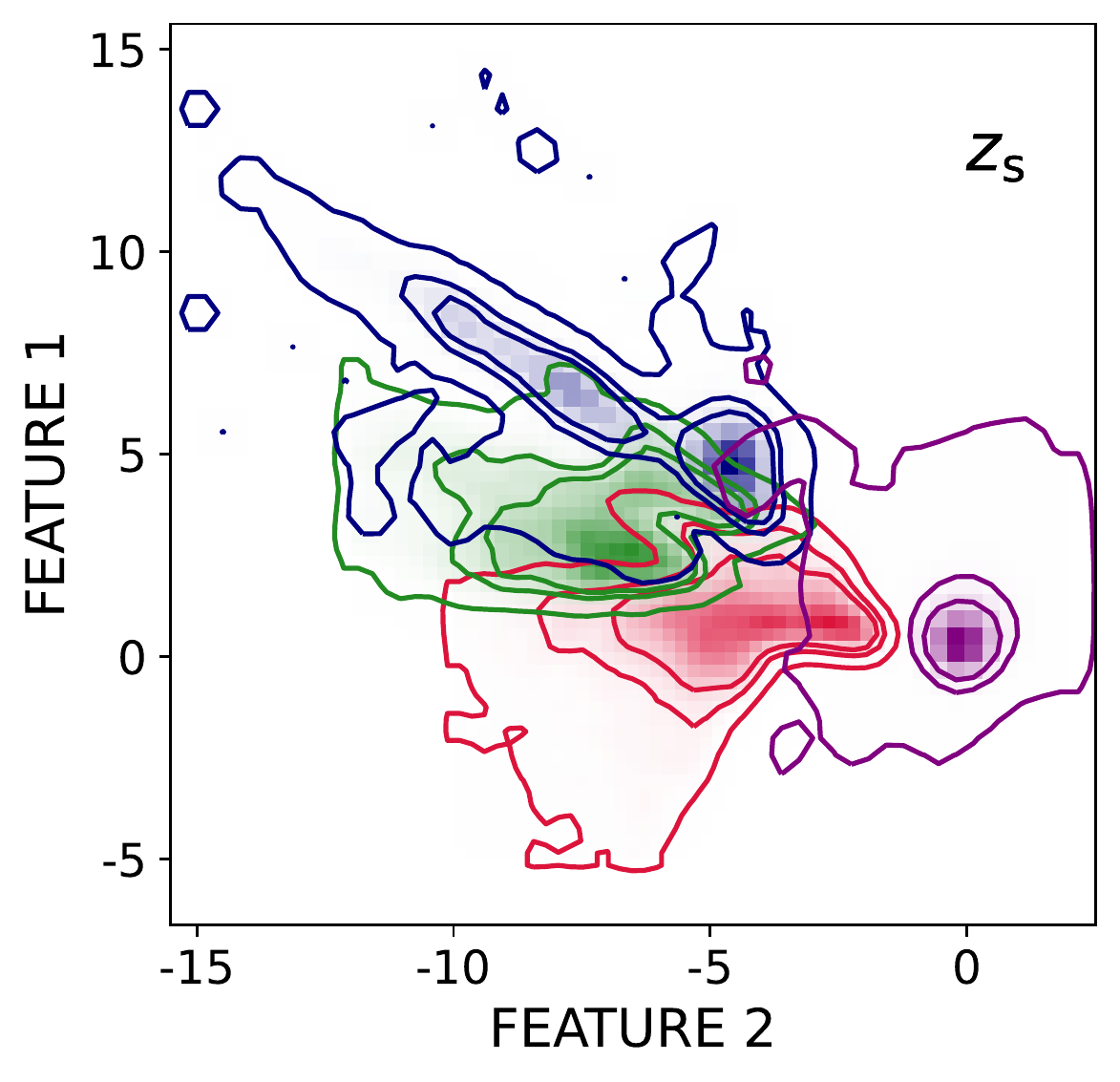}
\includegraphics[width= 0.45\textwidth]{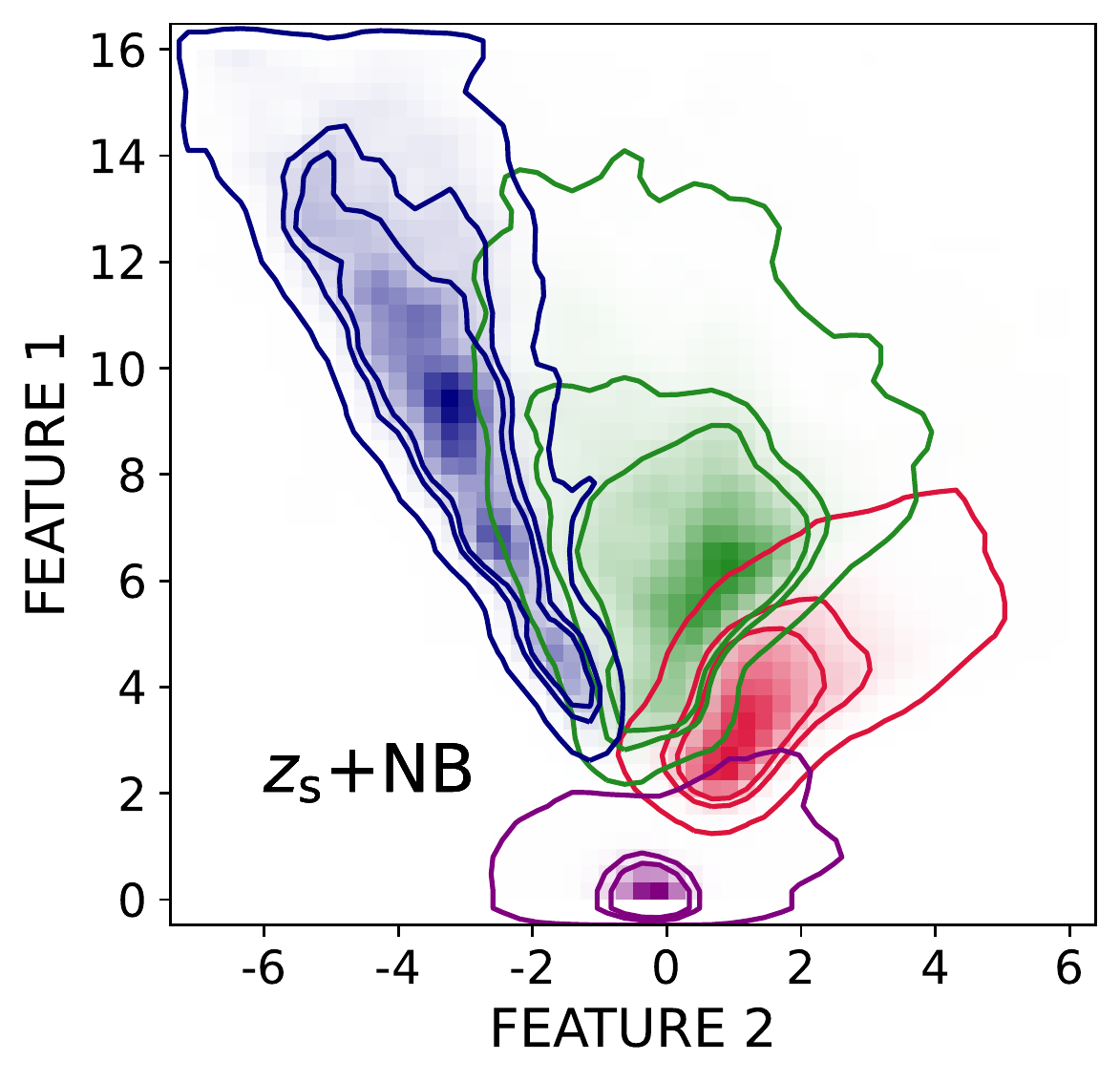}
\centering
\caption{Contours of the two-dimensional feature space coordinates for the $z_{\rm s}$ (\textit{left}) and $z_{\rm s}$+NB (\textit{right}) methods. The features from each of the methods are from independent training and cannot be compared. We can only compare the overlap of the different populations.  }
\label{fig:featurespacedist}
\end{figure*}

For this test, we modify the fiducial network architecture (see Sect.\,\ref{sec:method:arch} and Fig.\,\ref{fig:architecture}). In the modified network, we reduce the input dimension to two features, which are used to predict the photo-$z$ and reconstruct the narrow-band colours.
Encoding the galaxy information in a two-dimensional feature space simplifies its visualisation and brings it closer to the SOM colour-space representation, which we have already studied (Sect.\,\ref{sec:colourspace}). 
 
The galaxy representations in the two-dimensional feature space must encode all the information needed to make the photo-$z$ prediction. Furthermore, in the MTL network, those two numbers are also used to reconstruct the narrow-band photometry and thus must also encode the relevant information for this task. Therefore, comparing the feature space representation of the baseline network (decoding only to the photo-$z$) and the MTL network (also predicting the narrow-band photometry) helps us to better understand why the MTL improves the photo-$z$ estimates. 

As the network's feature space is not constrained, the network can encode the same galaxy differently in several independent trainings. Consequently, the coordinates assigned to each galaxy do not contain any valuable information by themselves and distances from different feature-space maps (e.g. the feature map of the $z_{\rm s}$ network and that of the $z_{\rm s}$+NB) cannot be directly compared. However, the overlap of different redshift populations in feature-space indicate potential degeneracies.

In Fig.\,\ref{fig:featurespacedist} we plot the 50\% , 68\%,  and 95\% contours of the feature-space coordinates for the $z_{\rm s}$ (left panel) and the $z_{\rm s}$+NB (right panel) methods. These are drawn using a test set of 70\,000 Flagship galaxies (Sect.\,\ref{sec:data:mocks}) to $i_{\rm AB}<25$, while the methods train on galaxies to $i_{\rm AB}<23$ (Sect.\,\ref{sec:depth_tests}).  We draw the contours for a selection the redshift bins to show the separation of high-redshift galaxies (red, blue, and green contours), where the MTL method significantly reduces the photo-$z$ scatter with respect to the $z_{\rm s}$ method (Fig.\,\ref{fig:depth_disp}, top and bottom left panels). We have also plotted the contours of a distant redshift population (purple contours) to show that this is further in feature space than the others.\\

There is a significant overlap amongst high-redshift populations in the $z_{\rm s}$ case (left panel). Particularly, the core of the green and blue contours overlap with the red-contour galaxies. We expect some overlap since the three contours are consecutive in redshift; however, the $z_{\rm s}$+NB method shows a cleaner separation between the three redshift populations. This indicates that the $z_{\rm s}$+NB has a better internal representation of the galaxies, where different redshift populations are further in feature space. The narrow-band reconstruction loss (Eq.\,\ref{eq:lossNB}) adds the low-resolution SED information to the training, which can potentially lead to an improved internal representation of galaxies in the two-dimensional feature space. Furthermore, MTL methods also include this information for galaxies without spectroscopic redshift, which effectively acts as a data augmentation technique. This is particularly important for high-redshift galaxies, for which we have very few examples in the spectroscopic sample (Fig.\,\ref{fig:cosmos_zbias}).

\subsection{MTL with other galaxy parameters}

\label{sec:MTLmech:extra}
\begin{figure}
\includegraphics[width= 0.5\textwidth]{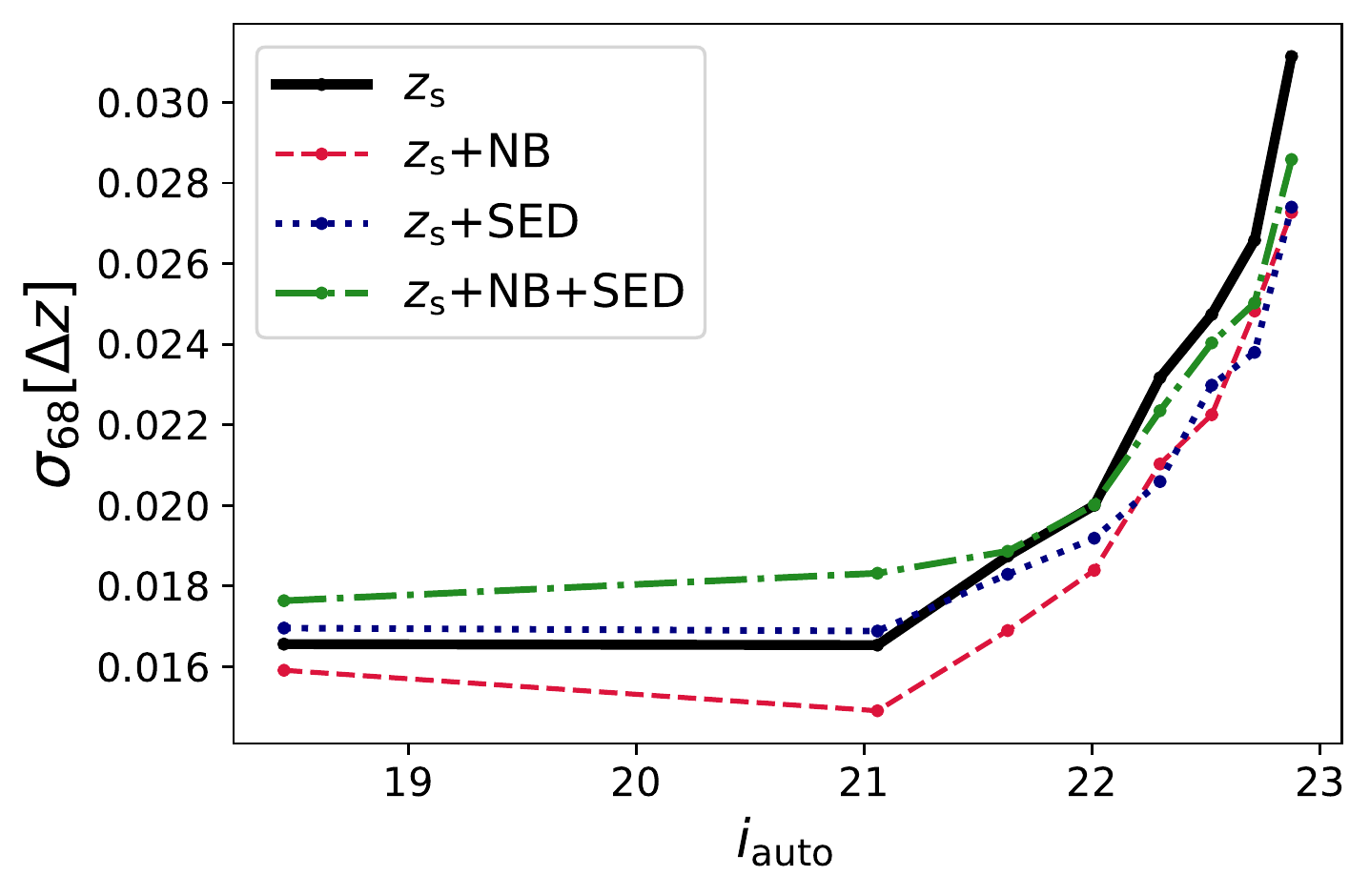}
\centering
\caption{ Photo-$z$ precision in the COSMOS field when the auxiliary task of predicting the galaxy SED is included in the training.} The galaxy SED prediction is addressed as a classification, where the true SED is a class between 1 and 47.
\label{fig:extraparams}
\end{figure}

 So far in this paper, we explored how photo-$z$ predictions benefit from MTL predicting PAUS narrow-band fluxes as an auxiliary task. However, MTL is a more general technique that could be exploited beyond narrow-band photometry reconstructions. While a conventional neural-network training searches for the function ($\phi$) that best predicts the photo-$z$ ($z$) given the broadband photometry ($f$), namely $\phi(z|f)$,  with MTL the optimisation is extended to the function that best predicts the photo-$z$ together with other related parameters ($x_{\rm i}$),
\begin{equation}
    \phi(z,x_{\rm 1},...,x_{\rm N}\,|\,\theta)\,,
\end{equation}
where $x_{\rm i}$  could be any galaxy parameter that correlates with the galaxy photo-$z$ such as the galaxy type.

Template-fitting photo-$z$ methods predict the joint probability distribution $p(z,t|f)$ of the redshift ($z$) and the galaxy type ($t$) and marginalise over the templates \citep[][]{Bpz}. In principle, this is closely related to what MTL does when it is required to predict both quantities at the same time. The network looks for the function that better generalises the prediction of both parameters (e.g. type and redshift), but makes independent predictions in which it `marginalises' over the parameter it is not predicting. 

Figure \ref{fig:extraparams} shows the photo-$z$ precision of data in the COSMOS field when the galaxy type is included as an MTL auxiliary task. The SED template is encoded as a discrete number between 1 and 47 as described in the COSMOS2015 catalogue. These correspond to 31 unique SEDs and 16 SEDs with different extinction laws. Including the SED template (dotted blue line) reduces the photo-$z$ scatter with respect to the baseline network (solid black line). However, MTL using PAUS narrow bands (dashed red line) still provides better photo-$z$ estimates. This result suggests that while the SED helps produce a better representation of the data in colour space (see Sect.\,\ref{sec:MTLmech:encoder}), PAUS narrow-band photometry contains information about the SED, as well as the emission lines or the extinction.

Figure\,\ref{fig:extraparams} also shows the photo-$z$ performance when both the SED and the narrow-band data are used as auxiliary tasks (green dashed-dotted line). We find that this degrades the photo-$z$ performance with respect to using the SED or the narrow-band photometry solely. In theory, using both the narrow-band photometry and the SED number should benefit the network. However, the information available in these two tasks is highly correlated, which can hinder the predictions. Understanding this better is ongoing research and further study is deferred to future work. 

We also explored MTL predicting galaxy parameters such as the star-formation rate, the galaxy mass, and the $E(B-V)$ extinction parameter as auxiliary tasks (not shown). However, none of these parameters improved the predicted photo-$z$s. Furthermore, including the near-infrared photometry did not improve the photo-$z$s either.

\subsection{Effect of narrow-band resolution}
\label{sec:nbres_out}
The improved photo-$z$ from predicting the narrow-band photometry can potentially result from a better internal description of the galaxy SED type. We test this hypothesis by evaluating the performance of the networks using MTL for different resolutions of the output predicted photometry.

\begin{figure}
\includegraphics[width= 0.5\textwidth]{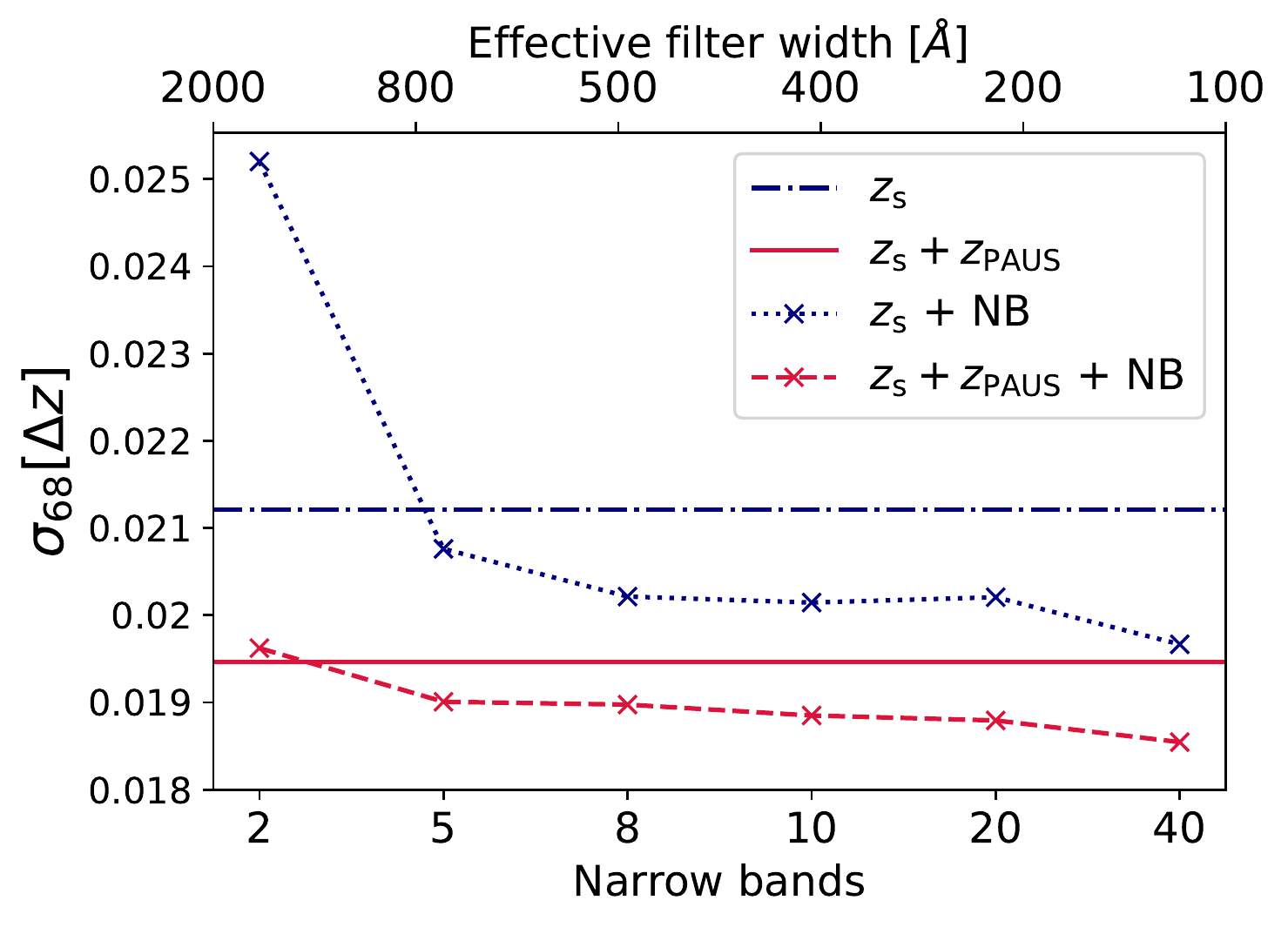}
\centering
\caption{Photo-$z$ precision as a function of number of bands in the predicted photometry for $z_{\rm s}$ +NB (dotted blue line) and  $z_{\rm s}+z_{\rm PAUS}$ +NB (dashed red line). The horizontal line corresponds to the $z_{\rm s}$ (dashed-dotted blue line) and $z_{\rm s}+z_{\rm PAUS}$ (solid red line), where MTL is not enabled.}
\label{fig:output_res}
\end{figure}

Figure \ref{fig:output_res} shows the photo-$z$ precision of the MTL methods as a function of the number of predicted narrow bands (i.e. the output photometry resolution). Assuming the MTL networks use the narrow-band photometry to improve the internal representation of galaxies, increasing the output photometry resolution effectively corresponds to turning on this mechanism. To obtain lower-resolution photometries, we take the mean of groups of consecutive narrow bands (e.g. 2, 4, and 10). Then, we train the $z_{\rm s}$+NB and $z_{\rm s}+z_{\rm PAUS}$+NB methods several times to predict the photo-$z$ and the narrow-band photometry with a different resolution in every training.

The horizontal flat lines in Fig.\,\ref{fig:output_res} indicate the photo-$z$ precision for the methods without MTL; $z_{\rm s}$ (dashed-dotted blue line) and $z_{\rm s}+z_{\rm PAUS}$ (solid red line). The dotted blue line and the dashed red line show the $z_{\rm s}$+NB and $z_{\rm s}+z_{\rm PAUS}$+NB performance for the different output photometry resolutions, respectively. As the output photometry resolution increases, the photometric redshift precision improves. This suggests that the MTL networks are using the narrow-band photometry prediction to improve the internal representation of the SED, and consequently the SED internal fitting, which has a direct impact in the photo-$z$ prediction. The narrow-band photometry contains important additional information about the SED type and galaxy parameters, which are useful when predicting the redshift. 

The $z_{\rm s}$+NB MTL recovering two-band photometry leads to predictions above the $z_{\rm s}$ line, which is the result without MTL. In this limit adding the photometry loss degrade the photo-$z$ results. We trained this network several times to ensure the result was correct, obtaining the same degrading in all cases.

\section{Discussion and conclusions}

Photometric redshifts are crucial for exploiting ongoing and future large galaxy broadband imaging surveys. While covering large sky areas, the broadband spectral resolution limits the redshift performance through colour-redshift degeneracies. The PAUS is a narrow-band imaging survey that can provide very precise photo-$z$ measurements for a combination of wide and deep fields. In this paper we have introduced a new method for improving broadband photo-$z$ estimates, using deep-learning techniques on PAUS narrow-band data.

Multi-task learning is a machine-learning training methodology that aims to improve the performance and generalisation power of a network by training it on several related tasks simultaneously. This forces the model to share representations among related tasks, exploiting their commonalities and enabling the network to generalise better on the original task. We implemented an MTL network that simultaneously predicts the photometric redshift and infers the narrow-band photometry from the broadband photometry (see Sect.\,\ref{sec:method}). The photo-$z$ network is therefore forced to share parameters that are also used to predict the narrow-band photometry, which improves the internal colour-space representation of the data.

In the COSMOS field for galaxies to $i_{\rm AB}<23$, our method reduces the photo-$z$ scatter by approximately 20\% (see Sect.\,\ref{sec:cosmosres:dispersion}) and the number of photo-$z$ outliers by from $\sim$ 1.1\% to $\sim$0.6\% (see Sect.\,\ref{sec:cosmosres:bias}). We also tested the potential of the method for fainter galaxies using \Euclid-like galaxy simulations. For this, we trained the network on a magnitude-limited sample with $i_{\rm AB}<23$ and evaluated it on a sample with $i_{\rm AB}<25$. The MTL predicts up to 16\% more precise photo-$z$s for galaxies with $i_{\rm AB}<25$ than the baseline network (see Sect.\,\ref{sec:depth_tests}).

We used SOMs to study the photo-$z$ performance in different colour-space regions, detecting a region that contains galaxies with degenerate photometry-redshift mappings. This region has a larger photo-$z$ variation within the SOM cells, suggesting that more than one galaxy population is assigned to the same colour-space location (see the left panel in Fig.\,\ref{fig:p+c_analysis}). This correlation results in a photo-$z$ mismatch between two galaxy populations, which affects broadband photo-$z$ estimates. \LC{Our MTL network improves the photo-$z$s in the degenerated colour-space regions using PAUS narrow-band data to learn the underlying colour-space distribution of galaxies.} 

This paper explores how to exploit data from narrow-band photometric surveys such as PAUS to improve broadband photo-$z$ estimates using machine learning. The key point of using MTL instead of, for example, just using the narrow-band photometry to obtain more precise photo-$z$s is that it only requires narrow-band photometry for the training galaxies and the photo-$z$ of any galaxy can be evaluated with only the broadband data. This enables us to exploit fields where we have narrow-band data to obtain better photo-$z$s in other fields where these are not available. PAUS photometry in the COSMOS field is publicly available, so current and future weak lensing surveys, such as \Euclid or the LSST, can readily benefit from this methodology to improve their photo-$z$ estimates. Moreover, MTL is a general machine-learning mechanism that enables fields with different types of photometry to be exploited in order to improve photo-$z$ predictions. While PAUS narrow-band photometry is a clear candidate, other surveys such as J-PAS \citep[][]{Jpas} or ALHAMBRA \citep[][]{Alhambra} provide more fields with interesting data to exploit for the benefit of photo-$z$ estimations.

\begin{acknowledgements}
The PAU Survey is partially supported by MINECO under grants CSD2007-00060, AYA2015-71825, ESP2017-89838, PGC2018-094773, PGC2018-102021, SEV-2016-0588, SEV-2016-0597, MDM-2015-0509, PID2019-111317GB-C31 and Juan de la Cierva fellowship and LACEGAL and EWC Marie Sklodowska-Curie grant No 734374 and no.776247 with ERDF funds from the EU Horizon 2020 Programme, some of which include ERDF funds from the European Union. IEEC and IFAE are partially funded by the CERCA and Beatriu de Pinos program of the Generalitat de Catalunya. Funding for PAUS has also been provided by Durham University (via the ERC StG DEGAS-259586), ETH Zurich, Leiden University (via ERC StG ADULT-279396 and Netherlands Organisation for Scientific Research (NWO) Vici grant 639.043.512), Bochum University (via a Heisenberg grant of the Deutsche Forschungsgemeinschaft (Hi 1495/5-1) as well as an ERC Consolidator Grant (No. 770935)), University College London, Portsmouth support through the Royal Society Wolfson fellowship and from the European Union's Horizon 2020 research and innovation programme under the grant agreement No 776247 EWC. The results published were also funded by the Polish National Agency for Academic Exchange (Bekker grant BPN/BEK/2021/1/00298/DEC/1), the European Union's  Horizon 2020 research and innovation programme under the Maria Skłodowska-Curie (grant agreement No 754510) and by the Spanish Ministry of Science and Innovation through Juan de la Cierva-formacion program (reference FJC2018-038792-I). The PAU data centre is hosted by the Port d'Informaci\'o Cient\'ifica (PIC), maintained through a collaboration of CIEMAT and IFAE, with additional support from Universitat Aut\`onoma de Barcelona and ERDF. We acknowledge the PIC services department team for their support and fruitful discussions. CosmoHub has been developed by the Port d'Informació Científica (PIC), maintained through a collaboration of the Institut de Física d'Altes Energies (IFAE) and the Centro de Investigaciones Energéticas, Medioambientales y Tecnológicas (CIEMAT) and the Institute of Space Sciences (CSIC\&IEEC), and was partially funded by the "Plan Estatal de Investigación Científica y Técnica y de Innovación" program of the Spanish government.
We gratefully acknowledge the support of NVIDIA Corporation with the donation of the Titan V  GPU used for this research.

\AckEC
\end{acknowledgements}

\section*{Data availability}
The PAUS raw data are publicly available through the ING group. A few reduced images are publicly available at \url{https://www.pausurvey.org.} 
The Flagship catalogue is a property of the Euclid Consortium.

\bibliographystyle{aa}
\bibliography{BBphotoz}

\section*{Affiliations}
$^{1}$ Institut de F\'{i}sica d'Altes Energies (IFAE), The Barcelona Institute of Science and Technology, Campus UAB, 08193 Bellaterra (Barcelona), Spain\\
$^{2}$ Port d'Informaci\'{o} Cient\'{i}fica, Campus UAB, C. Albareda s/n, 08193 Bellaterra (Barcelona), Spain\\
$^{3}$ Institute of Space Sciences (ICE, CSIC), Campus UAB, Carrer de Can Magrans, s/n, 08193 Barcelona, Spain\\
$^{4}$ Institut d'Estudis Espacials de Catalunya (IEEC), Carrer Gran Capit\'a 2-4, 08034 Barcelona, Spain\\
$^{5}$ Instituto de F\'isica Te\'orica UAM-CSIC, Campus de Cantoblanco, 28049 Madrid, Spain\\
$^{6}$ Ruhr University Bochum, Faculty of Physics and Astronomy, Astronomical Institute (AIRUB), German Centre for Cosmological Lensing (GCCL), 44780 Bochum, Germany\\
$^{7}$ Leiden Observatory, Leiden University, Niels Bohrweg 2, 2333 CA Leiden, The Netherlands\\
$^{8}$ Department of Physics and Astronomy, University College London, Gower Street, London WC1E 6BT, UK\\
$^{9}$ Instituci\'o Catalana de Recerca i Estudis Avan\c{c}ats (ICREA), 08010 Barcelona, Spain\\
$^{10}$ Centro de Investigaciones Energ\'eticas, Medioambientales y Tecnol\'ogicas (CIEMAT), Avenida Complutense 40, 28040 Madrid, Spain\\
$^{11}$ Institut de Ciencies de l'Espai (IEEC-CSIC), Campus UAB, Carrer de Can Magrans, s/n Cerdanyola del Vall\'es, 08193 Barcelona, Spain\\
$^{12}$ Universit\'e Paris-Saclay, CNRS, Institut d'astrophysique spatiale, 91405, Orsay, France\\
$^{13}$ Institute of Cosmology and Gravitation, University of Portsmouth, Portsmouth PO1 3FX, UK\\
$^{14}$ INAF-Osservatorio di Astrofisica e Scienza dello Spazio di Bologna, Via Piero Gobetti 93/3, 40129 Bologna, Italy\\
$^{15}$ Dipartimento di Fisica e Astronomia, Universit\'a di Bologna, Via Gobetti 93/2, 40129 Bologna, Italy\\
$^{16}$ INFN-Sezione di Bologna, Viale Berti Pichat 6/2, 40127 Bologna, Italy\\
$^{17}$ Max Planck Institute for Extraterrestrial Physics, Giessenbachstr. 1, 85748 Garching, Germany\\
$^{18}$ Universit\"ats-Sternwarte M\"unchen, Fakult\"at f\"ur Physik, Ludwig-Maximilians-Universit\"at M\"unchen, Scheinerstrasse 1, 81679 M\"unchen, Germany\\
$^{19}$ INAF-Osservatorio Astrofisico di Torino, Via Osservatorio 20, 10025 Pino Torinese (TO), Italy\\
$^{20}$ Dipartimento di Fisica, Universit\'a degli studi di Genova, and INFN-Sezione di Genova, via Dodecaneso 33, 16146, Genova, Italy\\
$^{21}$ INFN-Sezione di Roma Tre, Via della Vasca Navale 84, 00146, Roma, Italy\\
$^{22}$ INAF-Osservatorio Astronomico di Capodimonte, Via Moiariello 16, 80131 Napoli, Italy\\
$^{23}$ Instituto de Astrof\'isica e Ci\^encias do Espa\c{c}o, Universidade do Porto, CAUP, Rua das Estrelas, PT4150-762 Porto, Portugal\\
$^{24}$ Dipartimento di Fisica, Universit\'a degli Studi di Torino, Via P. Giuria 1, 10125 Torino, Italy\\
$^{25}$ INFN-Sezione di Torino, Via P. Giuria 1, 10125 Torino, Italy\\
$^{26}$ INAF-IASF Milano, Via Alfonso Corti 12, 20133 Milano, Italy\\
$^{27}$ INAF-Osservatorio Astronomico di Roma, Via Frascati 33, 00078 Monteporzio Catone, Italy\\
$^{28}$ INFN section of Naples, Via Cinthia 6, 80126, Napoli, Italy\\
$^{29}$ Department of Physics "E. Pancini", University Federico II, Via Cinthia 6, 80126, Napoli, Italy\\
$^{30}$ Dipartimento di Fisica e Astronomia "Augusto Righi" - Alma Mater Studiorum Universit\'a di Bologna, Viale Berti Pichat 6/2, 40127 Bologna, Italy\\
$^{31}$ INAF-Osservatorio Astrofisico di Arcetri, Largo E. Fermi 5, 50125, Firenze, Italy\\
$^{32}$ Centre National d'Etudes Spatiales, Toulouse, France\\
$^{33}$ Institut national de physique nucl\'eaire et de physique des particules, 3 rue Michel-Ange, 75794 Paris C\'edex 16, France\\
$^{34}$ Institute for Astronomy, University of Edinburgh, Royal Observatory, Blackford Hill, Edinburgh EH9 3HJ, UK\\
$^{35}$ Jodrell Bank Centre for Astrophysics, Department of Physics and Astronomy, University of Manchester, Oxford Road, Manchester M13 9PL, UK\\
$^{36}$ ESAC/ESA, Camino Bajo del Castillo, s/n., Urb. Villafranca del Castillo, 28692 Villanueva de la Ca\~nada, Madrid, Spain\\
$^{37}$ European Space Agency/ESRIN, Largo Galileo Galilei 1, 00044 Frascati, Roma, Italy\\
$^{38}$ Univ Lyon, Univ Claude Bernard Lyon 1, CNRS/IN2P3, IP2I Lyon, UMR 5822, 69622, Villeurbanne, France\\
$^{39}$ Observatoire de Sauverny, Ecole Polytechnique F\'ed\'erale de Lau- sanne, 1290 Versoix, Switzerland\\
$^{40}$ Mullard Space Science Laboratory, University College London, Holmbury St Mary, Dorking, Surrey RH5 6NT, UK\\
$^{41}$ Departamento de F\'isica, Faculdade de Ci\^encias, Universidade de Lisboa, Edif\'icio C8, Campo Grande, PT1749-016 Lisboa, Portugal\\
$^{42}$ Instituto de Astrof\'isica e Ci\^encias do Espa\c{c}o, Faculdade de Ci\^encias, Universidade de Lisboa, Campo Grande, 1749-016 Lisboa, Portugal\\
$^{43}$ Department of Astronomy, University of Geneva, ch. d'Ecogia 16, 1290 Versoix, Switzerland\\
$^{44}$ Department of Physics, Oxford University, Keble Road, Oxford OX1 3RH, UK\\
$^{45}$ INFN-Padova, Via Marzolo 8, 35131 Padova, Italy\\
$^{46}$ Universit\'e Paris-Saclay, Universit\'e Paris Cit\'e, CEA, CNRS, Astrophysique, Instrumentation et Mod\'elisation Paris-Saclay, 91191 Gif-sur-Yvette, France\\
$^{47}$ INAF-Osservatorio Astronomico di Trieste, Via G. B. Tiepolo 11, 34143 Trieste, Italy\\
$^{48}$ Aix-Marseille Universit\'e, CNRS/IN2P3, CPPM, Marseille, France\\
$^{49}$ Istituto Nazionale di Fisica Nucleare, Sezione di Bologna, Via Irnerio 46, 40126 Bologna, Italy\\
$^{50}$ INAF-Osservatorio Astronomico di Padova, Via dell'Osservatorio 5, 35122 Padova, Italy\\
$^{51}$ Institute of Theoretical Astrophysics, University of Oslo, P.O. Box 1029 Blindern, 0315 Oslo, Norway\\
$^{52}$ Jet Propulsion Laboratory, California Institute of Technology, 4800 Oak Grove Drive, Pasadena, CA, 91109, USA\\
$^{53}$ von Hoerner \& Sulger GmbH, Schlo{\ss}Platz 8, 68723 Schwetzingen, Germany\\
$^{54}$ Technical University of Denmark, Elektrovej 327, 2800 Kgs. Lyngby, Denmark\\
$^{55}$ Institut d'Astrophysique de Paris, UMR 7095, CNRS, and Sorbonne Universit\'e, 98 bis boulevard Arago, 75014 Paris, France\\
$^{56}$ Max-Planck-Institut f\"ur Astronomie, K\"onigstuhl 17, 69117 Heidelberg, Germany\\
$^{57}$ Department of Physics and Helsinki Institute of Physics, Gustaf H\"allstr\"omin katu 2, 00014 University of Helsinki, Finland\\
$^{58}$ NOVA optical infrared instrumentation group at ASTRON, Oude Hoogeveensedijk 4, 7991PD, Dwingeloo, The Netherlands\\
$^{59}$ Argelander-Institut f\"ur Astronomie, Universit\"at Bonn, Auf dem H\"ugel 71, 53121 Bonn, Germany\\
$^{60}$ Dipartimento di Fisica e Astronomia "Augusto Righi" - Alma Mater Studiorum Universit\`{a} di Bologna, via Piero Gobetti 93/2, 40129 Bologna, Italy\\
$^{61}$ Department of Physics, Institute for Computational Cosmology, Durham University, South Road, DH1 3LE, UK\\
$^{62}$  Universit\'e Paris Cit\'e, CNRS, Astroparticule et Cosmologie, 75013 Paris, France\\
$^{63}$ INFN-Bologna, Via Irnerio 46, 40126 Bologna, Italy\\
$^{64}$ Institute of Physics, Laboratory of Astrophysics, Ecole Polytechnique F\'{e}d\'{e}rale de Lausanne (EPFL), Observatoire de Sauverny, 1290 Versoix, Switzerland\\
$^{65}$ European Space Agency/ESTEC, Keplerlaan 1, 2201 AZ Noordwijk, The Netherlands\\
$^{66}$ Department of Physics and Astronomy, University of Aarhus, Ny Munkegade 120, DK-8000 Aarhus C, Denmark\\
$^{67}$ Space Science Data Center, Italian Space Agency, via del Politecnico snc, 00133 Roma, Italy\\
$^{68}$ Institute of Space Science, Bucharest, 077125, Romania\\
$^{69}$ Instituto de Astrof\'isica de Canarias, Calle V\'ia L\'actea s/n, 38204, San Crist\'obal de La Laguna, Tenerife, Spain\\
$^{70}$ Departamento de Astrof\'{i}sica, Universidad de La Laguna, 38206, La Laguna, Tenerife, Spain\\
$^{71}$ Dipartimento di Fisica e Astronomia "G.Galilei", Universit\'a di Padova, Via Marzolo 8, 35131 Padova, Italy\\
$^{72}$ Instituto de Astrof\'isica e Ci\^encias do Espa\c{c}o, Faculdade de Ci\^encias, Universidade de Lisboa, Tapada da Ajuda, 1349-018 Lisboa, Portugal\\
$^{73}$ Universidad Polit\'ecnica de Cartagena, Departamento de Electr\'onica y Tecnolog\'ia de Computadoras, 30202 Cartagena, Spain\\
$^{74}$ Universit\'e de Gen\`eve, D\'epartement de Physique Th\'eorique and Centre for Astroparticle Physics, 24 quai Ernest-Ansermet, CH-1211 Gen\`eve 4, Switzerland\\
$^{75}$ Kapteyn Astronomical Institute, University of Groningen, PO Box 800, 9700 AV Groningen, The Netherlands\\
$^{76}$ Infrared Processing and Analysis Center, California Institute of Technology, Pasadena, CA 91125, USA\\
$^{77}$ INAF-Osservatorio Astronomico di Brera, Via Brera 28, 20122 Milano, Italy\\
$^{78}$ Junia, EPA department, 59000 Lille, France

\begin{appendix}
\section{Self-organising maps}
\label{App:soms}
A SOM \citep{soms} is an unsupervised machine-learning algorithm trained to produce a low-dimensional (typically two-dimensional) representation of a multi-dimensional space. A two-dimensional SOM contains $N_{x} \times N_{y}$ cells with an associated vector of attributes ($\vec{w}^{k}$), where $N_{x}(N_y)$ is the dimension of the SOM on the $x(y)$-axis, and $k$ corresponds to the $k$th SOM cell. Each of these vectors has the same length as the input data.  

The SOM training phase is an iterative process during which the SOM cells compete amongst themselves to represent the training data. Initially, the cell vectors ($\vec{w}^{k}$) are randomly sampled from a uniform distribution, and these are updated after each iteration step ($t$). In every training iteration, each galaxy vector of measured attributes $\vec{x}$ (e.g. in our case the galaxy colours), is compared to all the SOM cells\rq\ vectors via a $\chi^2$ expression,
\begin{equation}
    \chi^2 \left( \vec{w}^{k}(t), \vec{x} \right) = \sum_{i}\ \left[ \frac{x_{i} - w^{ k}_{i}(t)}{\sigma_{i}}\right]^2\,,
    \label{eq:chi2_som}
\end{equation} where $i$ sums over galaxy attributes and $\sigma_i$ is the uncertainty associated with $x_{i}$. The evaluated galaxy is assigned to the cell with the lowest $\chi^2$, which updates its associated vector of attributes $\vec{w}^{k}(t)$ according to the matched galaxy features


Furthermore, in the SOM training procedure, the vector of features from cells neighbouring the best matching cell are also updated, clustering together galaxies with similar attributes. This is implemented with a neighbouring function $H(t,d)$, which depends on the distance ($d$) between the best matching cell and the updated one. The neighbouring function is commonly implemented as a Gaussian kernel with an iteration-dependent variance $\sigma_{\rm kernel}^2(t)$. Therefore, the vector of attributes for a particular cell $k$ after iteration $t+1$ is
\begin{equation}
    \vec{w}^{k}(t+1) = \vec{w}^{k}(t)\ +\ \alpha(t)\ H\left(t,|\vec{w}-\vec{x}|\right)\ \left(\vec{x} - \vec{w}^{k}(t) \right)\,,
\end{equation} where $\alpha(t)$ is the learning rate.

After a few iterations over the training sample, the result is a map of $(N_{x} \times N_{y}$) vectors in a two-dimensional space grouping together cells with similar features while preserving the topology of the multi-dimensional space.

\section{Redshift distributions, N({\it{z}}), and scatter plots}
\label{sec:cosmosres:nz}

\begin{figure*}
\includegraphics[width= 0.9\textwidth]{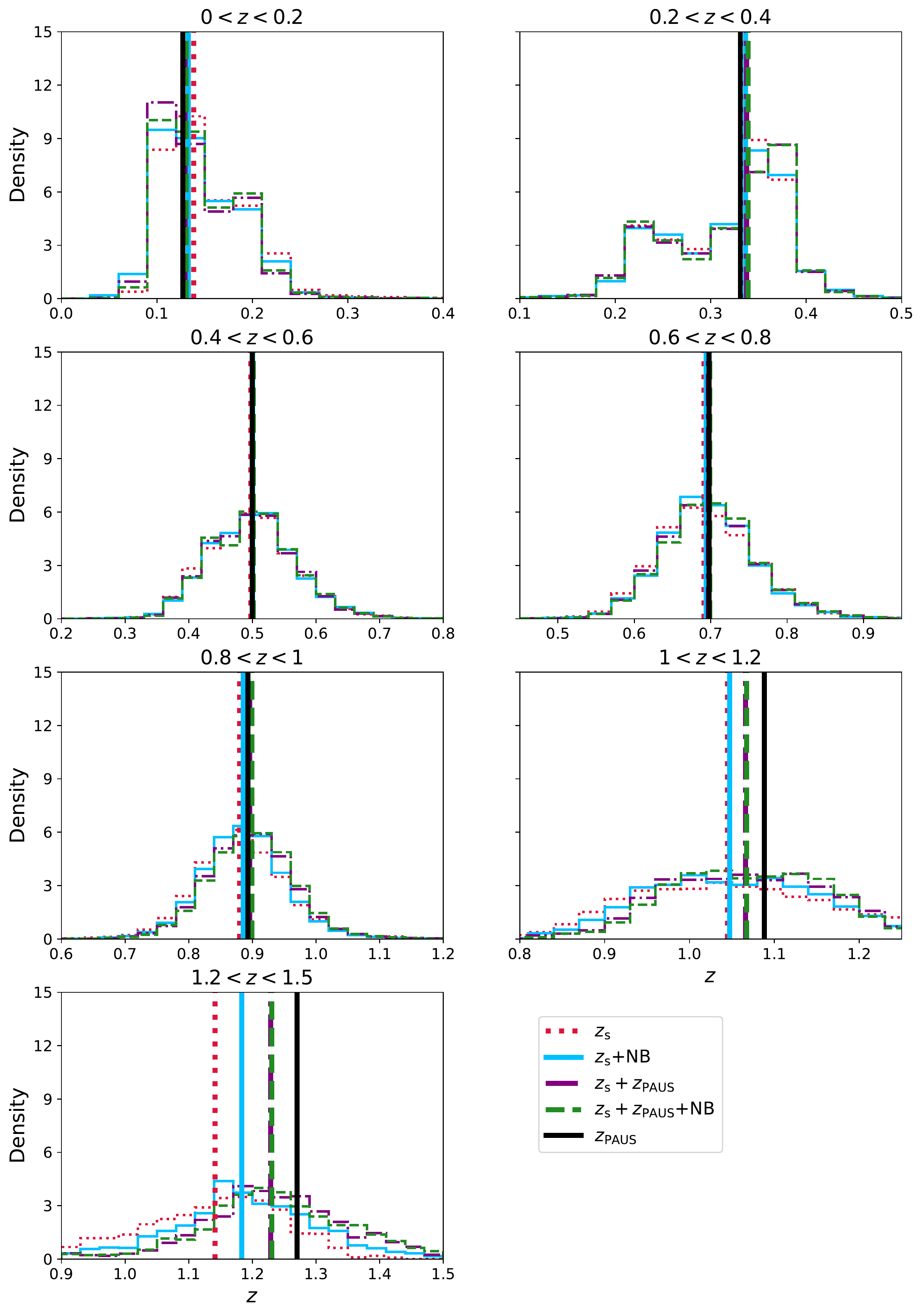}
\centering
\caption{ $N(z)$ estimates of the full COSMOS sample divided into seven tomographic bins over the redshift range $0<z<1.5$. Tomographic bins are defined using the spectroscopic redshifts and the PAUS+COSMOS high-precision photo-$z$s for galaxies without spectroscopy. \RR{The vertical solid black lines indicate the median ground-truth redshift, while the other vertical lines indicate the median redshifts of the $N(z)$ estimates.} Unseen lines are hidden by other overlapping lines.}
\label{fig:NofZ}
\end{figure*}

\begin{figure*}
\includegraphics[width= 0.9\textwidth]{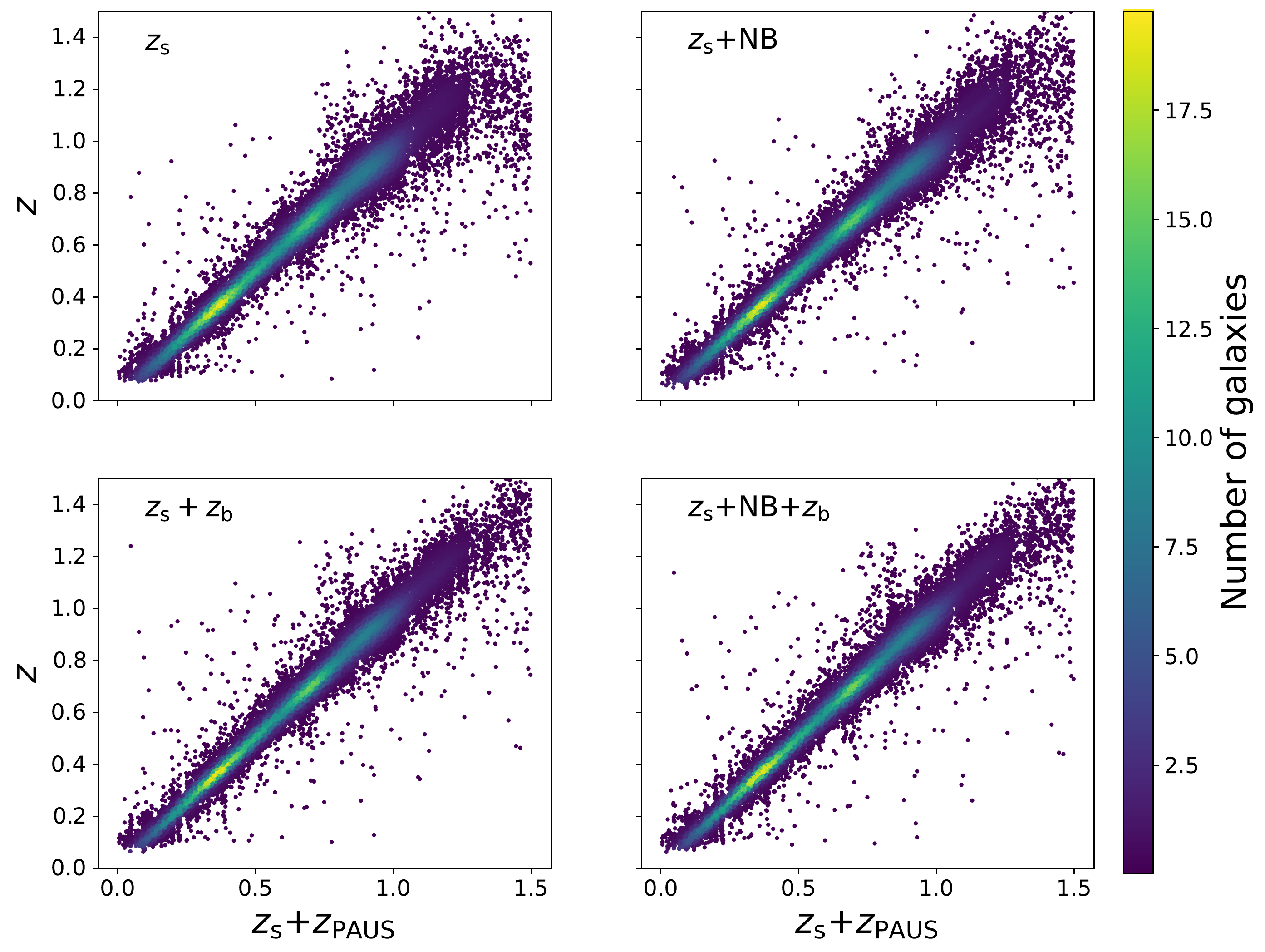}
\centering
\caption{Scatter plot of the 1:1 relation between the predicted photo-$z$ and the true redshift, which is a combination of spectroscopic redshift and PAUS+COSMOS photo-$z$s, in the complete COSMOS sample for the four methods in Sect.\,\ref{sec:method}.}
\label{fig:scatter}
\end{figure*}

Unbiased redshift distributions, $N(z)$, are crucial for a variety of science applications, with the most stringent requirements being in weak lensing \citep[e.g.][]{BB_hildebrant,Hoyle}. broadband photo-$z$s commonly suffer from biases due to degeneracies between colours and redshift, \citep[e.g.][]{color-redshift2,Masters1}.  

Figure \ref{fig:NofZ} shows $N(z)$ in tomographic redshift bins for $0<z_{\rm t}<1.5$ spaced by 0.2. The last tomographic bin is defined from $1.2 < z_{\rm t} < 1.5$  so that the number of galaxies in the bin is increased. The ground-truth redshift defining the tomographic bins ($z_{\rm t}$) is a combination of the spectroscopic redshift (when it is available) and the PAUS+COSMOS photo-$z$ elsewhere. The vertical solid grey line indicates the ground-truth median redshift of the tomographic bin, while the dashed coloured lines represent the median redshifts of the predicted photo-$z$s assigned to the bins. 

Multi-task learning  with photo-$z$ data augmentation ($z_{\rm s}+z_{\rm PAUS}$+NB) always provides equal or more accurate $N(z)$ than the baseline network ($z_{\rm s}$, black line). As expected from Fig.\,\ref{fig:cosmos_zbias}, the $N(z)$ values exhibiting the largest bias are those with $z_{\rm t}>1.2$, particularly the bin at $z_{\rm t} > 1.2$. In this bin, MTL together with the photo-$z$ data augmentation ($z_{\rm s}$+NB+$z_{\rm PAUS}$, green line), significantly shifts the median of the $N(z)$ towards the PAUS+COSMOS result.

Commonly, redshift distributions require a bias correction to reach the accuracy requirements of cosmological measurements. Techniques such as clustering redshifts are applied to correct such biases  \citep{clustering_z_1,clustering_z_2,clustering_z_3,clusteringz_5,SOM_hendrik}. MTL reduces the bias of the $N(z)$  already at the photo-$z$ prediction stage. Even if the MTL photo-$z$s still require some correction, the final redshift distributions would benefit from initially having less biased redshift distributions (if these redshift distributions are used to fit the clustering-$z$ data points).

\LC{Figure \ref{fig:scatter} shows the density scatter between the predicted photo-$z$s and the true redshift. Here, we are plotting the complete COSMOS sample, and we therefore use a combination of spectroscopic redshift and PAUS+COSMOS photo-$z$s as true redshift. The top left panel corresponds to baseline network ($z_{\rm s}$ method) and clearly shows higher photo-$z$ scatter and more outliers in the high-redshift region with respect to the other methods. The MTL method (top right, $z_{\rm s}$+NB) already reduces photo-$z$ scatter and number of high-redshift outliers. The methods including additional PAUS+COSMOS photo-$z$s in the training sample (bottom panels) further improve the photo-$z$ performance.}

\newpage
\section{Effect of training with photo-{\it z}s as ground-truth targets}
\label{App:photoz_targets}

In this work we have implemented and tested two training methodologies that rely on narrow-band photo-$z$ estimates as ground-truth targets (see methods $z_{\rm s}+z_{\rm PAUS}$ and NB+$z_{\rm s}+z_{\rm PAUS}$ in Sect.\,\ref{sec:method:arch}). Even if such photo-$z$s are overall very accurate, its implementation in the training could harm the photo-$z$ performance since these are less precise than spectroscopic redshifts and could potentially include outliers. In this section we explore the effect that less precise redshift labels (Sect.\,\ref{App:photozeffect:precision}) and the presence of outliers (Sect.\,\ref{App:photozeffect:out}) have on the photo-$z$ performance.
\begin{figure}
\includegraphics[width= 0.47\textwidth]{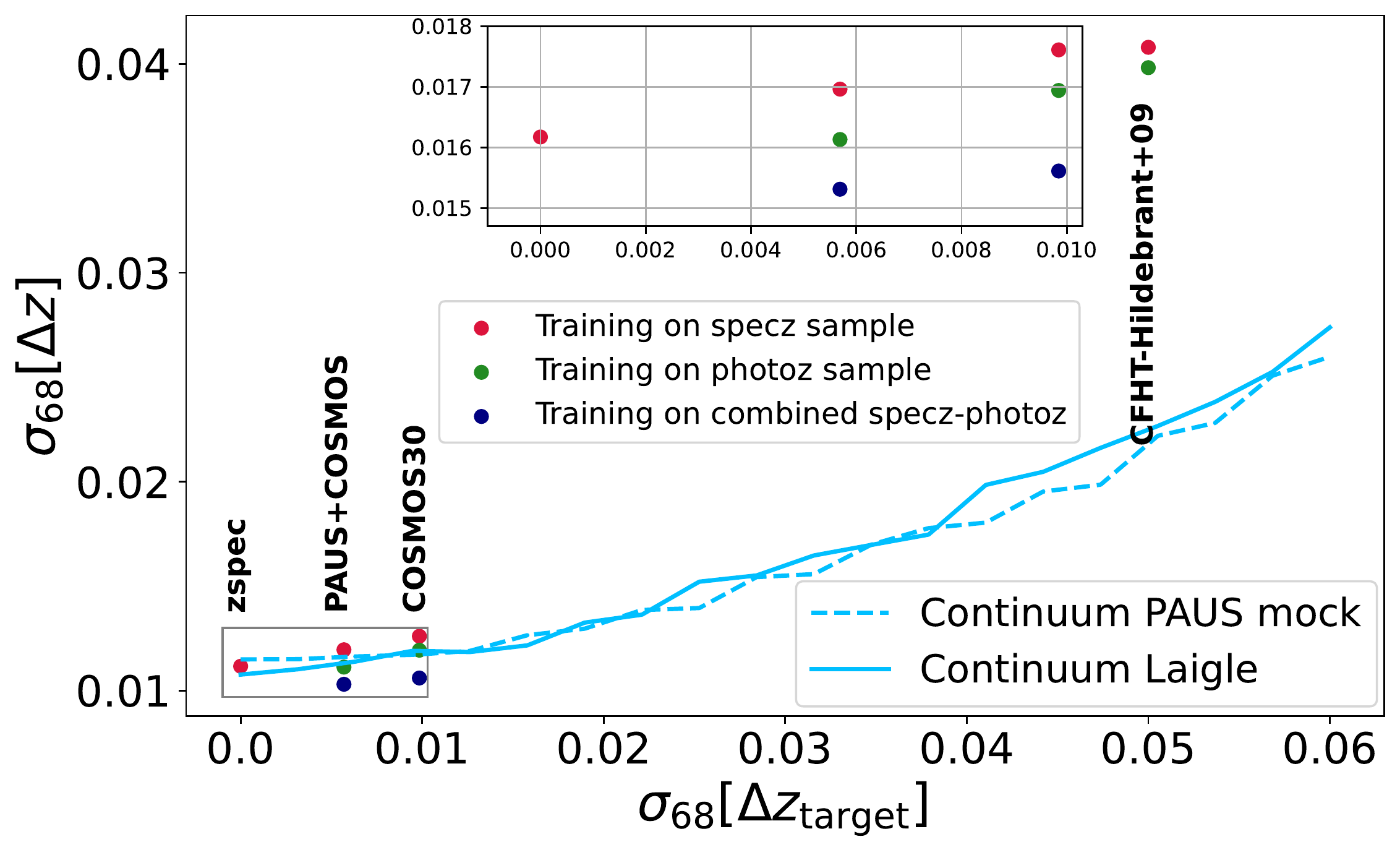}
\centering
\caption{Photo-$z$ performance as a function of the ground-truth redshift precision used for training the networks. The training redshifts are the spectroscopic redshifts, the PAUS+COSMOS photo-$z$s, COSMOS30, and a set of CFHT photo-$z$s in COSMOS. Red points correspond to training on the spectroscopic sample (around 6000 galaxies). The green and red points show the training sample extended to COSMOS galaxies with photo-$z$s (around 15\,000 galaxies). The blue lines show the expected photo-$z$ performance as a function of target redshift precision. The true redshifts, the spectroscopic redshift in the COSMOS2015 catalogue (solid blue line), and the simulated redshift in the PAUS mock (blue dashed line) are scattered with precision in 0.001 bins. The top inset zooms into the framed area in the main plot (lower-left corner)}
\label{fig:labelscatter}
\end{figure}

\subsection{Effect of the dispersion in the ground-truth photo-{\it z}}
\label{App:photozeffect:precision}

Figure \ref{fig:labelscatter} shows the photo-$z$ precision of a set of 1000 spectroscopic galaxies for four independent broadband networks (simply mapping colours to redshift), each of them trained with different ground-truth redshifts. The redshifts used for training are the spectroscopic redshifts (see Sect.\,\ref{sec:data:sampzs}), the PAUS+COSMOS photo-$z$s (see Sect.\,\ref{sec:data:samppz}), the COSMOS30 photo-$z$s \citep{Laigle}, which combine 30 photometric filters and estimates the photo-$z$ with \texttt{LePhare} \citep[][]{Lephare}, and a set of CFHT photo-$z$s from \citet{BB_hildebrant} combining six broad bands ($ugriz$) with photo-$z$ estimated with \texttt{BPZ} \citep{Bpz}. The input data are, in all cases, the CFHT $u$ band and the $BVriz$ Subaru broadband filters from COSMOS2015.

The red points in Fig.\,\ref{fig:labelscatter} show the redshift dispersion using a training sample of galaxies with spectroscopic redshift. We always keep the same training sample (which contains around 6000 galaxies) and change the labelled true redshifts in each independent training (spectroscopic catalogue, PAUS data, COSMOS30, and the CFHT catalogue). Using spectroscopic redshifts as ground-truth redshifts results in a dispersion of $\sigma_{\rm 68} = 0.016$. Replacing the spectroscopic redshift with the photo-$z$ from PAUS+COSMOS, COSMOS30, or CFHT yields $\sigma_{\rm 68} = 0.017$, $\sigma_{\rm 68} = 0.018$, and $\sigma_{\rm 68} = 0.046$, respectively. As the ground-truth redshifts become less precise, the machine-learning photo-$z$ performance degrades.

To obtain the green points (Fig.\,\ref{fig:labelscatter}), we extended the training sample to all galaxies in the COSMOS sample with a photo-$z$ estimate, which results in approximately 15\,000 galaxies when the four catalogues are merged. Then, three independent networks are trained using the PAUS+COSMOS, the COSMOS-30, and the CFHT photo-$z$s as true redshifts (the spectroscopic redshift is not used even if it is available). This provides a precision of $\sigma_{\rm 68} = 0.016$, $\sigma_{\rm 68} = 0.017$, and $\sigma_{\rm 68} = 0.045$ for the PAUS+COSMOS, the COSMOS30, and the CFHT photo-$z$s, respectively. The three networks improve the photo-$z$ precision with respect to training with spectroscopic redshifts only. Indeed, with the PAUS+COSMOS photo-$z$ labels we already reach the photo-$z$ precision with spectroscopic labels.

Finally, the blue points in the figure correspond to the networks trained with the same 15\,000 photo-$z$ galaxies as in the green points, but combining spectroscopic redshifts (if available) and photo-$z$s as ground-truth training redshifts. Combining spectroscopic redshifts with PAUS+COSMOS photo-$z$s yields  $\sigma_{\rm 68} = 0.015$, which improves upon the precision obtained with spectroscopic redshifts only. 

The light blue lines in Fig.\,\ref{fig:labelscatter} show the expected performance as a function of the ground-truth redshift precision. The solid line uses the COSMOS2015 $uBVriz$ broad bands and the dashed one uses simulated data from the PAUS mock described in Sect.\,\ref{sec:depth_tests}. In both cases, true redshifts (spectroscopic or simulated) are scattered with the corresponding dispersion in the abscissa.

Both networks (solid and dashed lines) are trained with 15\,000 galaxies to have a direct comparison with the previous results. We always use the scattered redshifts as ground-truth targets, in such a way that the lines should be compared with the green points since these are trained using only photometric redshifts. The results obtained with the PAUS+COSMOS and COSMOS30 match the expectation curves, but there is a significant mismatch with the CFHT photo-$z$s. This is potentially triggered by systematic errors or outliers in the CFHT photo-$z$ not represented in the blue curves.

\subsection{Effect of photo-{\it z} outliers in the training redshifts}
\label{App:photozeffect:out}

\begin{figure*}
\includegraphics[width= 0.98\textwidth]{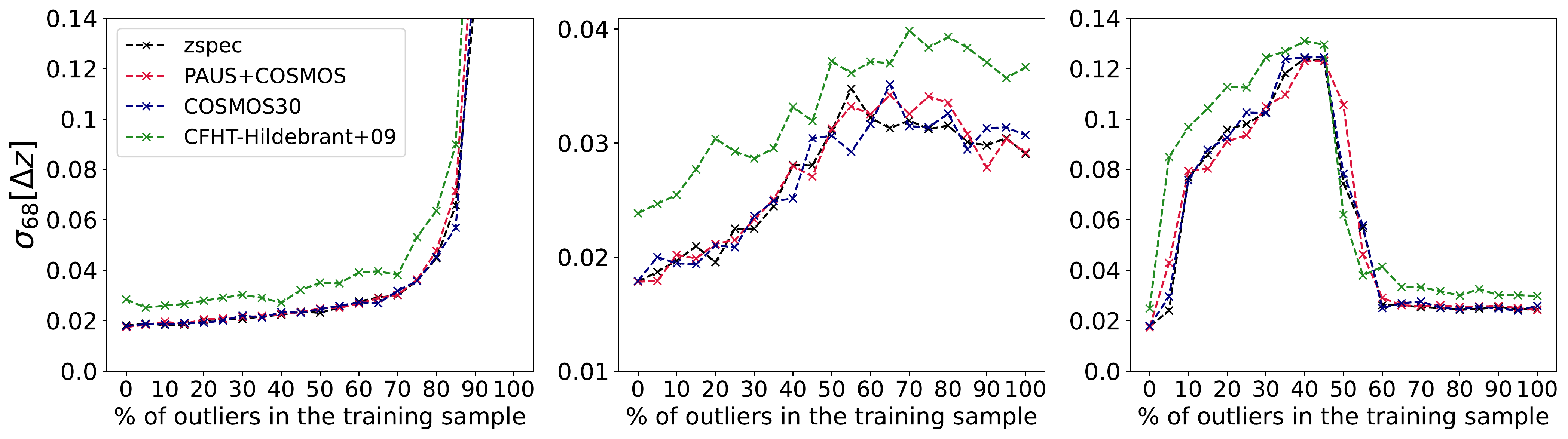}
\centering
\caption{Effect of outliers and systematic errors on the ground-truth redshift sample used during training. The training sample consists of 5000 spectroscopic galaxies with photometry from COSMOS2015. Each coloured line uses a different sample of redshifts as true redshifts, i.e. spectroscopic redshifts (black), PAUS+COSMOS photo-$z$s (red), COSMOS30 photo-$z$s (blue), and CFHT photo-$z$s (green). The ground-truth redshift of the selected fraction of training galaxies is replaced by a random redshift value sampled from $U(0,1.5)$ (\textit{left}), a 20\% higher redshift (\textit{centre}), and redshifts modified with Eq.\,\eqref{eq:zmod_el} (\textit{right}).}
\label{fig:labeloutliers}
\end{figure*}

\begin{figure*}
\includegraphics[width= 0.98\textwidth]{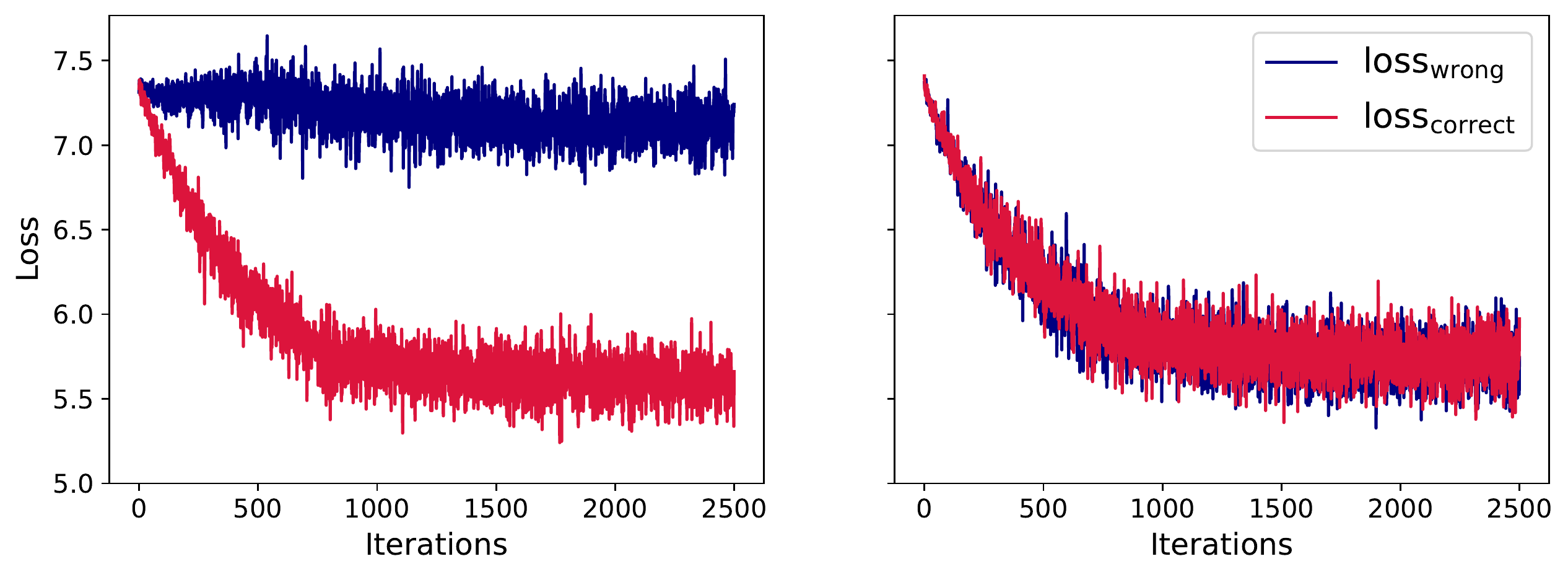}
\centering
\caption{Training loss function for galaxies with a wrong (blue) and a corrected (red) target redshift. The training sample consists of 5000 spectroscopic galaxies with photometry from COSMOS2015. In the left panel, the modified target redshifts are randomly switched to a value drawn from $U(0,1.5)$, while in the right panel the wrong redshift labels are generated with Eq.\,\eqref{eq:zmod_el}.}
\label{fig:lossoutliers}
\end{figure*}

Figure \ref{fig:labelscatter} showed a mismatch between the expected (solid blue curve) and photo-$z$ performance training a network with the CFHT photo-$z$s as ground-truth redshifts (rightmost red point). However, the expectation assumes that the CFHT photo-$z$s are not affected by other effects such as systematic errors or catastrophic outliers.

Figure \ref{fig:labeloutliers} shows the effect of outliers in the ground-truth targets of the training sample. The network is trained 20 independent times with 5000 COSMOS2015 spectroscopic galaxies including a fraction (monotonically increasing in each iteration) of labelled photo-$z$ outliers. This procedure is repeated for the spectroscopic redshifts (black line), the PAUS+COSMOS photo-$z$s (red line), the COSMOS30 photo-$z$s (blue line), and the CFHT photo-$z$s (green line).

In the left panel, the artificial outlier redshifts are swapped with a random value sampled from a uniform distribution $U(0,1.5)$ to simulate catastrophic outliers. The predicted photo-$z$ precision degrades as the fraction of target redshift outliers increases. This also affects the predicted $p(z)$, which become noisier and broader (not shown). However, and unexpectedly, the network can provide reasonable photo-$z$ estimates with up to 80\% of catastrophic outliers in the training sample. Furthermore, the network is able to make reliable photo-$z$ predictions of galaxies that have been used in the training sample with wrong target redshift values. This result holds when either spectroscopic redshifts or any of the photo-$z$s are used for training. 

The middle panel shows the effect of a systematic multiplicative shift in the training sample redshifts, where the selected targets are shifted to 20\% higher redshifts. In this scenario, the predicted photo-$z$ precision degrades faster than when outliers are random (left panel) but the network does never completely break. For an outlier fraction higher than 60\%, the precision settles at $\sigma_{\rm 68} = 0.03$, but the bias rapidly increases. Finally, the rightmost panel presents the effect of a systematic shifting the redshift ($z_{\rm mod}$) so that \ion{O}{iii} is confused with H${\rm \alpha}$ in the training redshifts:
\begin{equation}
z_{\rm mod} = \lambda_{\rm \ion{O}{iii}}/ \lambda_{\rm H\alpha}\; (1 + z_{\rm t}) -1\;,
\label{eq:zmod_el}
\end{equation} 
where $z_{\rm t}$ is the galaxy redshift.

The training degrades and breaks much faster than in the two previous cases, where with around 40\% of wrong target redshifts the network is not able to provide reliable predictions. As the fraction of affected target redshifts increases, the predicted $p(z)$ become more doubly peaked. Moreover, a plot of photo-$z$ versus spec-$z$ scatter displays two clear lines, one with the correct mapping and another shifted upwards (not shown), which is triggered by the training objects with the photo-$z$ artificially shifted to confuse the emission lines. Again, the effect of outliers is similar regardless of the redshifts used for training (spectroscopic or different-precision photo-$z$).

Contrary to expectations, the left panel of Fig.\,\ref{fig:labeloutliers} indicates that the network can learn the mapping between the galaxy photometry and redshifts with up to 80\% of catastrophic outliers in the training sample. Given that the training sample is composed of 5000 galaxies, this means that the network can effectively learn the colour-redshift relations from 1000 galaxies,  learning to ignore the remaining 4000 spurious galaxies. 

Figure~\ref{fig:lossoutliers} shows the cost function evolution of a network trained with wrong target redshifts for half of the training while keeping the rest to the correct redshift values. The cost function is split in two; one for those objects with correct redshift (red) and another for those with wrong redshifts (blue). In the left panel, the modified target redshifts are switched to a random value from a uniform distribution $U(0,1.5)$, as in the left panel of Fig.\,\ref{fig:labeloutliers}. The cost function of galaxies with the correct target redshift decreases, which indicates that the network is learning from them. In contrast, the cost function of incorrectly labelled galaxies remains constant along the training, showing that the network is not learning anything from them. Therefore, the network is effectively only learning from galaxies with correct target redshifts. Randomly swapping redshifts to different values breaks any correlation between the photometry and the redshifts. Hence, the network is only learning the colour-redshift mapping from galaxies with the correct target redshift. Nevertheless, having a large fraction of wrong labels adds noise to the training, broadening the predicted $p(z)$. 

The right panel of Fig.\,\ref{fig:lossoutliers} shows the loss function for the correct and the wrongly labelled training galaxies separately when the incorrect redshift labels are generated with Eq.\,\eqref{eq:zmod_el}. This introduces a new colour-redshift relation that forces the network to learn both from galaxies with wrong and correct target redshifts. This can also be noted in the $p(z)$ behaviour, which presents a double-peaked distribution (not shown).  Hence, Figs.\,\ref{fig:labeloutliers} and \ref{fig:lossoutliers} indicate that having catastrophic outliers in the training sample labels effectively adds noise to the photo-$z$ predictions. In contrast, a systematic bias in the training sample targets produces a bias in such predictions. 

\section{Robustness of the methods to outliers in the target redshifts}
\label{sec:app:flagship+}
\begin{figure*}[h!]
\includegraphics[width= 0.98\textwidth]{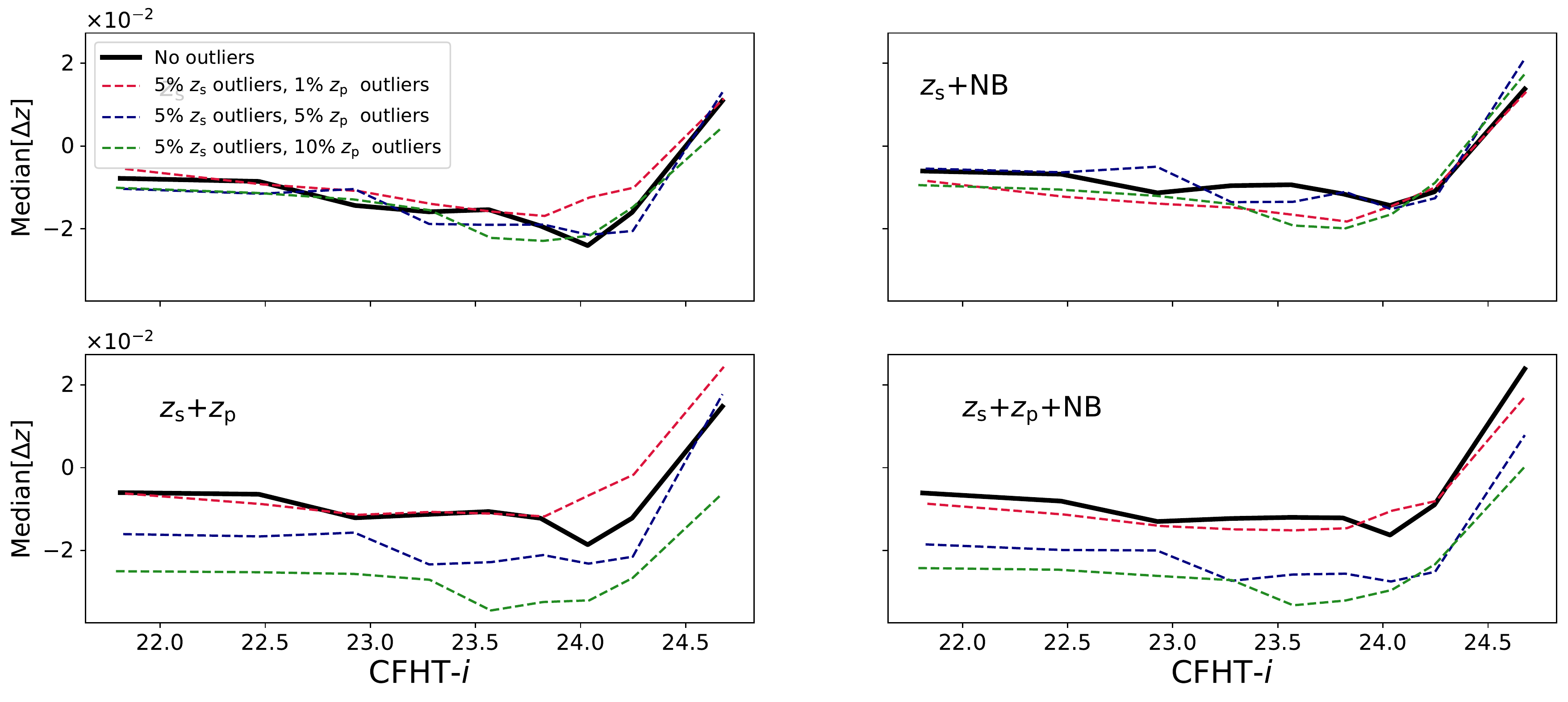}
\centering
\caption{Effect on the photo-$z$ predictions of different outlier rates in the target spectroscopic and high-precision photometric redshifts used as ground-truth targets to train the methods in Sect.\,\ref{sec:method}. In all cases, outliers have been included following Eq.\eqref{eq:zmod_el}. }
\label{fig:outliers_fs2}
\end{figure*}
\LC{In this appendix we study the robustness of our training methodologies to outliers in the redshifts used as ground-truth to train the network. In Sect.\,\ref{sec:depth_tests}, in order to simulate the PAUS+COSMOS photo-$z$s used to train the $z_{\rm s}$+$z_{\rm PAUS}$ and the $z_{\rm s}$+$z_{\rm PAUS}$+NB methods, we scatter the true redshift from the simulations to a similar precision of PAUS+COSMOS photometric redshifts. This process assumes that photometric redshift errors are purely Gaussian; however, real data also have non-Gaussian errors and photo-$z$ outliers, which can be caused by, for example, noisy photometry, emission-line confusions, and other artefacts in the data.}

\LC{Systematic outliers in the ground-truth target redshifts have a much stronger impact than random catastrophic outliers (Figs. \ref{fig:lossoutliers} and \ref{fig:labeloutliers} ). Therefore, we assume the most adverse scenario where all outliers are systematically shifted according to Eq.\eqref{eq:zmod_el}.} 

\LC{Figure \ref{fig:outliers_fs2} shows the impact that outliers in the target redshifts have in the performance of the methods. We studied four different cases: a sample without outliers in the training sample (black solid line), with a 5\% of outliers in the spectroscopic redshifts and 1\% of outliers in the photo-$z$s (red dashed line), 5\% of spectroscopic redshift outliers and 5\% photo-$z$s outliers (blue dashed line), and 5\% of spectroscopic redshift outliers and 10\% photo-$z$s outliers (green dashed line)}.

\LC{Adding the 5\% of spectroscopic redshift outliers already has an impact on the predicted photo-$z$ bias of the baseline method ($z_{\rm s}$, top left panel). The effect of spectroscopic redshift outliers in the training sample is mitigated by the MTL network (top right panel), which is not affected by training-photo-$z$ outliers since these are not used during the training (method $z_{\rm s}$ + NB, Sect.\,\ref{sec:method:arch}). We also observe that training with photo-$z$ samples with up to $\sim$5\% of outliers also mitigates the effect of outliers in the spectroscopic sample. This is expected since adding more training data reduces the relative importance of an outlier in the training sample. However, the bottom plots also show that training samples with more than 10\% of systematic photo-$z$ outliers degrade the photo-$z$ performance.}

\LC{We also studied the robustness of the dispersion and the outlier rate to the target-redshift outliers and these two metrics are much less affected by the presence of target-redshift outliers.}

\newpage
\section{Further studies of multi-task training}
\begin{figure}[h]
\includegraphics[width= 0.48\textwidth]{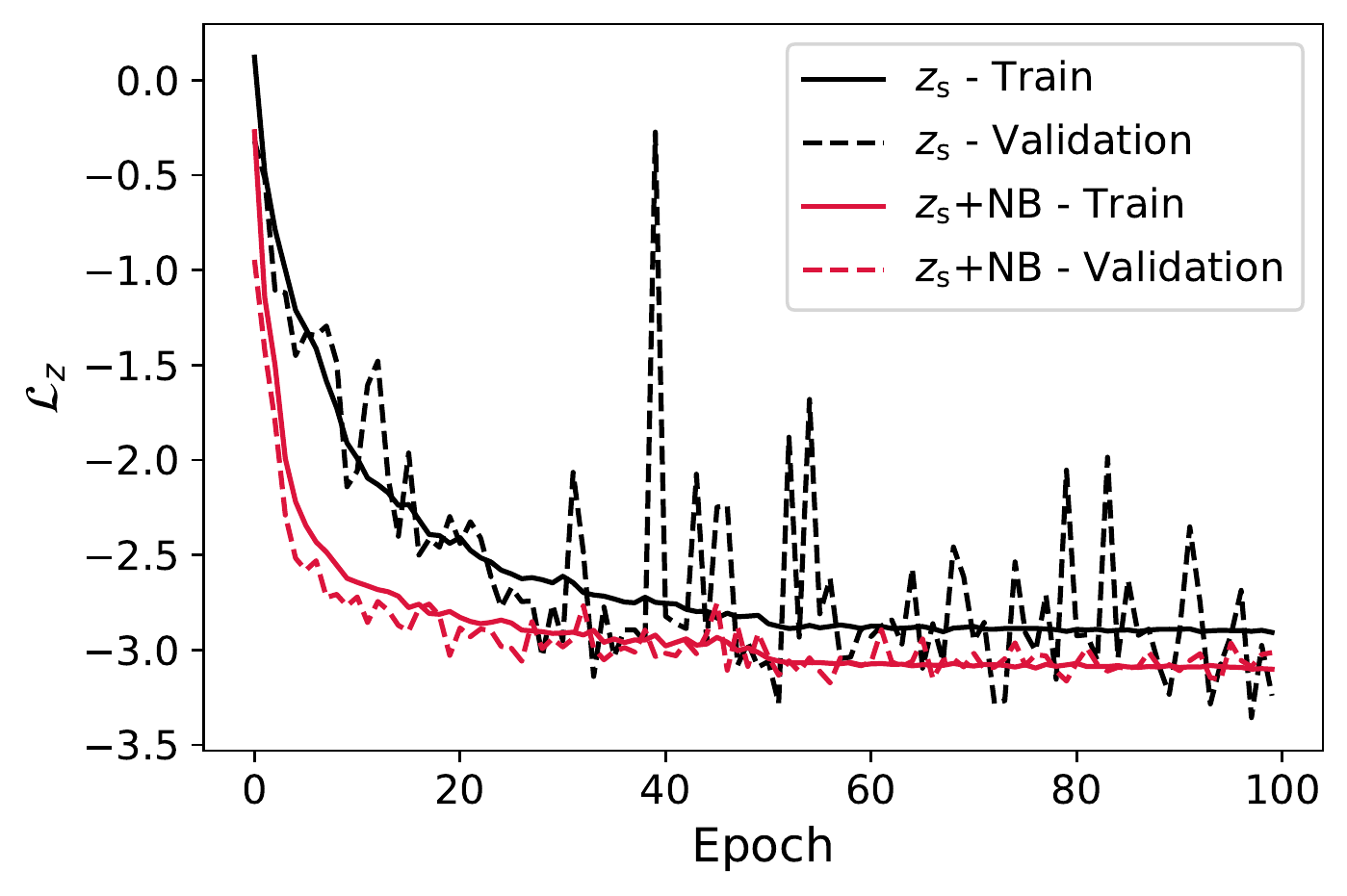}
\centering
\caption{Training (solid lines) and validation (dashed lines) loss for the $z_{\rm s}$ (black) and $z_{\rm s}$+NB (red) methods. All methods are trained for 100 epochs with an initial learning rate of $10^{-3}$ and the same initial conditions.}
\label{fig:lossminimum}
\end{figure}

\LC{In this section we aim to give a more technical view of the network functioning. In Fig. \ref{fig:lossminimum} we show the evolution of the photo-$z$ prediction loss function (Eq. \ref{eq:lossz})  with time for the baseline method ($z_{\rm s}$, black line) and the MTL method ($z_{\rm s}$+NB, red line). In both cases, the solid line corresponds to the training loss, while the dashed line is the validation loss. The networks have been trained for 100 epochs with an initial learning rate of $10^{-3}$, which decreases to $10^{-4}$ after 50 epochs. In this test, unlike previous tests, we initialise the two networks with the same weights. Still we observe training with the two different losses leads to a lower photo-$z$ loss ($\mathcal{L}_{\rm z}$). We also find adding the narrow-band loss stabilises the photo-$z$ loss in the validation sample, meaning the network better generalise with the additional narrow-band loss.}
\end{appendix}
\end{document}